\documentclass[a4paper,fleqn,usenatbib]{mnras}

\usepackage[T1]{fontenc}
\usepackage{ae,aecompl}
\usepackage[center]{caption}
\usepackage{multirow}

\usepackage{graphicx}	
\usepackage{amsmath}	
\usepackage{amssymb}	

\title[Galaxy environment and Multiple Populations]{Role of host galaxy in the formation of multiple stellar populations: Analysis of NGC 1786 and NGC 1898}

\author[V.Viswajith et al.]{
Viswajith Vanaraj$^{1,2}$\thanks{E-mail: vanaraj@uni-potsdam.de}
Florian Niederhofer$^{2}$
Paul Goudfrooij$^{3}$
\\
$^{1}$Universit\"{a}t Potsdam, Institut f\"{u}r Physik und Astronomie, Karl Liebknecht Stra{\ss}e 24/25, 14476 Potsdam, Germany\\
$^{2}$Leibniz-Institut f\"{u}r Astrophysik Potsdam, An der Sternwarte 16, 14482 Potsdam, Germany\\
$^{3}$Space Telescope Science Institute, 3700 San Martin Drive, Baltimore, MD 21218, USA\\
}

\date{Accepted XXX. Received YYY; in original form ZZZ}

\pubyear{2020}

\begin{document}
\label{firstpage}
\pagerange{\pageref{firstpage}--\pageref{lastpage}}
\maketitle

\begin{abstract}
Even after decades of research, the origin of multiple stellar populations in globular clusters remains enigmatic. The question as to whether the galaxy environment plays a role in their formation remains unanswered. To that extent, we analysed two classical ($>$ 10 Gyr old) Large Magellanic Cloud globular clusters namely, NGC 1786 and NGC 1898, using imaging data from \textit{Hubble Space Telescope} to compare and contrast them with ancient Galactic globular clusters to assess systematic differences that might exist between their abundance variations. We calculated their Red Giant Branch width, subtracted the effect of metallicity and compared it with the available data on Galactic globular clusters by plotting them against initial and current cluster mass. We see that the two clusters follow the same general trend as that of the Galactic globular clusters and Galactic globular clusters from different progenitors follow the same general trend as one another, indicating that the galaxy environment may only play a minor role in the formation of multiple stellar populations within globular clusters.
\end{abstract}




\section{Introduction}

Chemical inhomogeneities among stars in globular clusters (GCs) have been known to exist for nearly a century \citep[]{Lindblad1922, Popper1947}. In the last few decades, huge strides made in astronomical instrumentation have enabled us to study these inhomogeneities in great depth using spectroscopy and photometry. It has revealed significant star-to-star abundance variations in light elements such as C, N, Na, O, Al and Mg which has laid to rest the traditional assumption that the GCs were simple stellar populations (SSPs). The inhomogeneities are not random but exhibit certain (anti-)correlations in light elements such as N-C, Na-O and sometimes Al-Mg (anti-)correlations \citep{carretta2009, carretta2015}. This phenomenon is known as ``multiple populations" (MPs) (see \citealt{bastianlardo2018} for a review).

MPs seem to be an inherent property of Galactic GCs older than 10 Gyr though there are numerous exceptions \citep{milone2014, salinasstrader2015, dotter2018, milone2020}. Similar abundance variations have also been found in classical extragalactic GCs within Fornax dSph \citep{Larsen2014}, Large Magellanic Cloud (LMC) \citep{mucciarelli2009,gilligan2019} and Small Magellanic Cloud (SMC) \citep{dalessandro2016, niederhofer2017a}. So, MPs were thought to be exclusive to classical GCs (> 10 Gyr old). But recently, MPs have also been detected in intermediate age GCs (2-6 Gyr) in Magellanic Clouds (MCs). The youngest GCs with confirmed multiple populations are approximately 2 Gyr old \citep{hollyhead2019,martocchia2018b, martocchia2018a,martocchia2021}.

A defining feature of such GCs is that they host two stellar populations- the primordial population or first generation (1G) which consists of stars with a similar chemical composition as that of halo stars with the same metallicity and the secondary population or second generation (2G) which consists of stars enhanced in certain elements exhibiting the above mentioned anti-correlations with respect to the primordial population. 2G stars are also found to be enriched in Helium \citep{zennaro2019, lagioia2018, milone2018}. Some clusters additionally host an extended primordial population indicating Fe abundance variations and presence of binary systems within that population \citep{milone2015b,antona2016,marino2019}.

Though much progress has been made in the study of MPs in the last two decades, their origins remain enigmatic. Several scenarios have been proposed to explain their formation in different environments  \citep{gieles2018,bekki2019}. To understand the origin and formation of MPs, we need to know the impact of various parameters (e.g. age, mass, environment etc.) on the manifestation of MPs. Colour-magnitude diagram (CMD) splits the two populations based on abundance variations depending on the filters used. Filters sensitive to molecular bands corresponding to the above mentioned light elements will be able to discriminate between the two populations due to the underlying abundance variations in those elements. For example, \textit{(m\textsubscript{F336W}--m\textsubscript{F438W})--(m\textsubscript{F438W}--m\textsubscript{F343N})}, which is the colour in the Hubble Space Telescope (HST) WFC3/UVIS filters F336W, F438W and F343N, is useful to identify MPs with different C and N abundances due to the presence of strong NH absorption lines in F336W and F343N filters and CH absorption features in F438W filter \citep{marino2008,david2008,milone2012,milone2013,piotto2015,niederhofer2017a, niederhofer2017b}. Since large colour spreads are observed in the Red Giant Branch (RGB) with constant C$+$N$+$O sum which otherwise could cause split sequences in some filters \citep{pietrinferni2009}, the correlation of RGB width with the global parameters (metallicity, mass, age etc.) is a useful tool to understand the phenomenology of MPs. The difference between RGB width observed in SSPs and MPs is that the observed spread in SSPs can be explained exclusively by observational errors.

\citet{milone2017} analysed 57 Galactic GCs from the \textit{HST UV legacy survey} \citep{piotto2015} and the correlation of their RGB width with various global GC parameters. The RGB width was calculated in the colour \textit{m\textsubscript{F275W}--m\textsubscript{F814W}} and in the pseudo-colour \textit{C\textsubscript{F275W,F336W,F438W}} defined by \textit{(m\textsubscript{F275W}--m\textsubscript{F336W})--(m\textsubscript{F336W}--m\textsubscript{F438W})}. It was found that the intrinsic RGB width correlates with metallicity and after removing the dependence on metallicity, significant correlations between RGB width and mass and luminosity were observed. The role of age and galaxy environment couldn't be discerned since the sample was restricted to classical Galactic GCs older than 11 Gyr \citep{dotter2010}.

\citet{lagioia2019} analysed the above sample along with additional Galactic and extragalactic GCs. The RGB width was calculated in the pseudo-colour \textit{C\textsubscript{F336W,F438W,F814W}}. As noted by them, there is a large number of observations in these bands, in the archives of both HST and ground-based telescopes. Furthermore, this combination of filters requires significantly lesser observation time than the one used by \citet{milone2017} with the same S/N ratio due to the fast decline of UV flux in the RGB stars and the low transmittance of F275W filter, which enables us to observe distant clusters with lesser difficulties. It was found that there was no correlation between intrinsic RGB width and age of the cluster while there was a significant correlation between intrinsic RGB width and mass of the cluster. They also found that intrinsic RGB width and metallicity subtracted RGB width, which they call `normalized' RGB width, of the extragalactic GCs is systematically lower than that of Galactic GCs, indicating that MC GCs might have smaller internal light element abundance variations than Galactic GCs. However, the 7 extragalactic GCs used by \citet{lagioia2019}, with the exception of NGC 121 are much younger than Galactic GCs. In order to learn about the role played by the host galaxy environment in the manifestation of MPs, we need to analyse extragalactic GCs in the same age range as that of Galactic GCs so that we can study the effect of their respective host galaxies. Hence, in this study, we analyse two classical LMC GCs, namely, NGC 1898 and NGC 1786 and look at the correlation between their normalized RGB width in pseudo-colour \textit{C\textsubscript{F336W,F438W,F814W}} and mass to compare them with Galactic GCs and see if they exhibit systematic abundance variations.

This paper is divided as follows: in section \ref{section2}, we describe the dataset and the photometric data analysis techniques; in section \ref{section3}, we explain the procedure to derive the intrinsic RGB width and the associated error of the GCs; in section \ref{section4}, we look at the normalized RGB width and their correlation with mass and compare them with the Galactic GCs; in section \ref{section5}, we discuss our results in association with other studies; in section \ref{section6}, we present the summary of our results and draw final conclusions.

\section{Observations and Data Reduction} \label{section2}

\subsection{Observations} \label{section21}
\paragraph*{} We retrieved observations of NGC 1898 and NGC 1786 from the HST MAST Archive collected with WFC3/UVIS in the UV band F336W and in the optical bands F438W and F814W. These bands are analogous to U, B and I bands in Johnsons-Cousins respectively. We have also used the observations of NGC 121 and Lindsay 1 used by \citet{lagioia2019} to aid in comparison of our RGB width values considering that there are a few differences (photometry pipeline, PSF computation, quality cuts for selection of stars etc.) between our method and those employed by them. A brief summary of the observations is provided in Table \ref{tab:table1}.
\begin{table*}
	\caption{Observation Data Set}
	\label{tab:table1}
	\begin{tabular}{|l|c|c|c|l|c}
		\hline
		\textbf{Cluster} & \textbf{Date} & \textbf{Camera} & \textbf{Filter} & \textbf{No. x Exposure time (s)} & \textbf{Proposal} \\
		\hline
		NGC 1898 & 2014 Jan 10 & WFC3/UVIS & F336W & $2\times 1035$ & 13435 \\
		& & & F438W & $2\times 200$ & \\
		& & & F814W & 100 & \\
		NGC 1786 & 2014 Jun 13 & WFC3/UVIS & F336W & $2\times 1015$ & 13435 \\
		& & & F438W & $2\times 200$ & \\
		& & & F814W & 100 & \\
		NGC 121 & 2014 May 16, 2014 Oct 16 & WFC3/UVIS & F336W & $4\times 1061$ & 13435 \\
		& & & F438W & $4\times 200$ & \\
		& & & F814W & $2\times 100$ & \\
		Lindsay 1 & 2003 Jul 11, 2005 Aug 21 & ACS/WFC & F555W & $2\times 20 + 480 + 4\times 496$ & 9891, 10396 \\
		& & & F814W & $2\times 10 + 290 + 4\times 474$ & \\
		& 2014 June 19 & WFC3/UVIS & F336W & $500 + 4\times 1200$ & 14069 \\
		& & & F438W & 500 + 800 + 1650 + 1850 & \\
		& & & F343N & $120 + 2\times 460$ & \\
		\hline
		
	\end{tabular}
\end{table*}

%
%
\subsection{Data Reduction using DOLPHOT} \label{section22}
We use DOLPHOT 2.0 \footnote{\url{http://americano.dolphinsim.com/dolphot/}} \citep{dolphin2000} to preprocess the \textit{flt} images and run Point Spread Function (PSF) photometry. \textit{flt} images are debiased and flat-field corrected multi-extension \textit{fits} images. DOLPHOT provides pre-computed PSFs for every filter of all cameras onboard HST, pixel area maps (PAMs) and also ready-made modules specific for each camera. It also enables us to perform photometry on images obtained from different instruments simultaneously which was useful in the analysis of Lindsay 1 (see Section \ref{section26}). From here on, we summarise the steps involved in preprocessing and PSF photometry of WFC3/UVIS images performed using DOLPHOT, described in detail in the general DOLPHOT manual and DOLPHOT/WFC3 manual. The preprocessing of ACS/WFC images are similar but with different task names, as described in DOLPHOT/ACS manual. As such, the images from each camera must be preprocessed separately.

\textit{wf3mask} is used to mask the pixels flagged as bad in the data quality extension of the FITS file (UVIS images are 6-extension files) and multiply by the PAM to render the images in the units of electrons. PAM corrects for the geometric distortion in \textit{flt} images and matches their output count to those of drizzled images. Then \textit{splitgroups} task is used to convert the images into single-extension FITS files by splitting the images into UVIS 1 and UVIS 2 chips. This is necessary since we use a drizzled image as our reference image. This is followed by employing \textit{calcsky} task with the recommended parameters in the DOLPHOT manual for obtaining sky images. This is followed by initiating the DOLPHOT routine for PSF photometry. In the parameter file for the routine, we include the split single extension FITS images and include the drizzled reference image. We have consistently used the F438W drizzled image as the standard drizzled image. The image alignment is carried out by setting \textit{UseWCS=1} and the Charge Transfer Efficiency (CTE) correction for the images is done by setting \textit{WFC3useCTE=1}. The rest of the parameters are set according to the recommendations in the manual. The magnitudes are calibrated to the VEGAMAG system.

\subsection{Selection of stars} \label{section23}
\begin{figure*}
	
	\centering
	\includegraphics[width=0.325\linewidth, height=1\columnwidth]{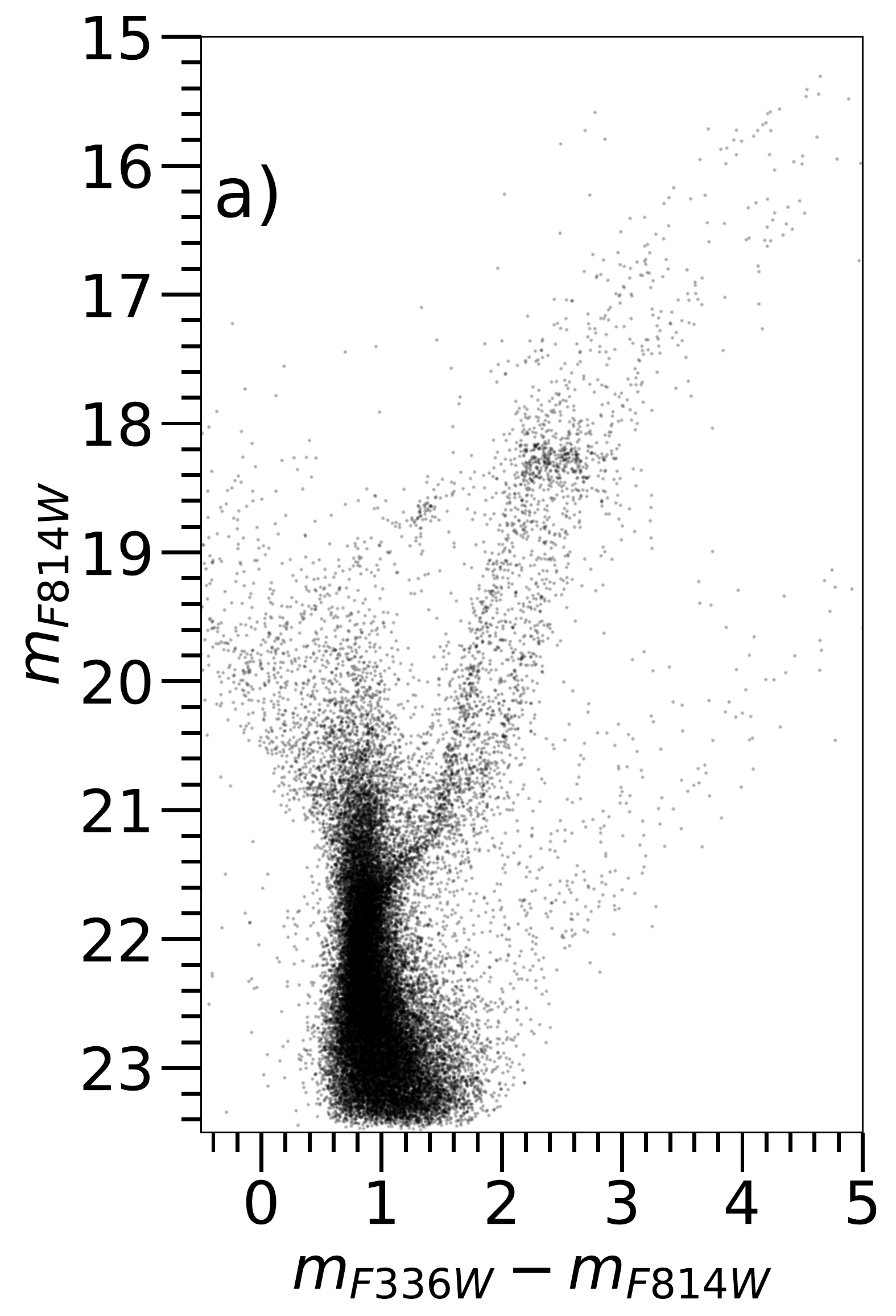}
	\centering
	\includegraphics[width=0.325\linewidth, height=1\columnwidth]{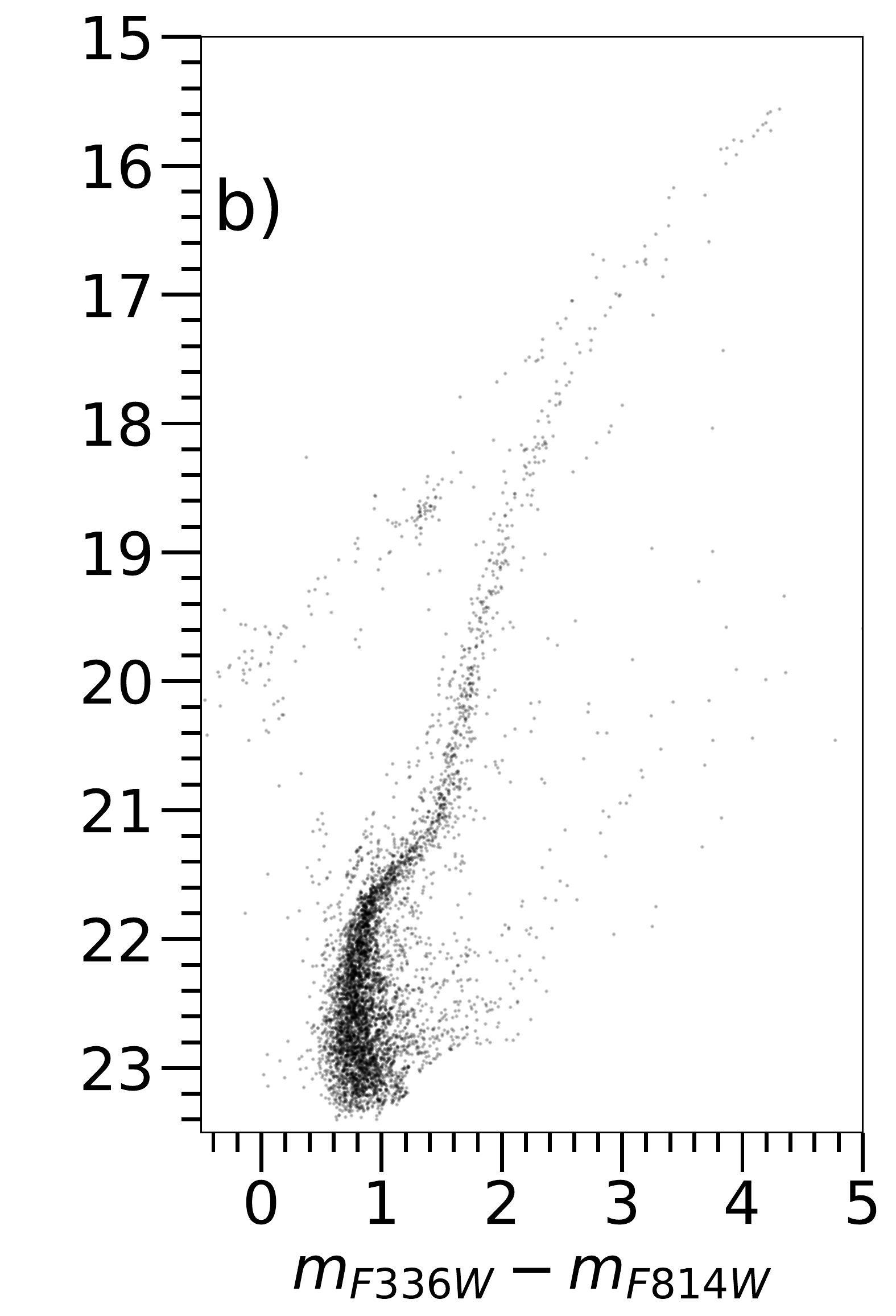}
	\centering
	\includegraphics[width=0.325\linewidth, height=1\columnwidth]{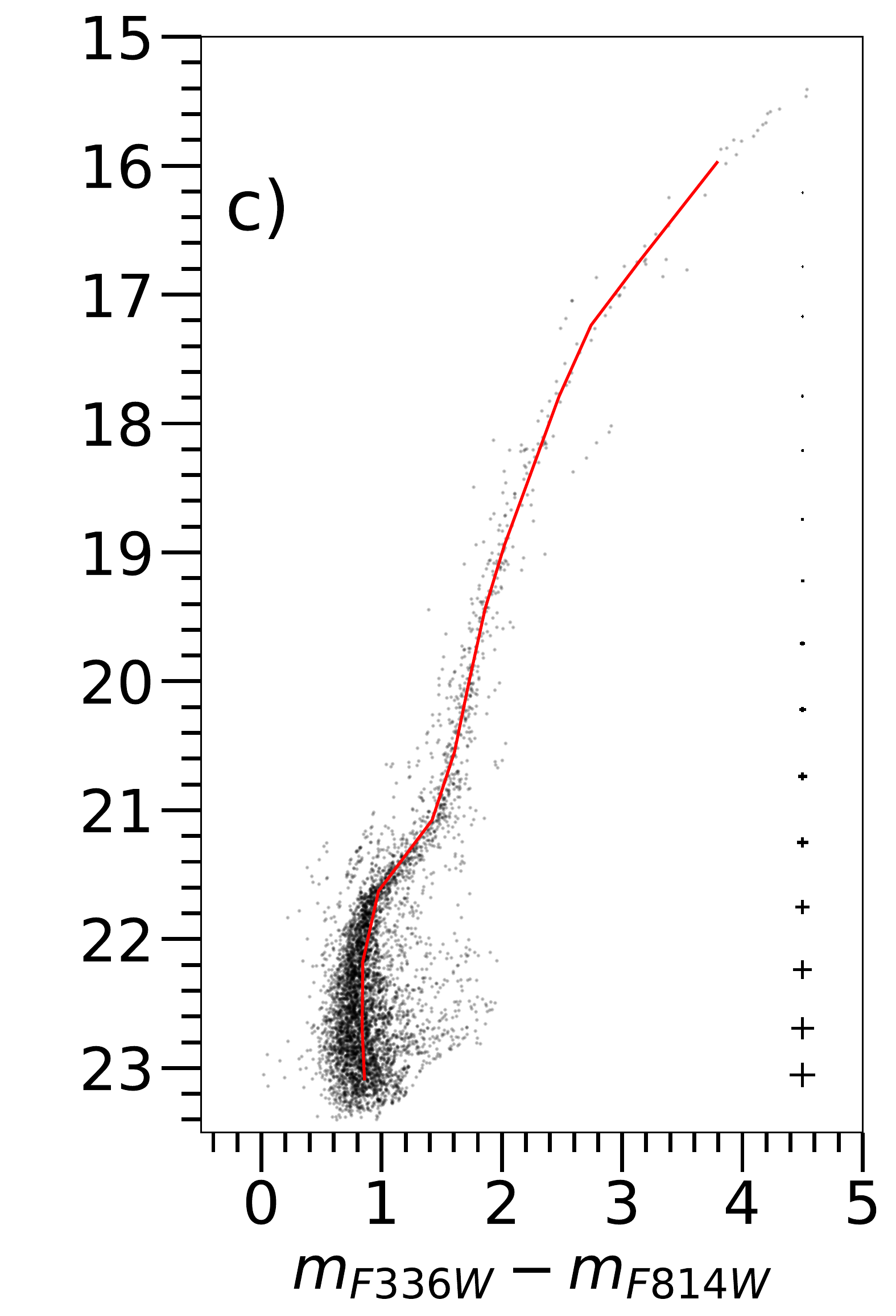}
	\caption{\textit{m\textsubscript{F814W}} vs \textit{m\textsubscript{F336W}$-$m\textsubscript{F814W}} CMD of NGC 1898. \textit{(a):} CMD containing stars selected through quality cuts. \textit{(b):} CMD of the cluster core region after field-star subtraction. \textit{(c):} same as (b) after visual selection using TOPCAT with the fiducial line fit along which artificial stars were generated. The photometric error bars are represented on the right side.}
	\label{fig:resol}
\end{figure*}

\begin{figure*}
	
	\centering
	\includegraphics[width=0.325\linewidth, height=1\columnwidth]{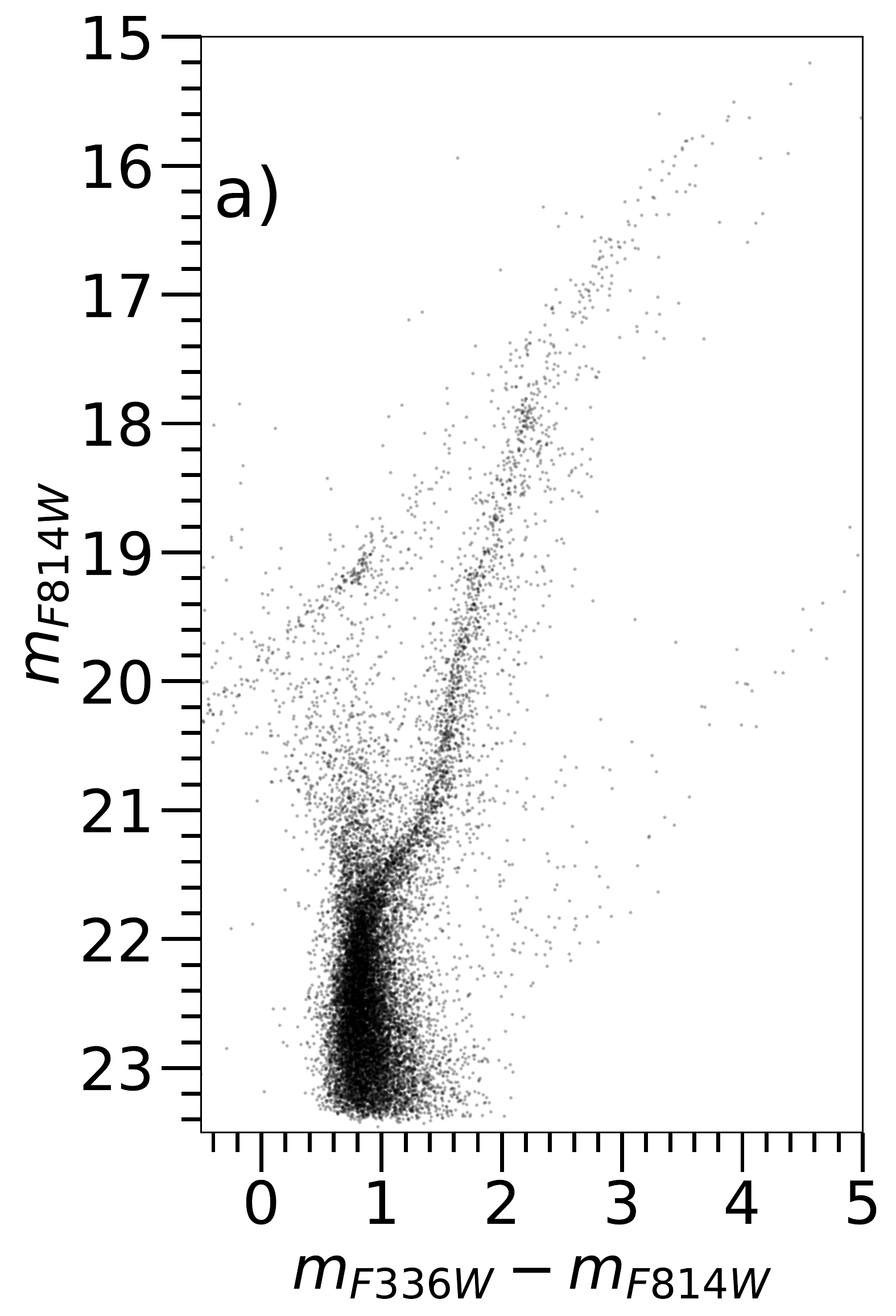}
	\centering
	\includegraphics[width=0.325\linewidth, height=1\columnwidth]{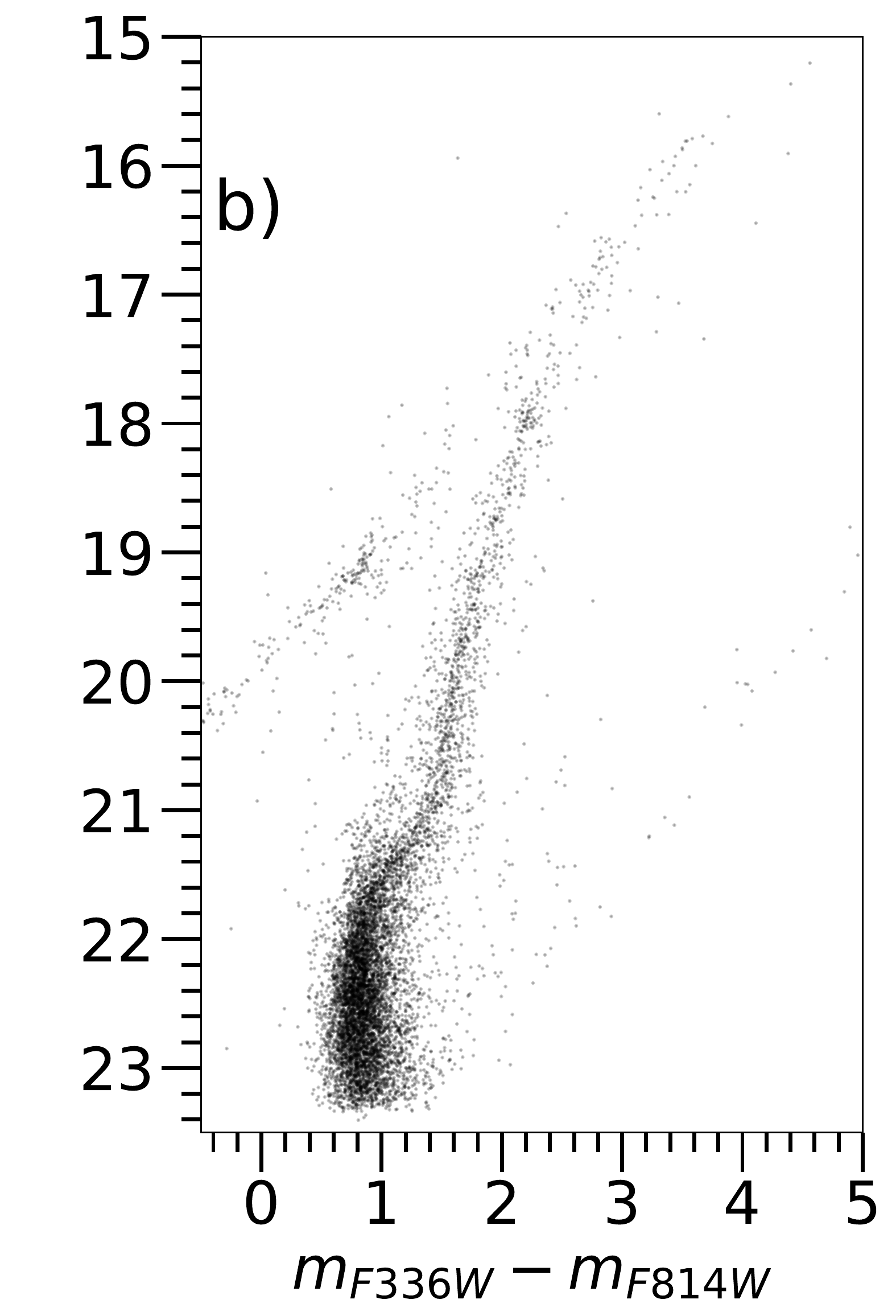}
	\centering
	\includegraphics[width=0.325\linewidth, height=1\columnwidth]{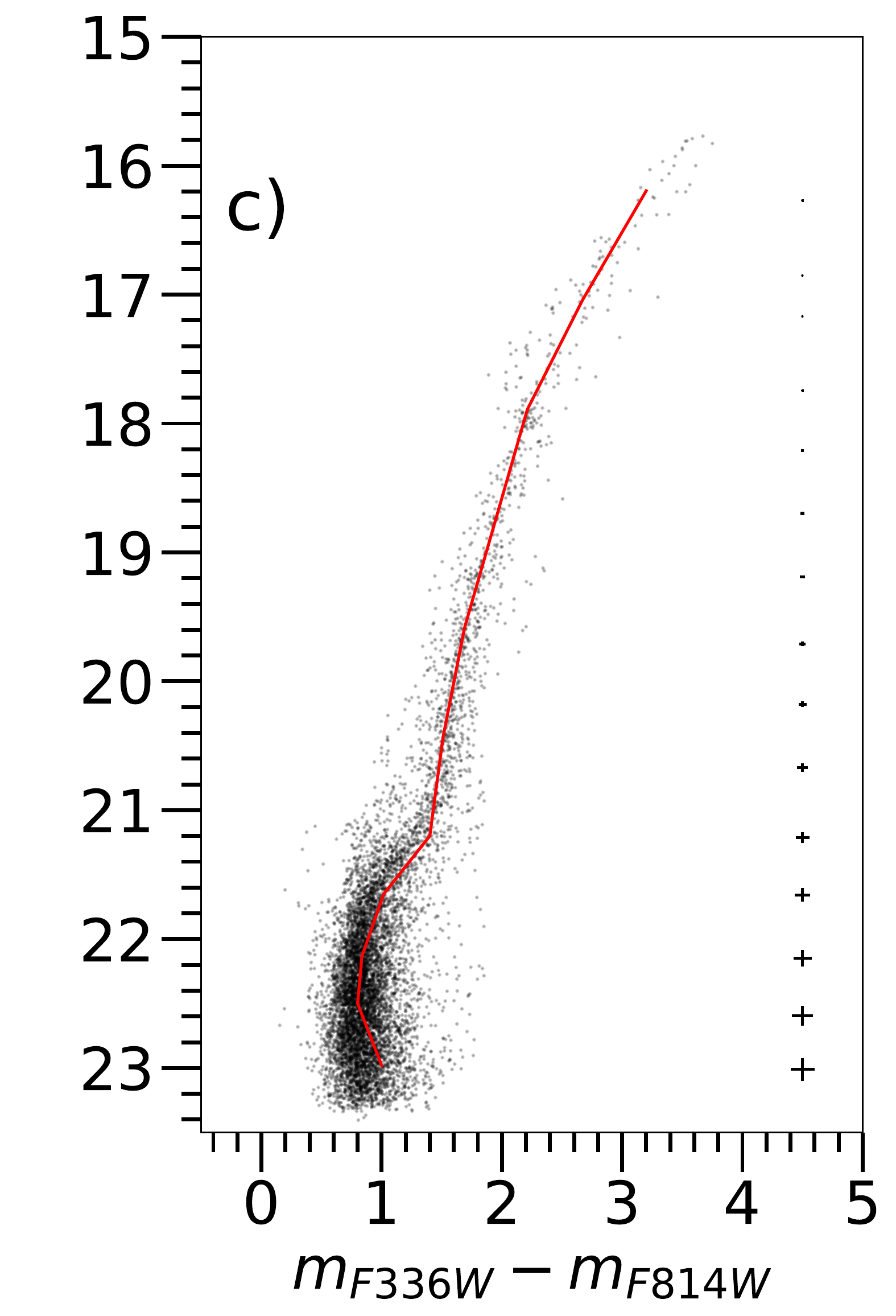}
	\caption{Same as Figure \ref{fig:resol}, for NGC 1786.}
	\label{fig:resol1}
\end{figure*}

\paragraph*{} The output is subject to various quality cuts. We selected based on object type, selecting those stars that are classified as good and those too faint for PSF determination, leaving out those that are elongated, extended or too sharp. We also selected based on error flags, selecting only those stars that have been extremely well recovered from the image and those with photometry apertures extending off the chips. This was followed by filtering out the stars that have a magnitude uncertainty greater than 0.1 mag and with sharpness that lies outside the $\pm 0.1$ interval. Beyond that, we only select those stars that are detected in all the three filters used in this study. Panel (a) of Figures~\ref{fig:resol} and~\ref{fig:resol1} shows the stars selected through quality cuts. This is followed by a statistical removal of field stars. We defined a rectangular reference field region having the same sky area as the circular cluster core region (722 pixel radius for both the clusters) and residing at the edge of the field-of-view, as far away as possible from the cluster area. The reason for choosing equal areas is that the distribution of field stars is assumed to be homogenous and hence its density will be the same across the image. Since the tidal radii of the clusters are greater than 20 pc ($\approx$20 pc for NGC 1786 and $\approx$35 pc for NGC 1898, \citet{mclaughlin2005} assuming a King's profile) and the field of view of WFC3/UVIS is $\approx$38 pc (assuming a distance of 50 kpc to the LMC), a certain fraction of stars is still present that is subtracted as field stars. But, as \citet{niederhofer2015} points out, it is not a serious issue as the oversubtraction mostly affects the well populated regions of the CMD and keeps the overall structure unchanged. We make use of \textit{m\textsubscript{F336W}} vs \textit{m\textsubscript{F336W} $-$ m\textsubscript{F814W}} CMD for field star subtraction. For every star found in the reference field CMD, the star nearest to it in the cluster CMD was eliminated if the difference in magnitude and colour between the two stars was lesser than 0.5 mag and 0.25 mag respectively. Panel (b) of figures~\ref{fig:resol} and~\ref{fig:resol1} shows the \textit{m\textsubscript{F814W}} vs \textit{m\textsubscript{F336W} $-$ m\textsubscript{F814W}} CMD of the cluster core region after field-star removal. Quality cuts and statistical subtraction of field stars don't get rid of undesirable stars entirely altogether. Certain stars to the red of the RGB branch were recovered in the photometry performed on artificial stars (see section \ref{section24}) in the same regions of the CMD as that of the observed stars but on visual inspection of images, were found to be problematic. Figures \ref{fig:vred} and \ref{fig:vredd} show some of the outliers in the RGB marked with red dots. Panel (a) in both figures show the location of these stars before the field star subtraction. It could be seen clearly that field stars form a second RGB diverging out of Sub-Giant Branch. Panel (b) shows that some of the very red stars are retained after the field star subtraction. Panel (c) is the CMD of artificial stars and it could be seen that there are outlying RGB stars at roughly the same colours and magnitudes as in the panel (b) which indicates that they agree with each other. When looked at their position on the images, some of the stars were placed in ring shaped artefacts, some of them were placed in the diffraction spikes of brighter stars and the rest of the stars, though look apparently distinguished in the image, are considerably off the fiducial sequence of the cluster. So, to be consistent and prevent overestimation, we selected through visual inspection using \textit{TOPCAT} only those stars that lie along and close to the fiducial sequence of the CMD in the observed data (See \ref{section24} for artificial stars). For Lindsay 1, we eliminated only a single star, after looking at its position on the observational CMD and the lack of any stars in that region of the synthetic CMD generated after the photometry performed on artificial stars and hence ruled it to be a field star.

\subsection{Artificial stars} \label{section24}

\begin{figure*}
\centering
\includegraphics[width=0.325\linewidth, height=1\columnwidth]{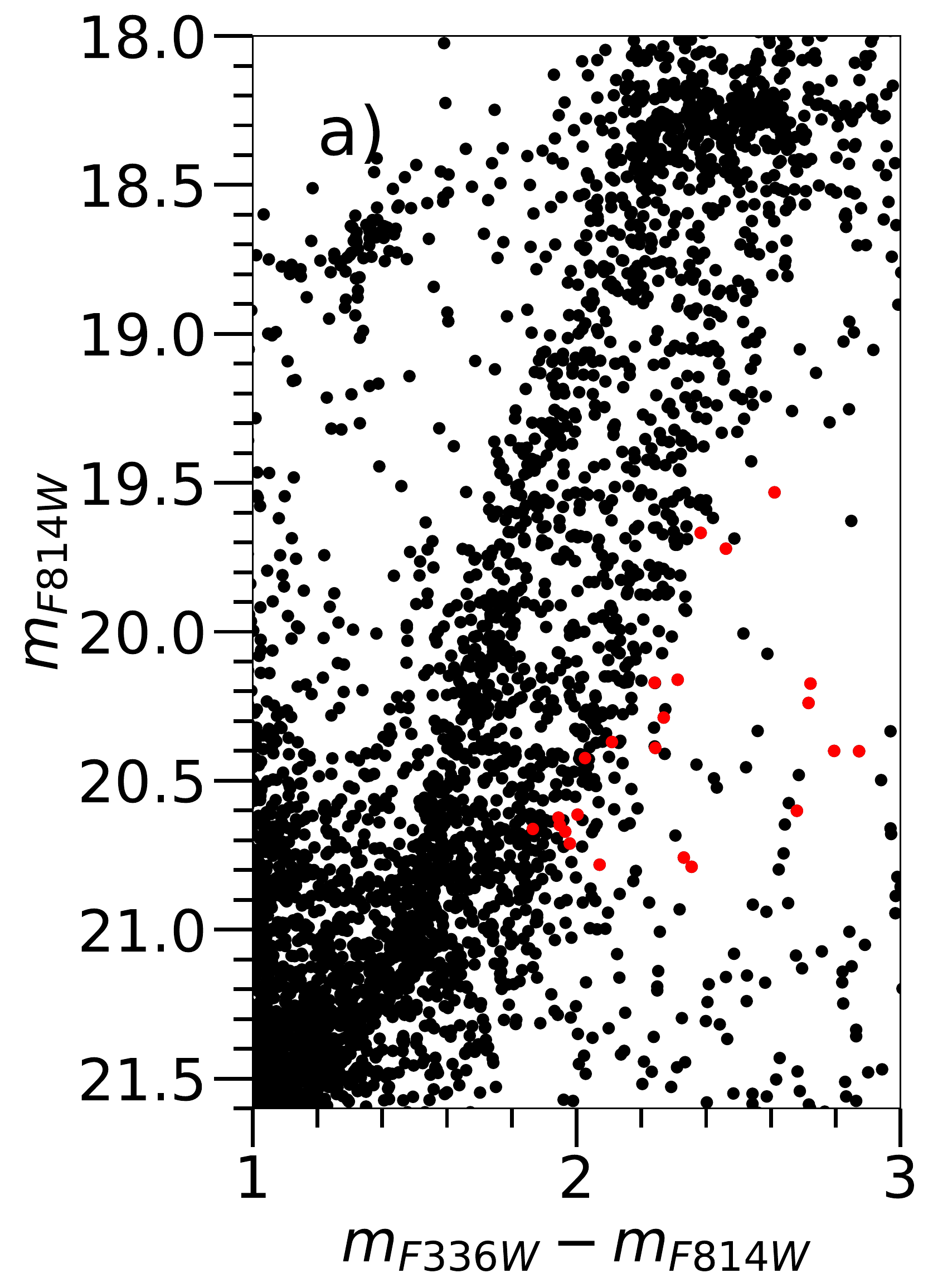}
\centering
\includegraphics[width=0.325\linewidth, height=1\columnwidth]{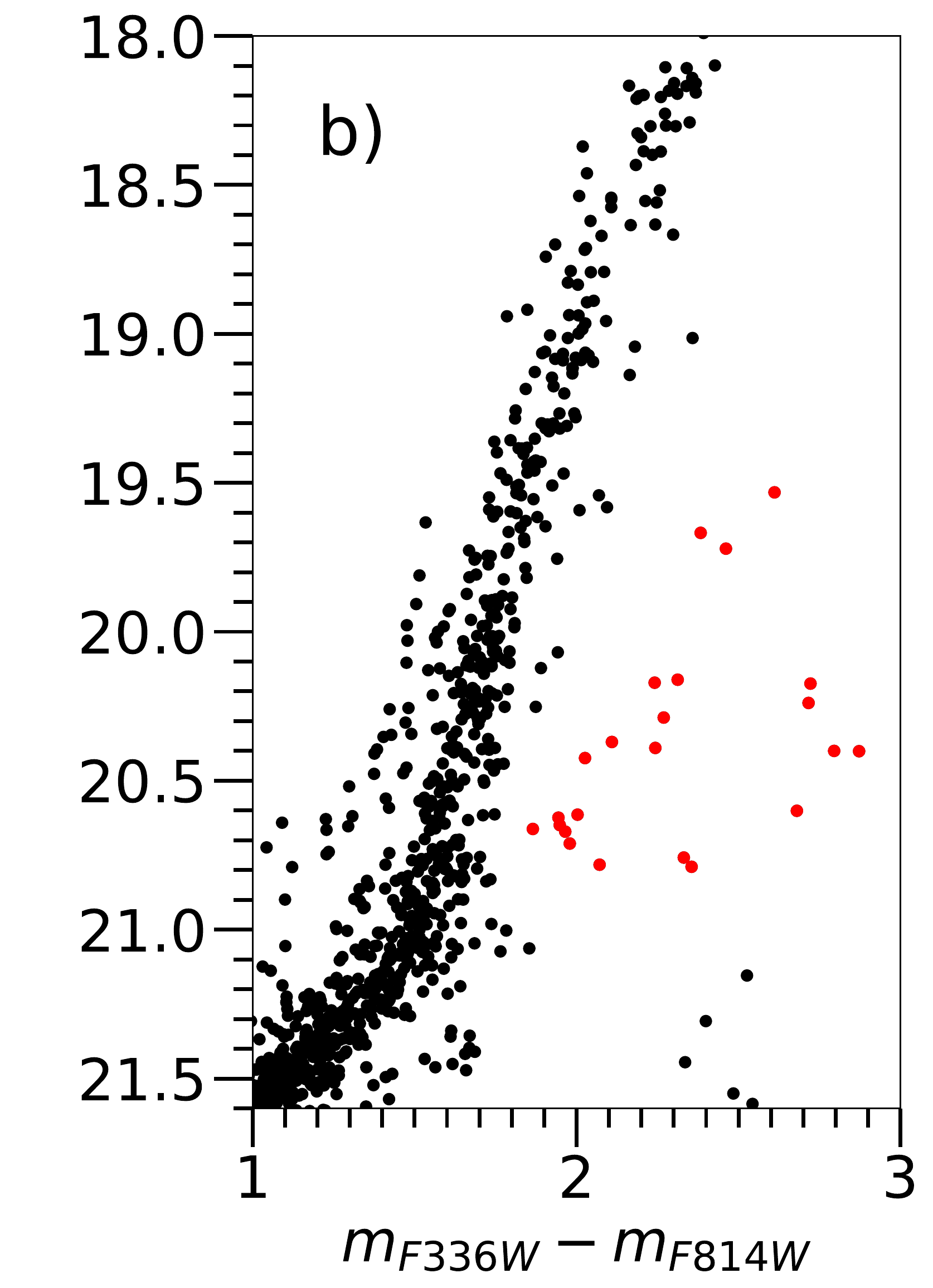}
\centering
\includegraphics[width=0.325\linewidth, height=1\columnwidth]{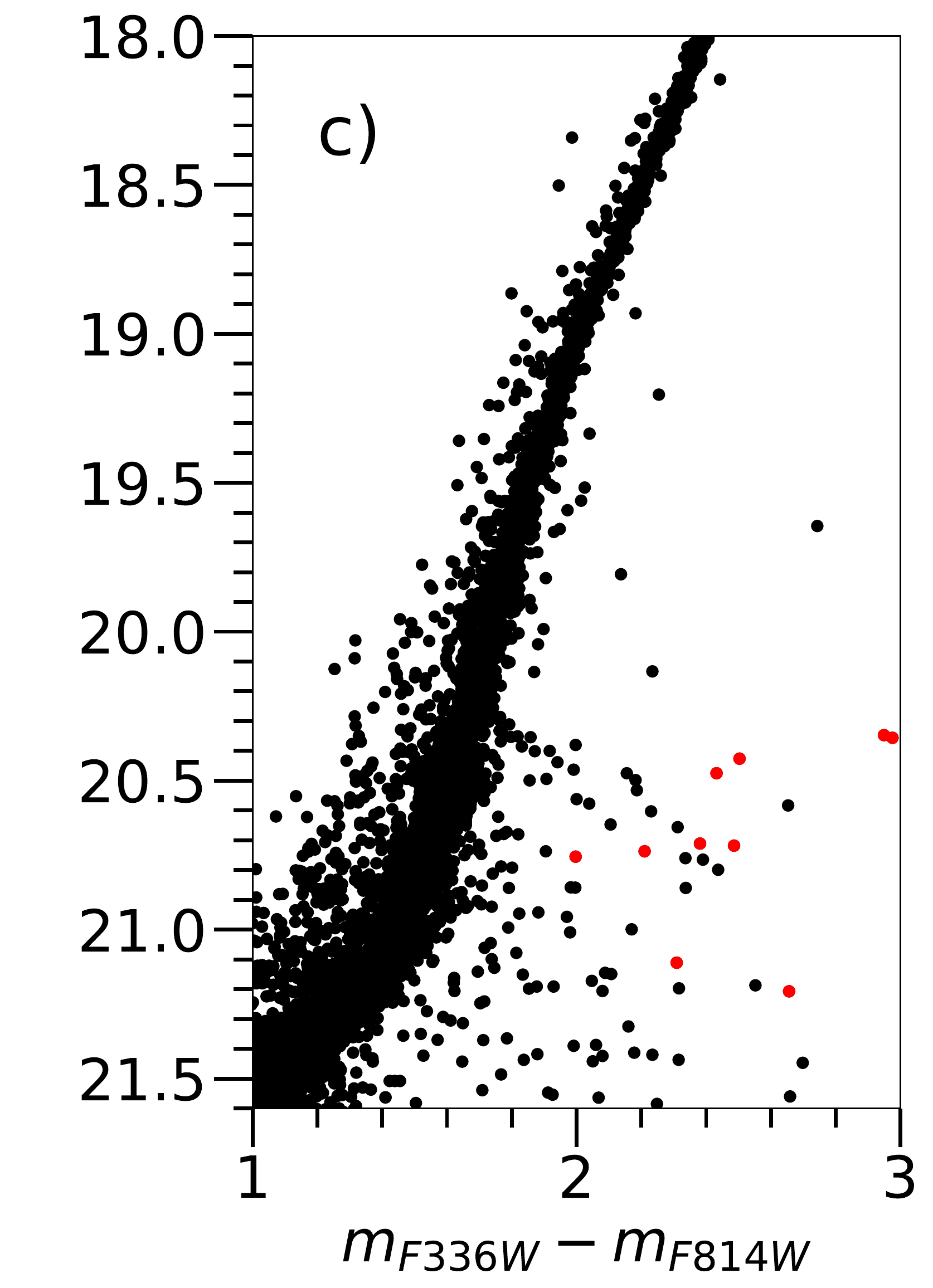}
\caption{\textit{m\textsubscript{F814W}} vs \textit{m\textsubscript{F336W}$-$m\textsubscript{F814W}} CMD of NGC 1898. \textit{(a):} Enlarged section showing the lower RGB before field star subtraction. We can see two strands of RGB diverging from sub-giant branch. Eliminated stars are marked in red \textit{(b):} RGB branch after field-star subtraction. \textit{(c):} CMD of the artificial stars with the eliminated stars marked in red.}
\label{fig:vred}
\end{figure*}

\begin{figure*}
	\centering
	\includegraphics[width=0.325\linewidth, height=1\columnwidth]{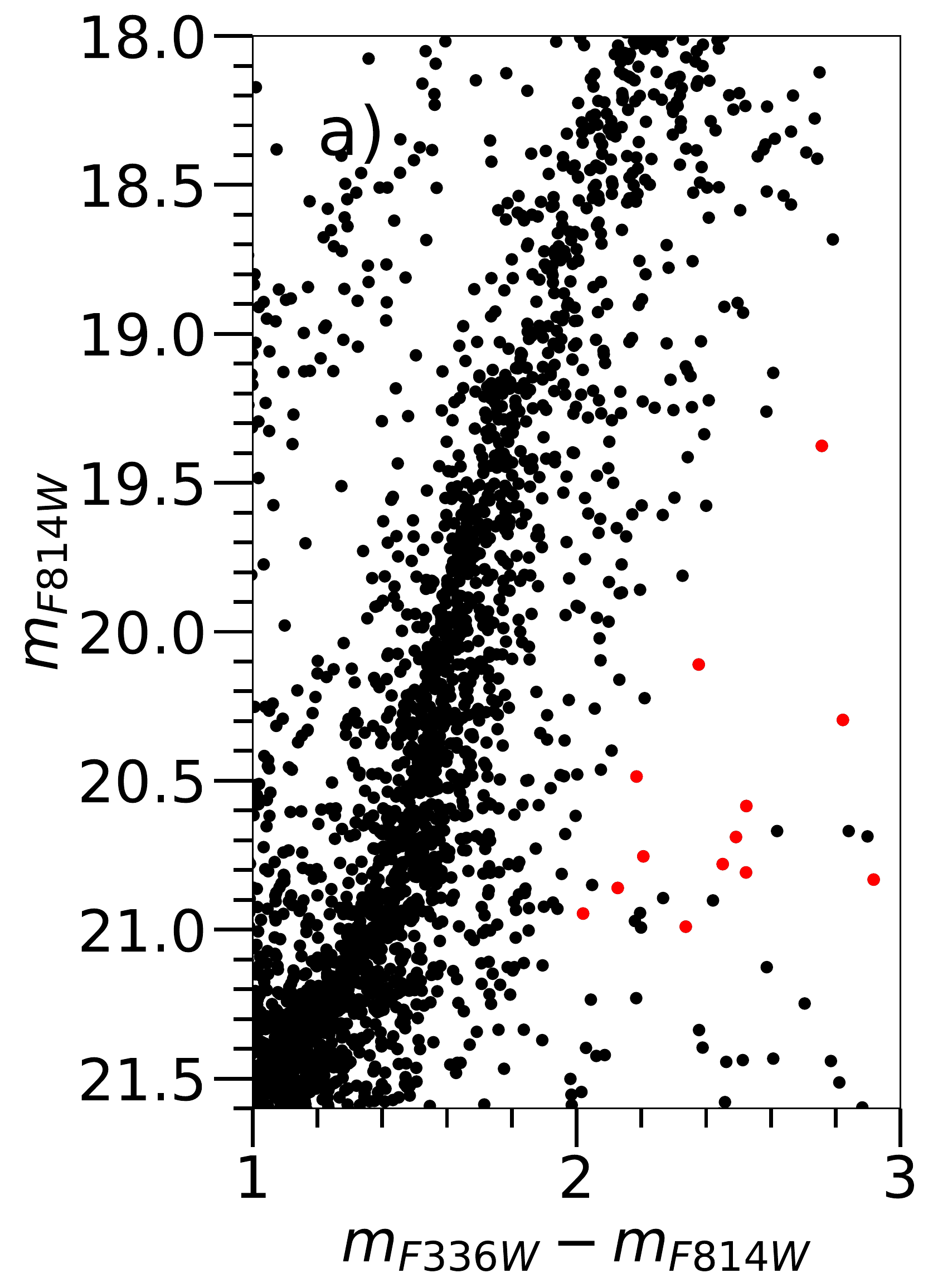}
	\centering
	\includegraphics[width=0.325\linewidth, height=1\columnwidth]{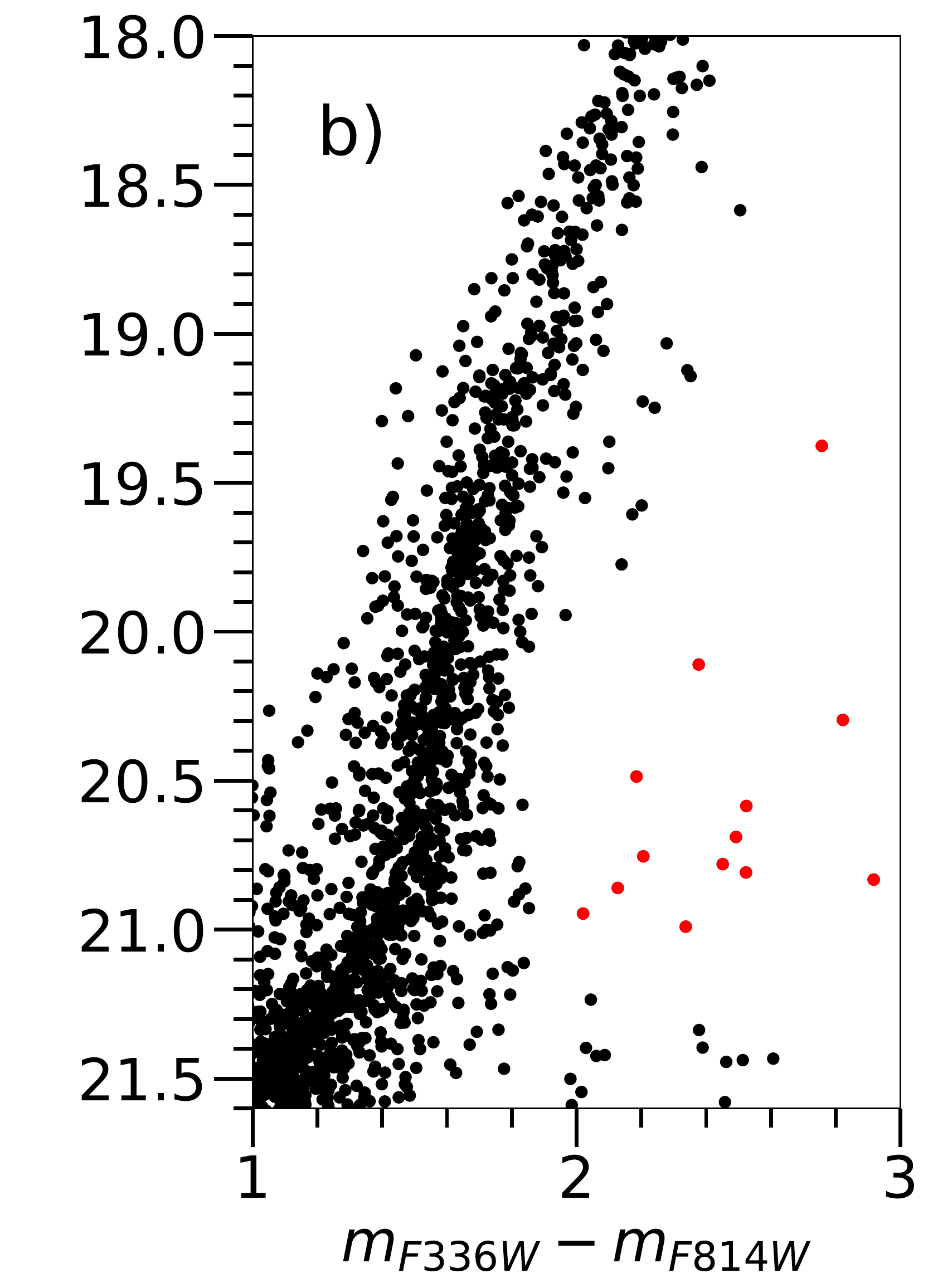}
	\centering
	\includegraphics[width=0.325\linewidth, height=1\columnwidth]{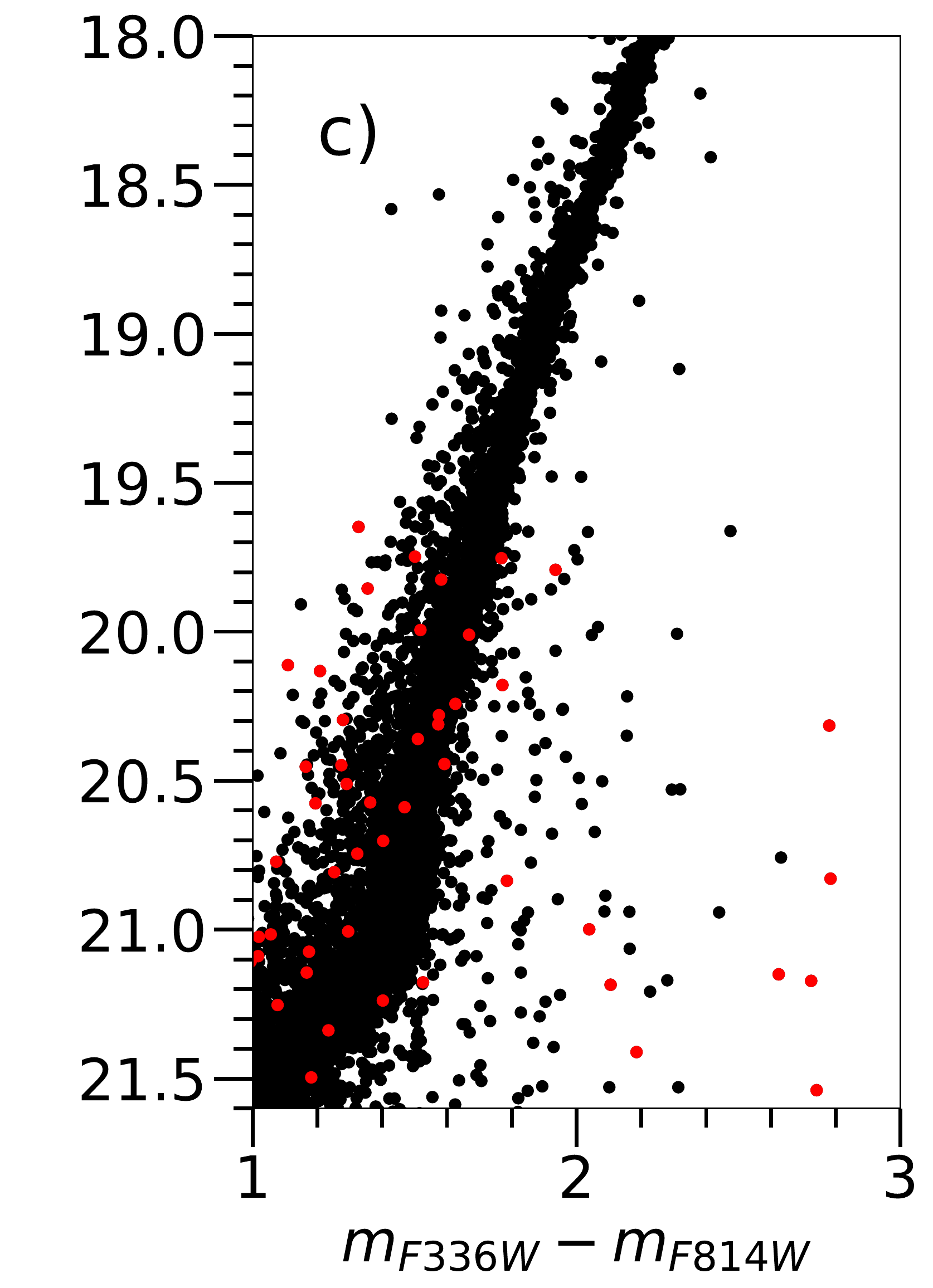}
	\caption{Same as Figure \ref{fig:vred}, for NGC 1786.}
	\label{fig:vredd}
\end{figure*}

\paragraph*{} Artificial stars (ASs) were used to produce synthetic pseudo-colour vs magnitude diagrams, compare them to observational CMDs and also estimate the photometric error. We follow the method described in detail in \citet{anderson2008} and is described in summary here. We fit a fiducial line on the cluster CMD and generate 50,000 stars with random magnitudes that lie on the fiducial line, with proportionate distribution in each bin. Here, for every random F814W magnitude, we generate a F336W magnitude and F438W magnitude that lie on the fiducial line of F336W$-$F814W vs F814W and F438W$-$F814W vs F814W CMD respectively. The coordinates of the stars are generated by defining a circular cluster area and generating coordinates within the circumference of the circle. We introduced an offset of 50 units on each axis of the coordinate map relative to the observational CMD. The artificial stars are added to the science images one at a time to prevent crowding. The photometry has been performed with the same parameters as for the real images and the same quality cuts were applied as well. The CMD of artificial stars too show outliers as shown in panel (c) of figures \ref{fig:vred} and \ref{fig:vredd}. Due to the huge number of stars in the artificial catalogue, elimination through visual inspection proved to be difficult. So, we eliminated the stars whose difference in output and input magnitude was greater than 1.5 mag in any of the three filters. Though this doesn't eliminate all of the undesirable stars, it has the advantage of getting rid of a sizeable number of them without eliminating a significant number of well measured stars.

\subsection{Differential Reddening Correction and PSF zero point variation} \label{section27}

\begin{figure*}
	
	\centering
	\includegraphics[width=0.24602\linewidth, height=1.1\columnwidth]{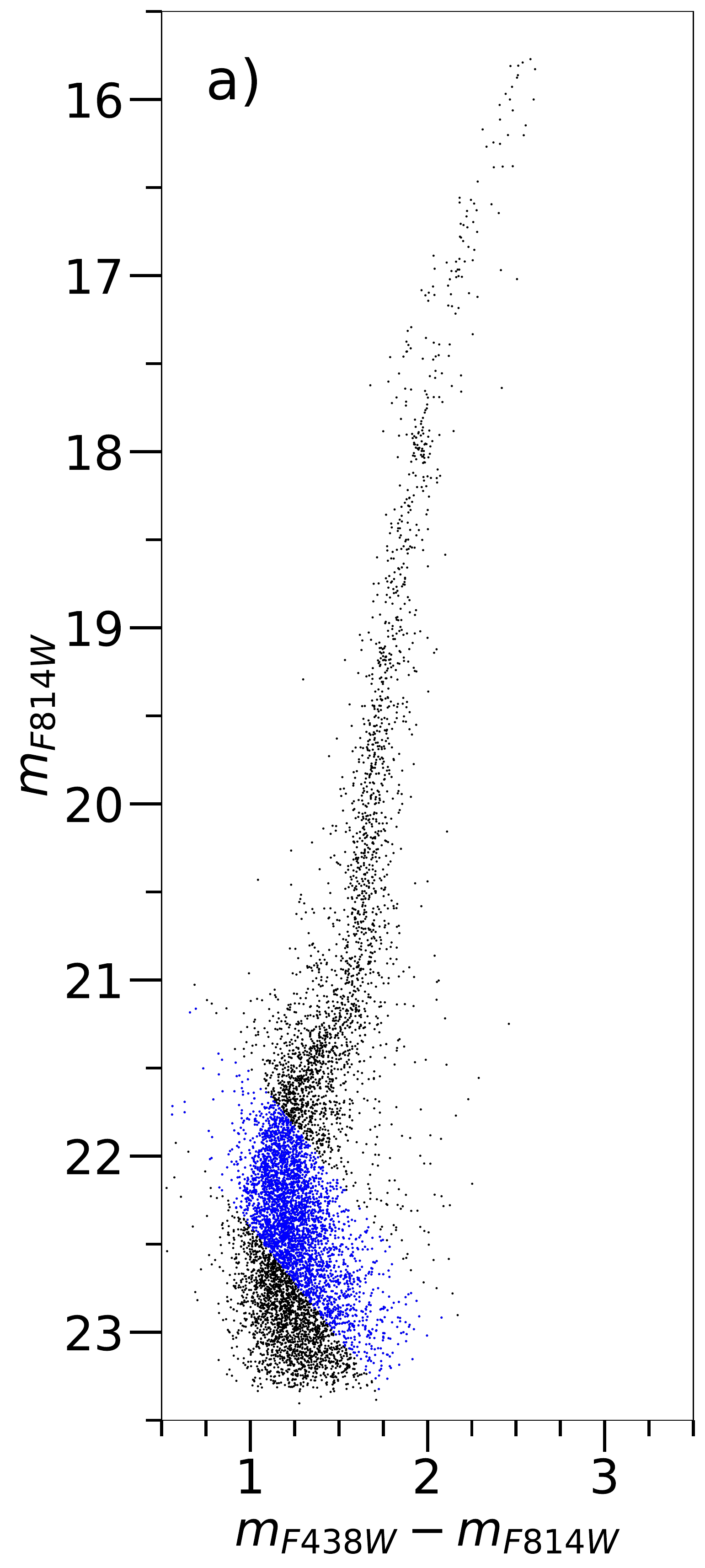}
	\centering
	\includegraphics[width=0.24602\linewidth, height=1.1\columnwidth]{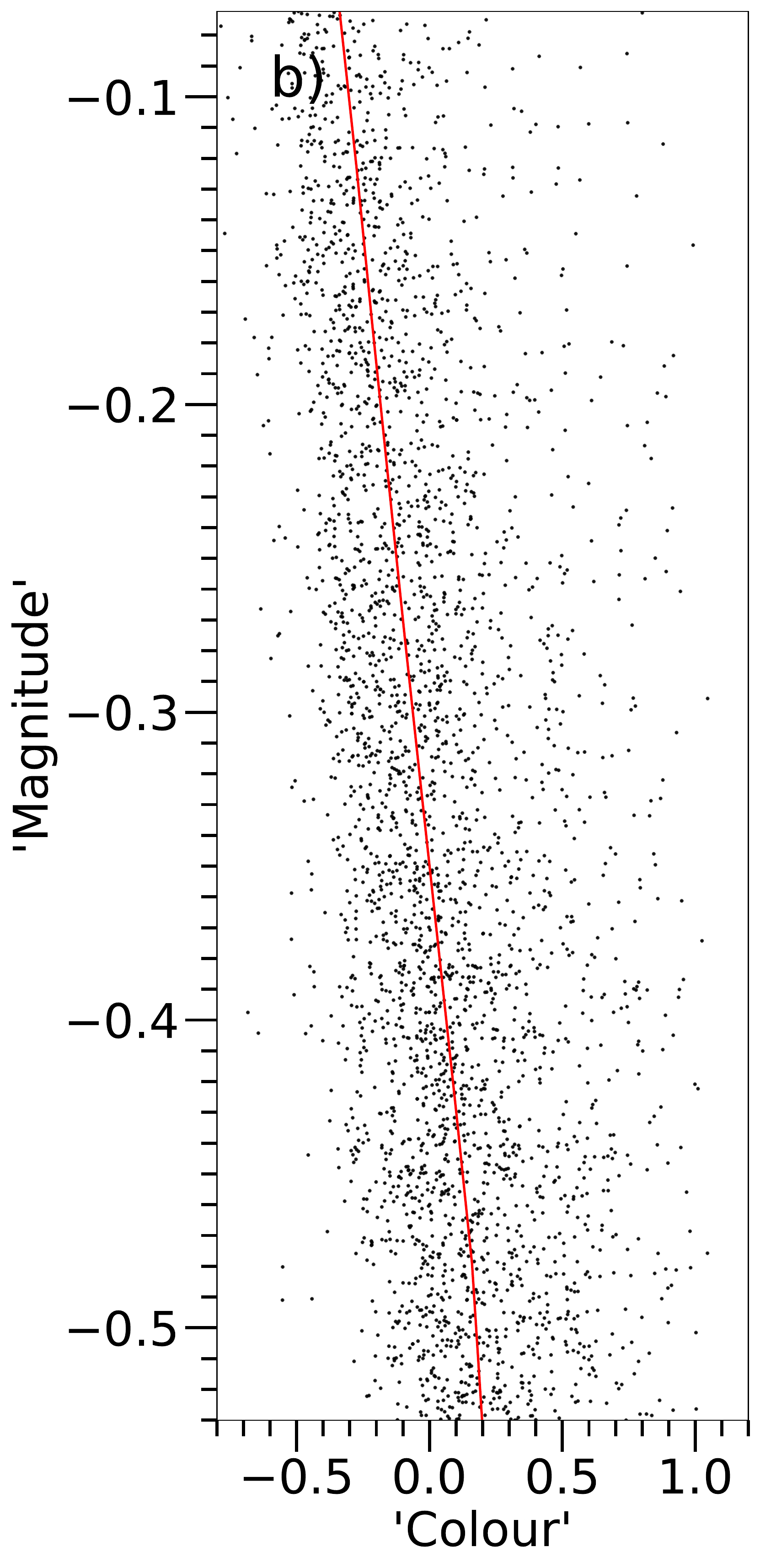}
	\centering
	\includegraphics[width=0.24602\linewidth, height=1.1\columnwidth]{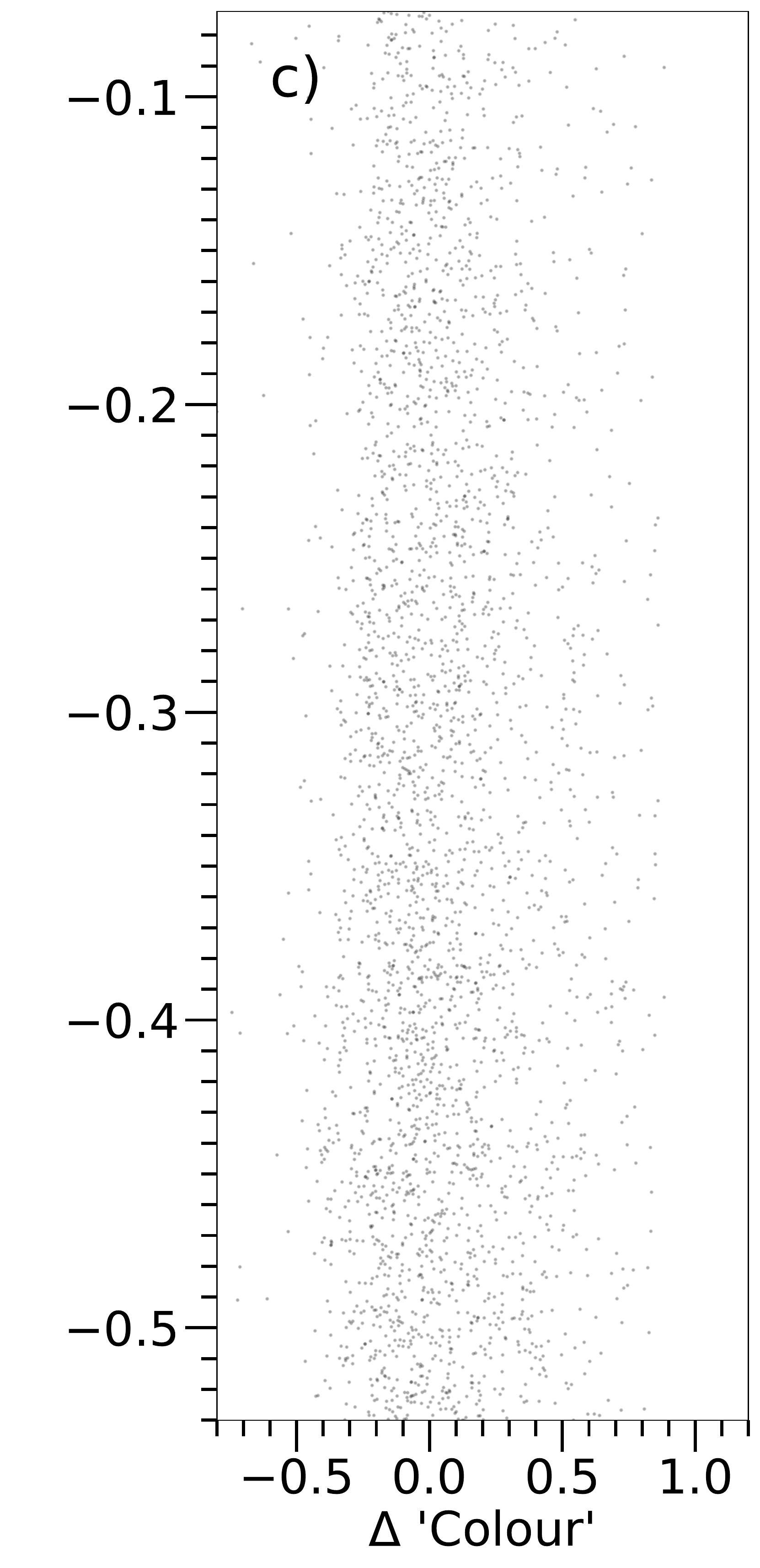}
	\centering
	\includegraphics[width=0.24602\linewidth, height=1.1\columnwidth]{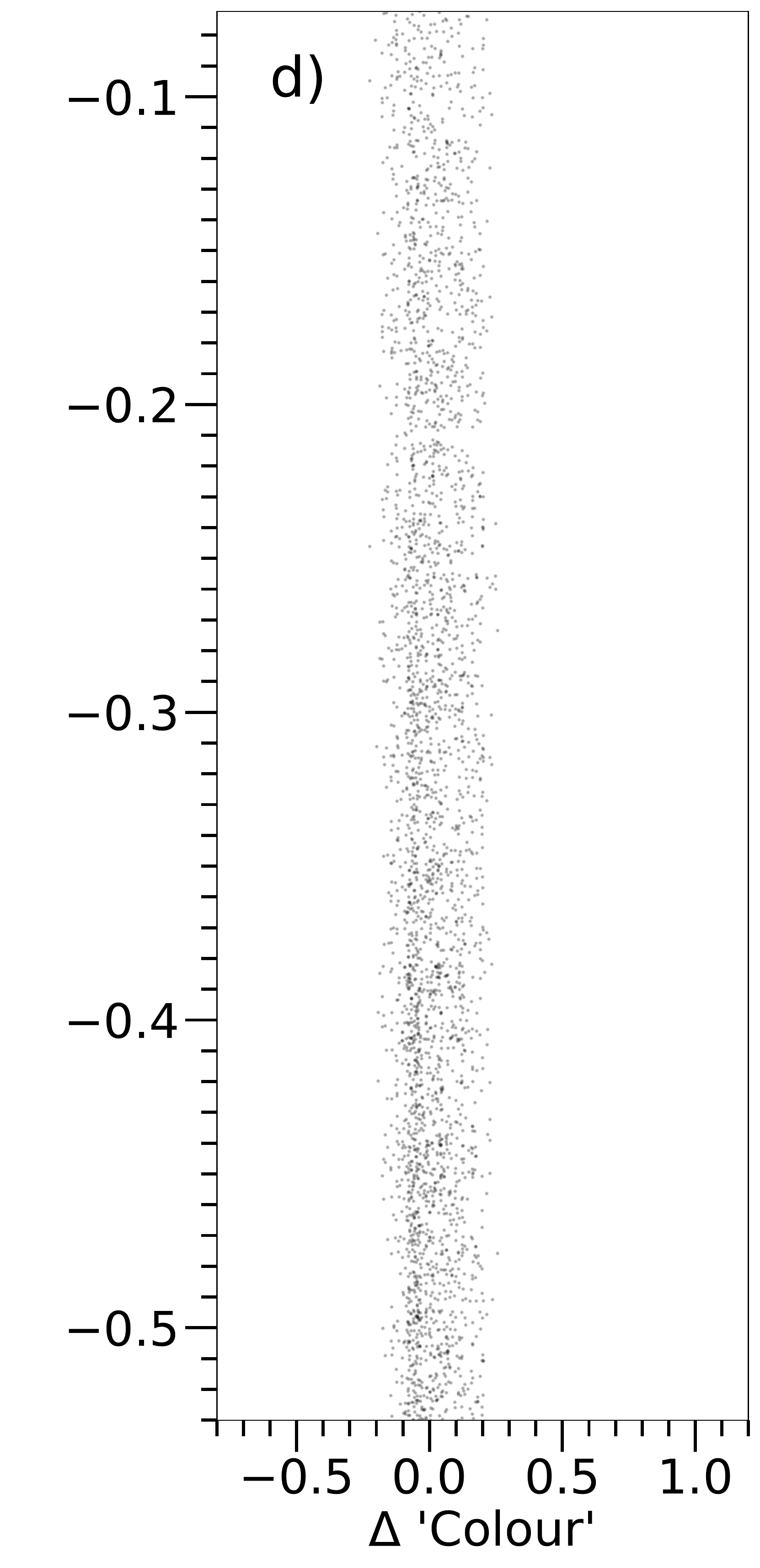}
	\centering
	\includegraphics[width=0.27\linewidth, height=1.1\columnwidth]{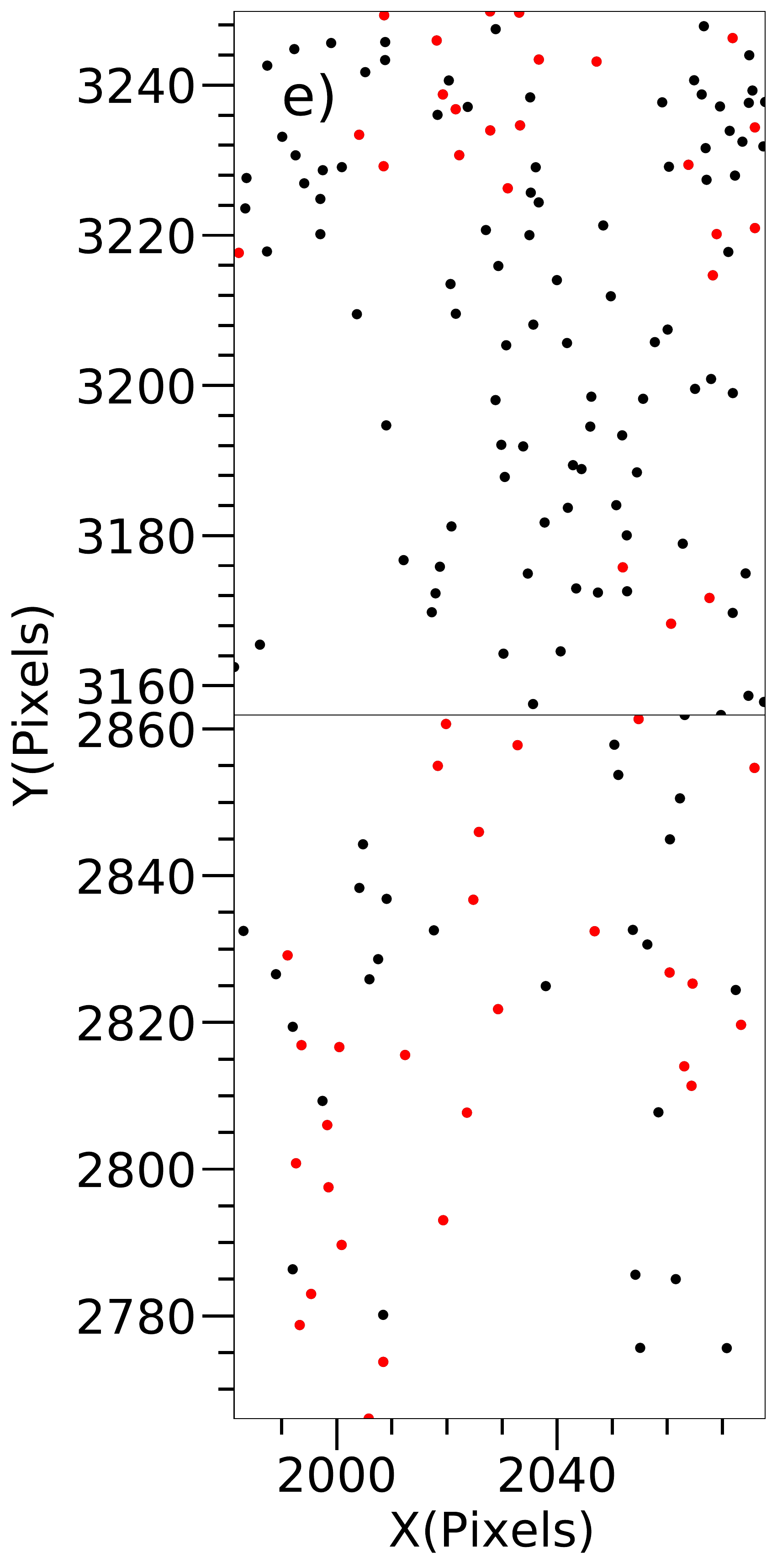}
	\centering
	\includegraphics[width=0.22\linewidth, height=1.1\columnwidth]{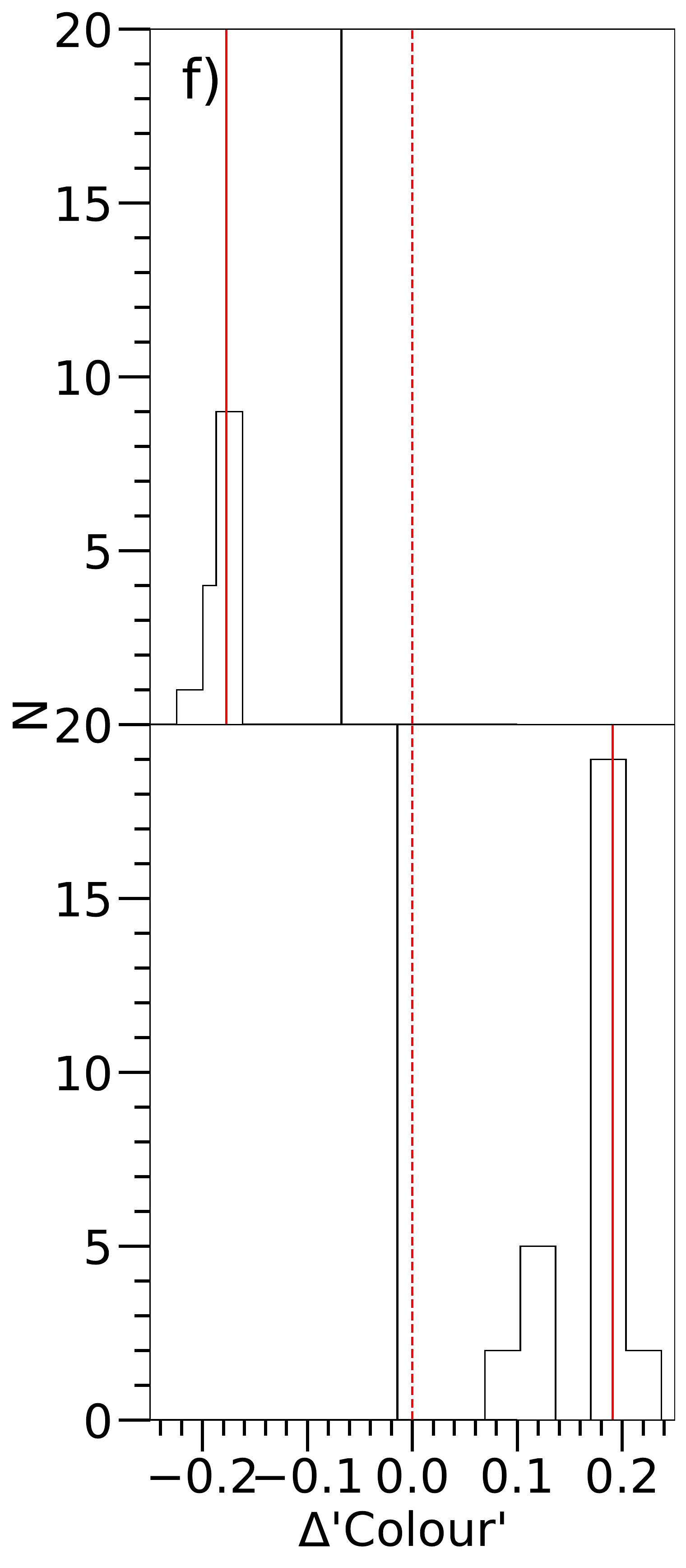}
	\centering
	\includegraphics[width=0.24602\linewidth, height=1.1\columnwidth]{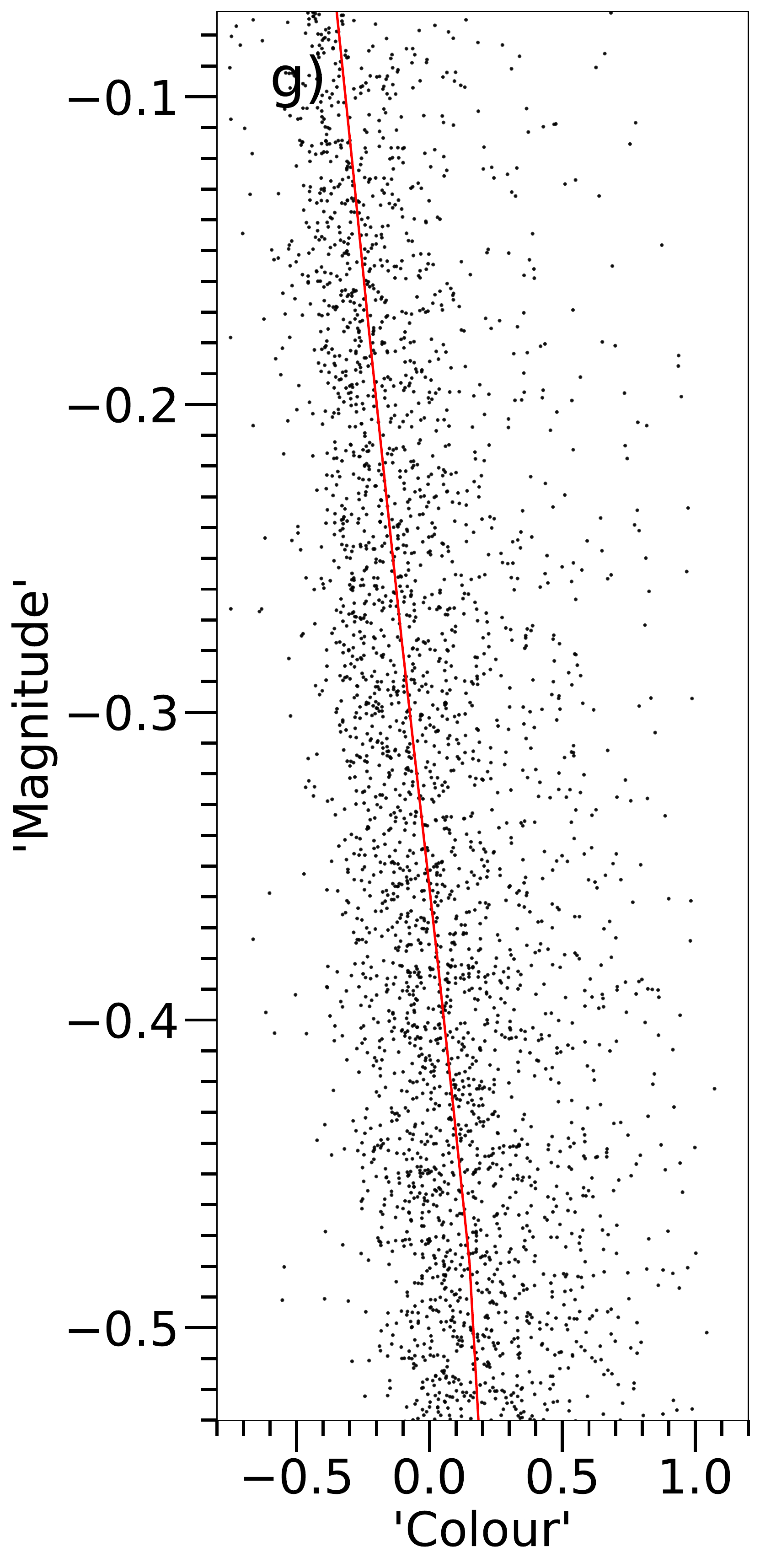}
	\centering
	\includegraphics[width=0.24602\linewidth, height=1.1\columnwidth]{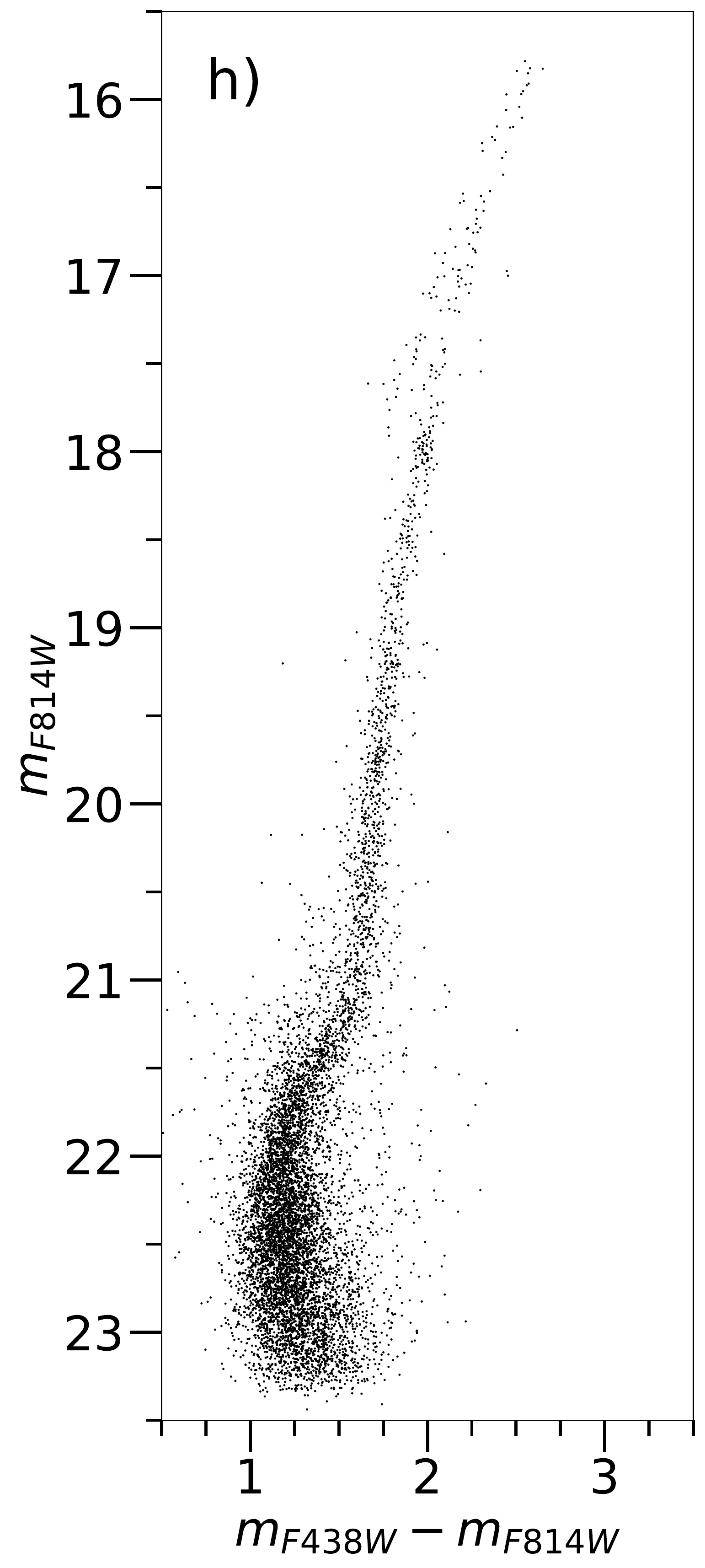}
	\caption{Method for correcting DR illustrated using NGC 1786. \textit{(a):} \textit{m\textsubscript{F438W}}$-$\textit{m\textsubscript{F814W}} vs m\textsubscript{F814W} CMD. The blue dots represent the selected reference stars in the translated CMD. \textit{(b):} The translated CMD with the reddening direction parallel to the x-axis. The red curve represents the fiducial line fit of the reference stars. \textit{(c):} The translated CMD after the subtraction of the `Colour' of fiducial line from the `Colour' of each star at the same magnitude. \textit{(d):} The $\Delta$`Colour' of each star in \textit{(c)} is replaced by $\Delta$`Colour' of its 70 nearest neighbours in the pixel-coordinate map. \textit{(e):} Two bins, one in the core of the cluster and the other in the outskirts, in the 15$\times$15 grid division of the pixel coordinate map are shown. The red dots indicate the reference stars in the bins. \textit{(f):} Histogram of the $\Delta$`Colour' distributions in the two bins shown in \textit{(e)}. The continuous red line and black line indicates the median `Colour' of the reference stars before and after tophat convolution respectively. \textit{(g):} The translated CMD after the tophat smoothed median $\Delta$`Colour' of the reference stars have been subtracted from the `Colour' of stars in all the cells of the 15$\times$15 grid. The red line indicates the fiducial line of the new CMD.\textit{(h):} The CMD after it is rotated back to the original reference frame.}
	\label{fig:reddproc}
\end{figure*}

\begin{figure*}
	
	\centering
	\includegraphics[ height=\columnwidth]{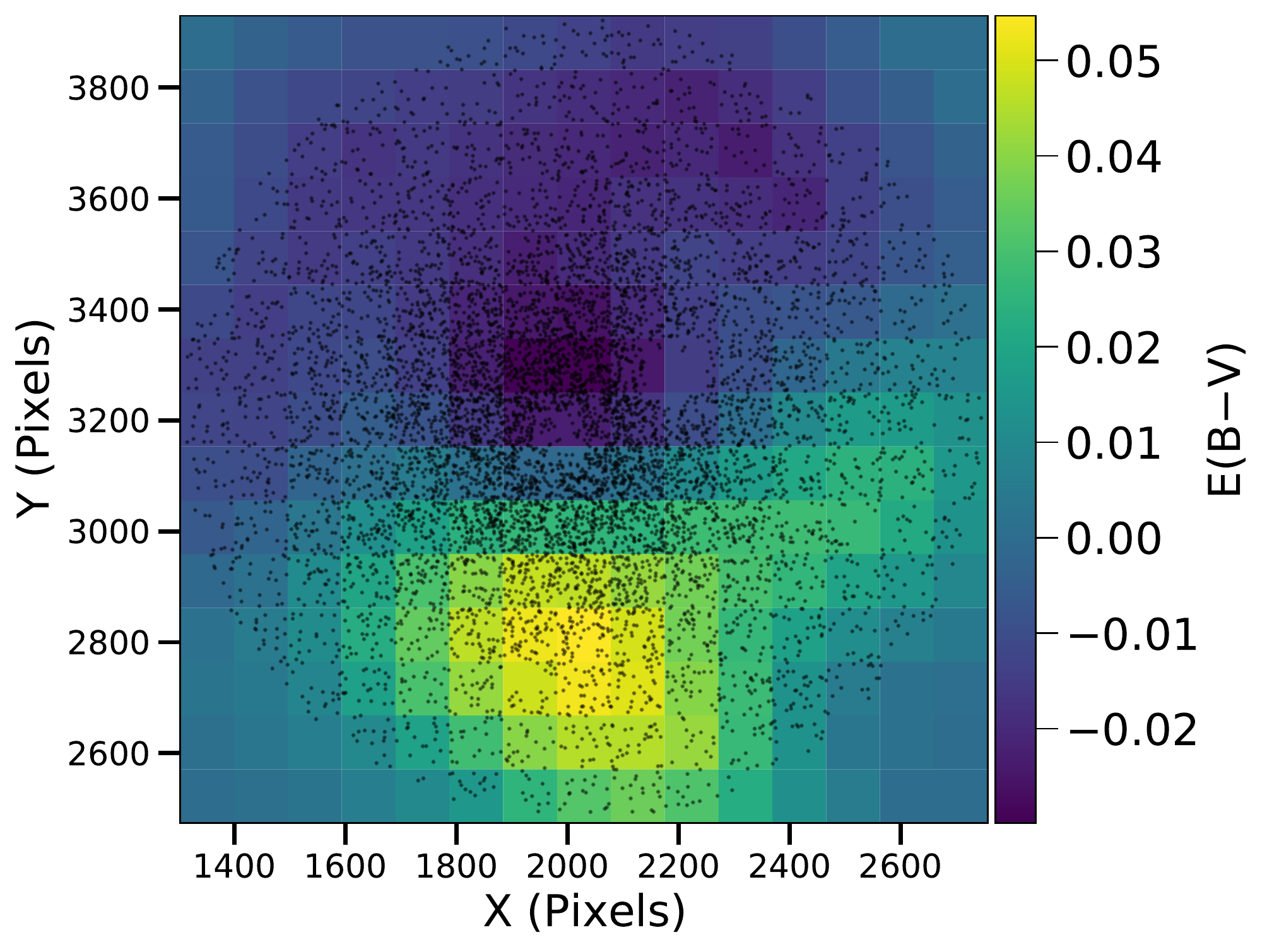}
	\caption{Differential Reddening map of NGC 1786.}
	\label{fig:reddmap}
\end{figure*}

\paragraph*{} We corrected for the effects of differential reddening (DR) and PSF zero point variation following approximately the method described in \citet{milone2012}. The procedure is described here. The optical CMD \textit{m\textsubscript{F814W}} vs \textit{m\textsubscript{F438W} $-$ m\textsubscript{F814W}} was used for this purpose and the procedure is demonstrated in Figure \ref{fig:reddproc} using NGC 1786. A new reference frame is adopted in which the origin of the CMD was shifted arbitrarily to a point near the MSTO magnitude and then the CMD was rotated counterclockwise through an angle $\theta$ defined by:
\begin{align*}
\theta=arctan\frac{A\textsubscript{F438W}-A\textsubscript{F814W}}{A\textsubscript{F814W}}
\end{align*}
where A\textsubscript{\textit{x}} denotes the extinction in filter \textit{x}. The new reference frame is easier to work with since the horizontal axis represents the direction of the reddening vector rather than the oblique reddening line in the non-translated CMD. The horizontal axis of the new reference frame shall henceforth be called `Colour' and the vertical axis `Magnitude'. We selected the MS stars  between 0.55 and 0.2 mag along the `Magnitude' axis as reference stars, divided them into 6 bins and fit a fiducial line by fitting a quadratic spline to the median `Colour' points determined with a 3-sigma clip in each bin. Panels (a) and (b) of Figure \ref{fig:reddproc} shows the \textit{m\textsubscript{F438W}$-$m\textsubscript{F814W}} CMD and the MS reference stars in the translated CMD respectively. The reference stars are marked as blue dots in Panel (a) and the fiducial line of the MS reference stars is represented by a continuous red curve in Panel (b). For each reference star (target), the distance from the fiducial line ($\Delta$`Colour') is determined, then substituted with the median of $\Delta$`Colour' of the 70 nearest reference stars excluding the target star as demonstrated in the Panels (c) and (d) of Figure \ref{fig:reddproc}. Then we divided the pixel coordinate map of the cluster into $15\times15$ grid with each cell consisting of $97\times97$ pixels. This is followed by determining the median $\Delta$`Colour' of the reference stars in each cell of the $15\times15$ grid and then smoothing using 2D tophat filter. Panel (e) consists of two sub-panels. The top panel shows a cell in the $15\times15$ grid that lies in the core of the cluster and the bottom show one that lies towards the outskirts of the cluster. The reference stars are marked as red dots in both panels. Panel (f) shows the $\Delta$`Colour' histogram in these two cells with the red and black lines representing the median of the reference stars before and after tophat convolution respectively. The smoothed median $\Delta$`Colour' of each cell in the $15\times15$ grid is subtracted from `Colour' of each star in the cell. A fiducial line was then fit to the reference stars in the translated CMD and it was near identical to the one fit to the non-corrected reference stars at the start of this procedure, as demonstrated in Panel (g) of Figure \ref{fig:reddproc}, indicating the goodness of the fit. The smoothed median $\Delta$`Colour' of the cells in the grid are then converted to E(B$-$V) and then to filter specific extinction using the following relation for total-to-differential absorption obtained from Aaron Dotter \citep{dotter2016} by \citet{lagioia2019}:
\begin{align*}
E(B-V)=\frac{A\textsubscript{F336W}}{5.100}=\frac{A\textsubscript{F438W}}{4.182}=\frac{A\textsubscript{F814W}}{1.842}
\end{align*}
It is then followed by subtracting the extinction magnitudes from the observed magnitudes in the respective filters. The corrected \textit{m\textsubscript{F438W}-m\textsubscript{F814W}} vs \textit{m\textsubscript{F814W}} is shown in Panel (h) of Figure \ref{fig:reddproc}. DR map of NGC 1786 is shown in Figure \ref{fig:reddmap}.

\begin{figure*}
	
	\centering
	\includegraphics[width=0.49\linewidth, height=\columnwidth]{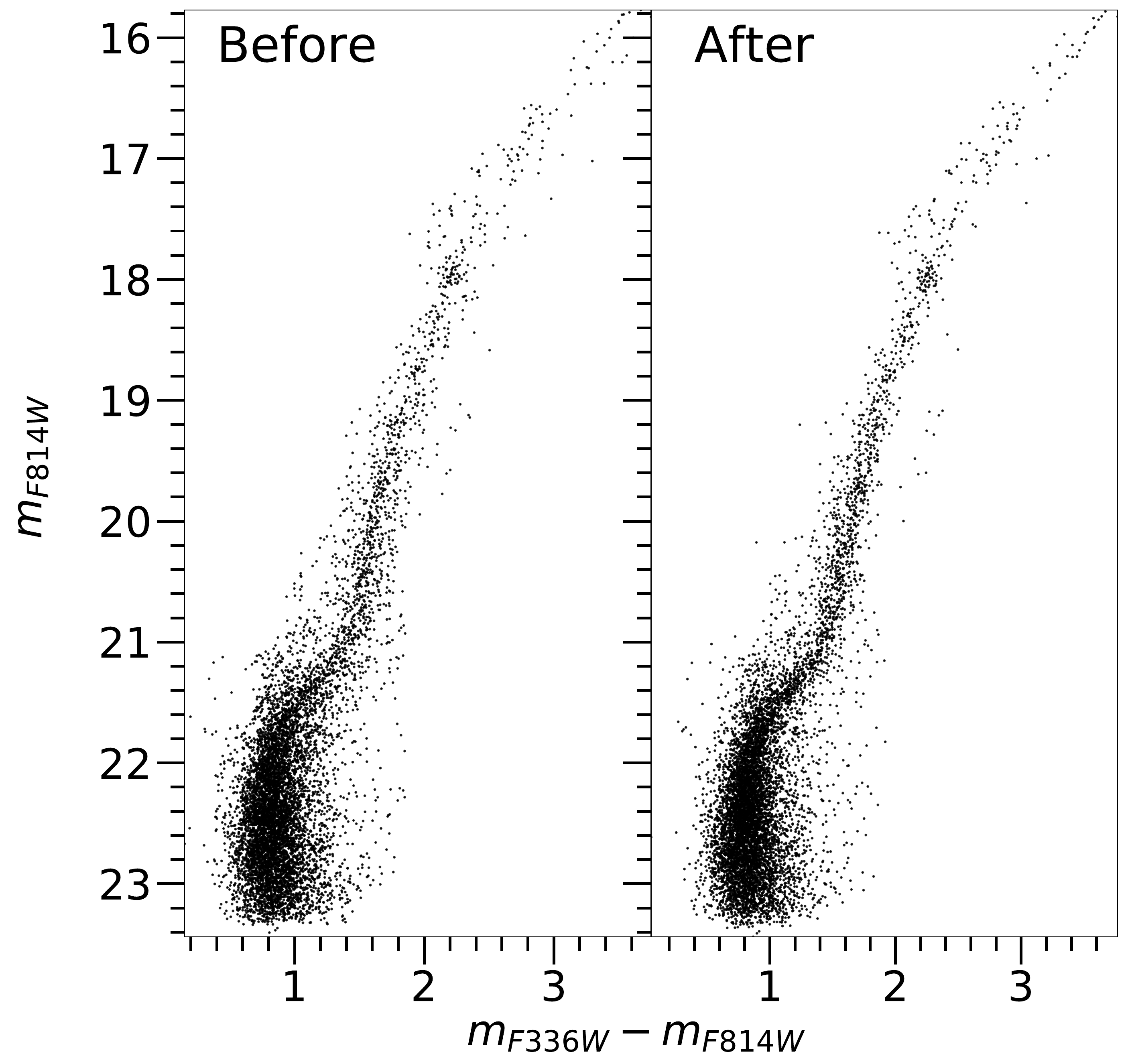}
	\centering
	\includegraphics[width=0.49\linewidth, height=\columnwidth]{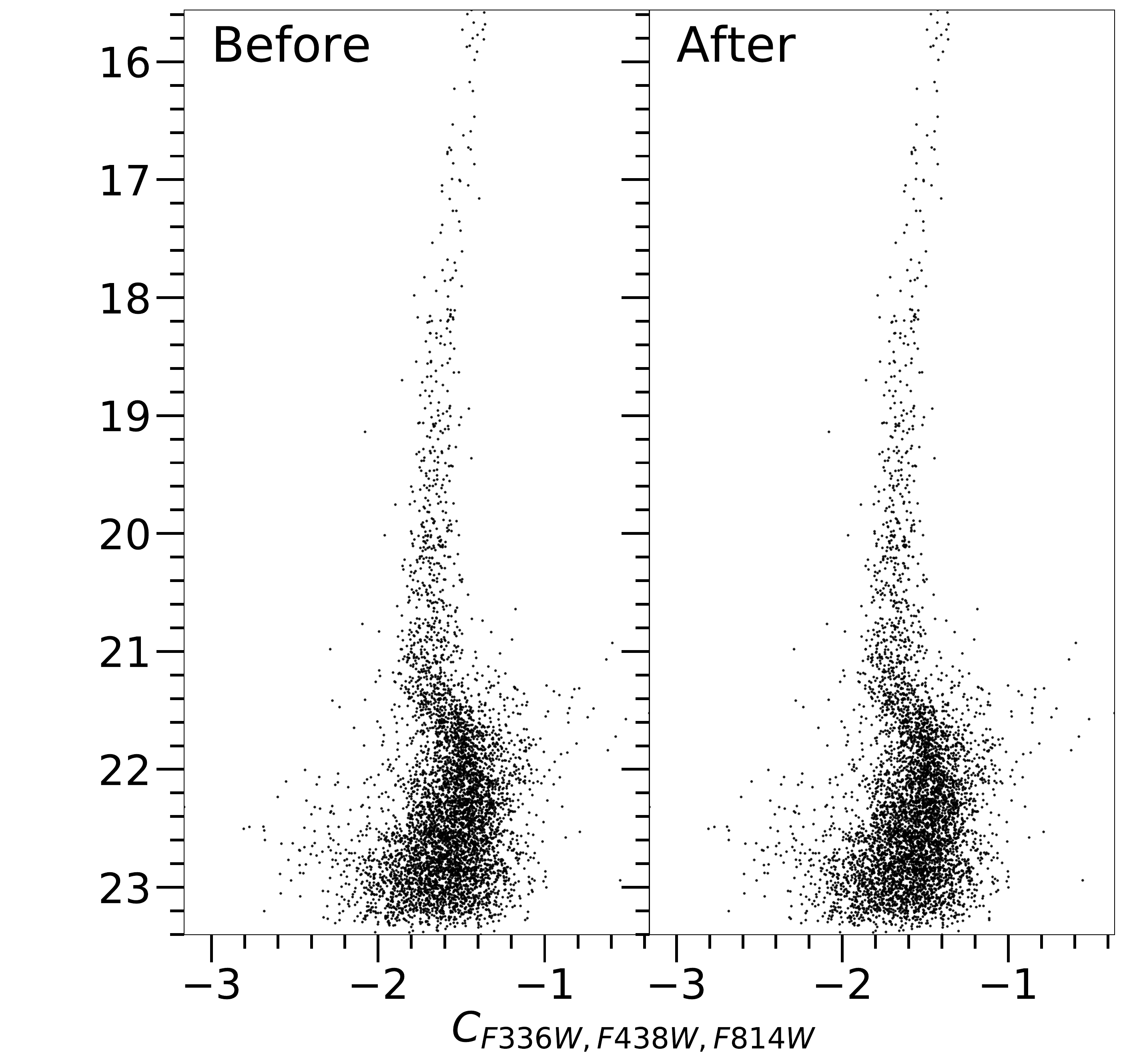}
	\caption{Left panel shows \textit{m\textsubscript{F336W}-m\textsubscript{F814W} vs m\textsubscript{F814W}} CMD of NGC 1786 before and after DR correction. Right panel shows \textit{C\textsubscript{F336W,F438W,F814W}} vs \textit{m\textsubscript{F814W}} pseudo-CMD of NGC 1898 before and after PSF zero point correction.}
	\label{fig:beforeafter}
\end{figure*}

\paragraph*{}The average E(B$-$V) of NGC 1898 is less than 0.10 mag \citep{mclaughlin2005} and since it is negligible, we applied corrections only to photometric zero point variations due to slight unmodelled PSF variations. The procedure for correction is almost the same as the correction for DR but along the direction of pseudo-colour \textit{C\textsubscript{F336W,F438W,F814W}} instead of the direction of reddening vector and without shifting the origin of the CMD. Figure \ref{fig:beforeafter} depicts the results of DR correction and PSF zero point correction of NGC 1786 and NGC 1898 using \textit{m\textsubscript{F336W}-m\textsubscript{F814W} vs m\textsubscript{F814W}} CMD and \textit{C\textsubscript{F336W,F438W,F814W}} vs \textit{m\textsubscript{F814W}} pseudo-CMD respectively.

\subsection{NGC 121 and Lindsay 1} \label{section26}

\paragraph*{} Since we wanted to quantify the possible offsets between our results and those of \citet{lagioia2019} due to differences in our method, we attempted to reproduce their results using our techniques. We selected two SMC clusters for analysis: NGC 121 and Lindsay 1. For these two clusters, they performed photometry on images taken with F555W and F814W filters from ACS/WFC and F343N filter from WFC3/UVIS apart from those taken with the three WFC3/UVIS filters used in this analysis. In this study, we performed photometry on the exact dataset for Lindsay 1 as used in \citet{lagioia2019} but restricted ourselves to the three standard filters for NGC 121 without considering even the images taken with ACS/WFC F814W filter. In the case of Lindsay 1, the selection criteria for stars after photometry didn't include the presence (or lack thereof) of magnitudes of stars in the ACS/WFC F555W filter and WFC3/UVIS F343N filter. So, the difference lies in the inclusion (or lack thereof) of ACS/WFC F814W filter. This approach helps us to decide the importance of including ACS/WFC F814W images in future studies depending on whether we are able to obtain a similar intrinsic RGB width as \citet{lagioia2019}.

\subsection{Cluster Parameters} \label{section28}

\paragraph*{} For Galactic GCs, we obtained current cluster mass ($M_{c}$) from \citet{baumgardt2018}, initial cluster mass ($M_{ini}$) from \citet[private communication]{baumgardt2019}, metallicities ([Fe/H]) from \citet[2010 version]{harris1996}, ages from \citet{dotter2010,dotter2011}, \citet{vandenberg2013} and \citet{milone2014}. For SMC clusters (Lindsay 1, Lindsay 38, Lindsay 113, NGC 121, NGC 339 and NGC 416) and LMC cluster NGC 1978, $M_{ini}$ were obtained from \citet{milone2020}, total cluster masses were obtained from \citet{glatt2011} and \citet{chantereau2019}, ages from \cite{lagioia20199}, \citet{glatt2008} and \citet{milone2009}.

\paragraph*{} For NGC 1898 and NGC 1786, the parameter values and their sources are listed in Table \ref{tab:table2}. We derived their initial masses using the methods and programs presented in \citet{goudfrooij2014, goudfrooij2011}. The initial masses and current masses of Galactic GCs used in this study derived by \citet{baumgardt2019} are updated versions based on \textit{Gaia} early data release 3 (eDR3, \citealt{gaia2021}). Although the initial masses of Galactic and MC GCs used in this paper are up to date, there are a lot of uncertainties associated with their determination due to the insufficient knowledge of the evolution of these galaxies and their tidal fields, mass loss and initial mass segregation in these clusters (see Section 2.2 in \citealt{milone2020}). The initial mass data of GCs used in this study assume zero initial mass segregation.

\section{Measurement of intrinsic RGB width} \label{section3}

In this section, we describe the method to deduce the intrinsic RGB width of the clusters in our analysis in the pseudo-colour \textit{C\textsubscript{F336W,F438W,F814W}}, as outlined in \citet{lagioia2019}.

\begin{figure*}
	
	\centering
	\includegraphics[width=0.24602\linewidth, height=1.1\columnwidth]{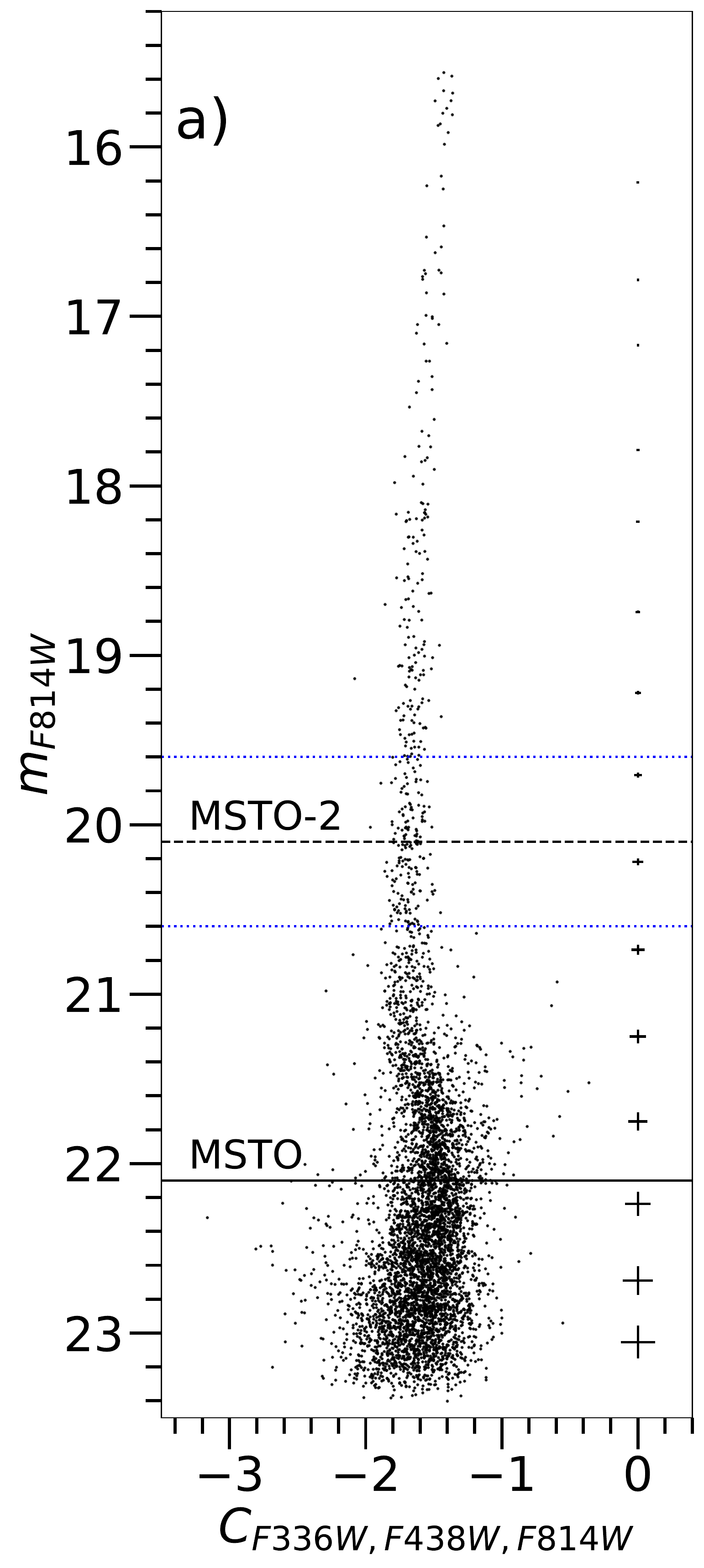}	
	\centering
	\includegraphics[width=0.24602\linewidth, height=1.1\columnwidth]{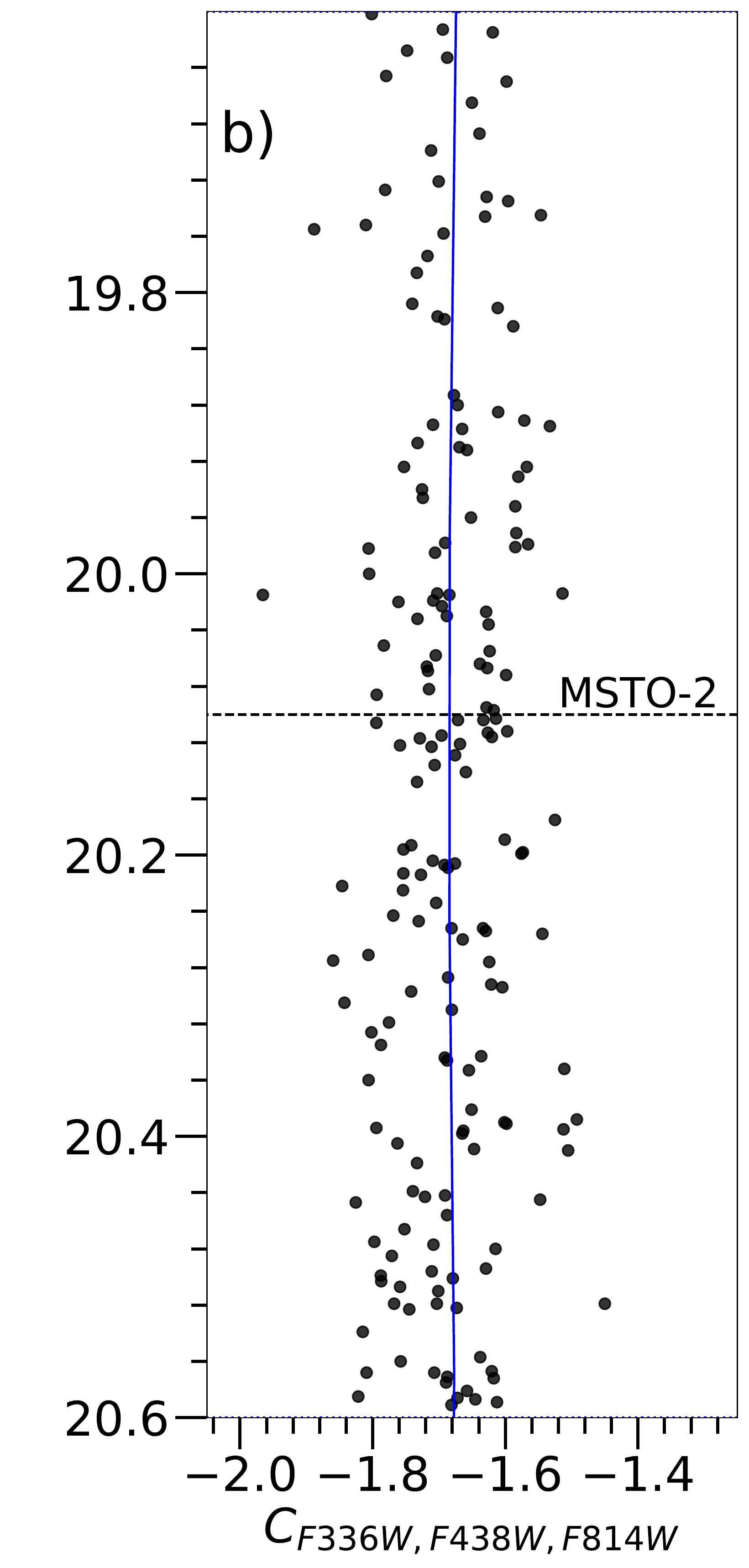}
	\centering
	\includegraphics[width=0.24602\linewidth, height=1.1\columnwidth]{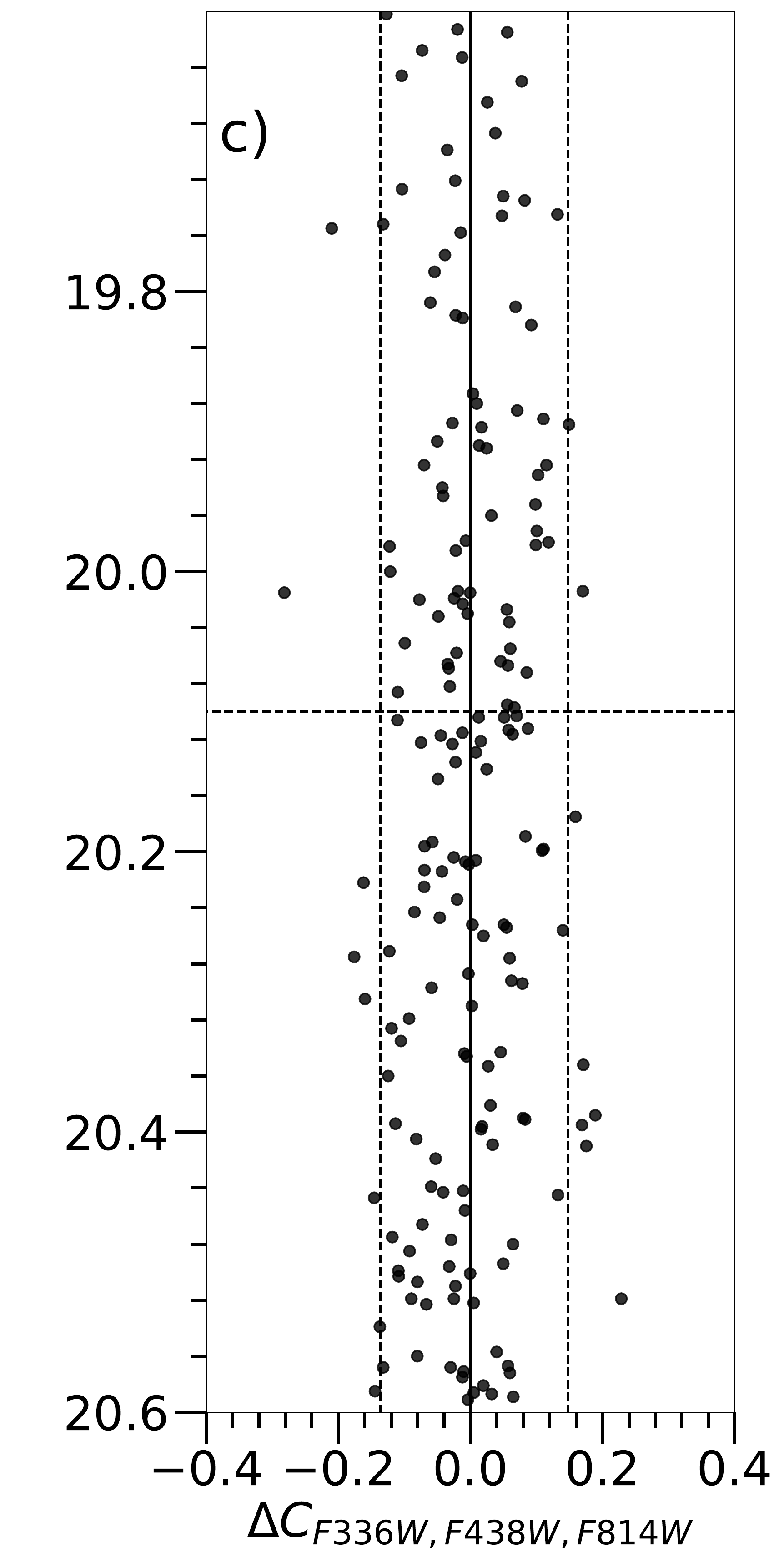}
	\centering
	\includegraphics[width=0.24602\linewidth, height=1.1\columnwidth]{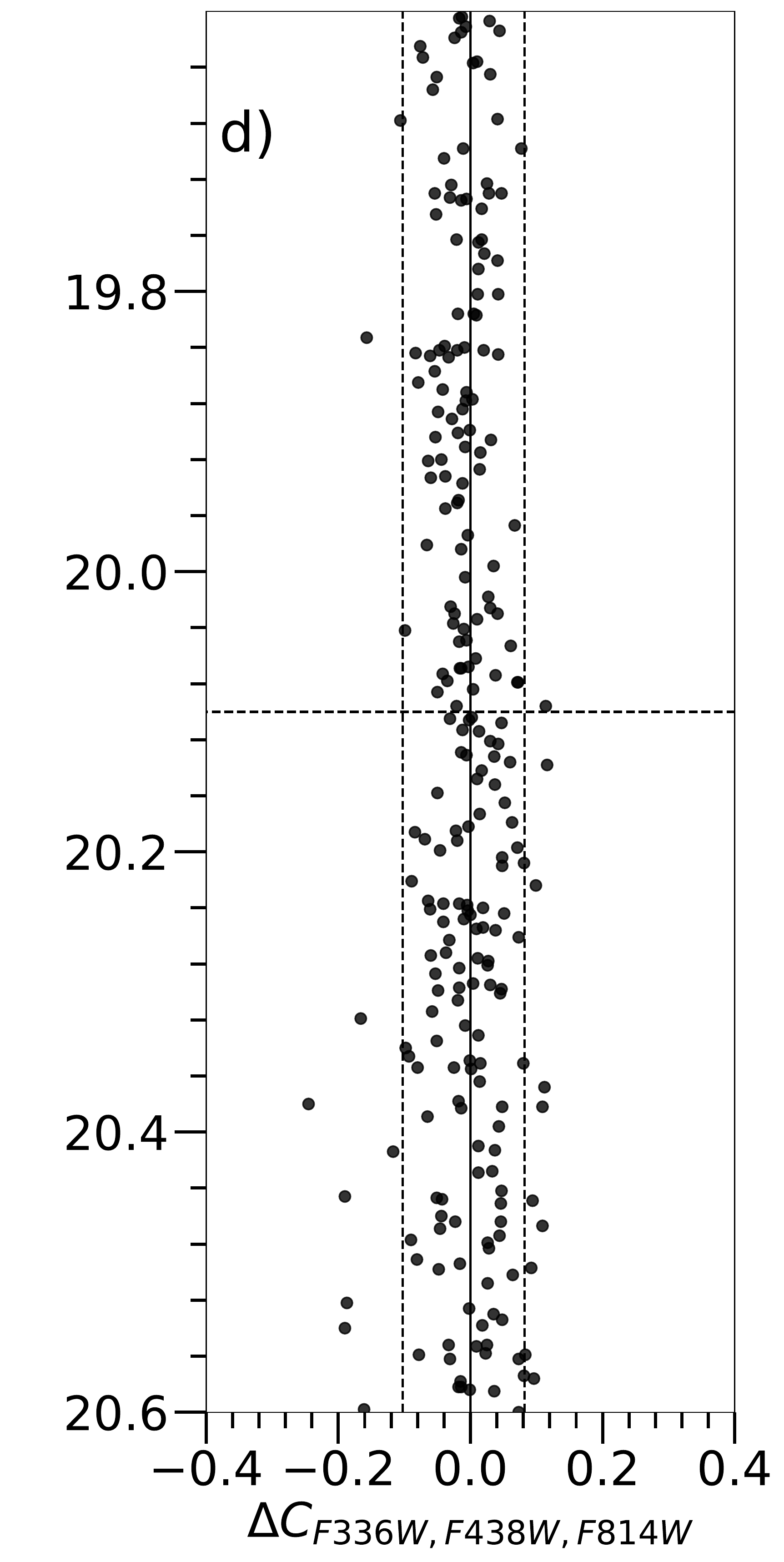}
	\caption{Method to determine the intrinsic RGB width of NGC 1898. \textit{(a):} \textit{m\textsubscript{F814W}} vs \textit{C\textsubscript{F336W,F438W,F814W}} pseudo-CMD. The solid back line represents \textit{m\textsubscript{MSTO}} while the broken black line represents \textit{m\textsubscript{MSTO}}$-$2 mag. The broken blue lines represent the selected magnitude interval. The photometric error bars are represented on the right side. \textit{(b):} Fiducial line fit for the selected 1 mag interval. \textit{(c):} Verticalized CMD in the 1 mag interval. The broken black lines represent the 4th and 96th percentile of distribution of pseudo-colour difference, representing \textit{W\textsubscript{obs}}. \textit{(d):} Width due to photometric uncertainties as reconstructed from AS test.}
	\label{fig:1898}
\end{figure*}

\begin{figure*}
 	
 	\centering
 	\includegraphics[width=0.24602\linewidth, height=1.1\columnwidth]{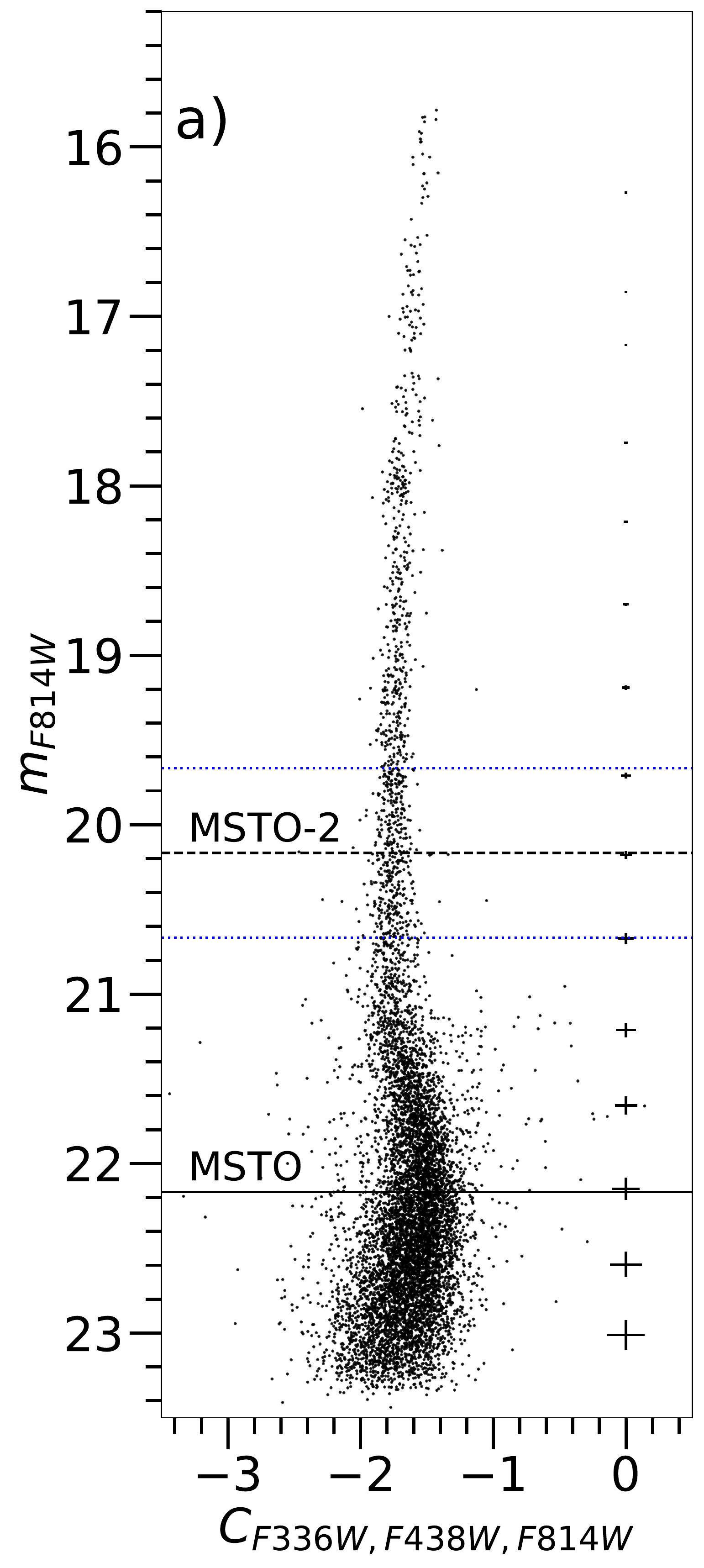}
 	\centering
 	\includegraphics[width=0.24602\linewidth, height=1.1\columnwidth]{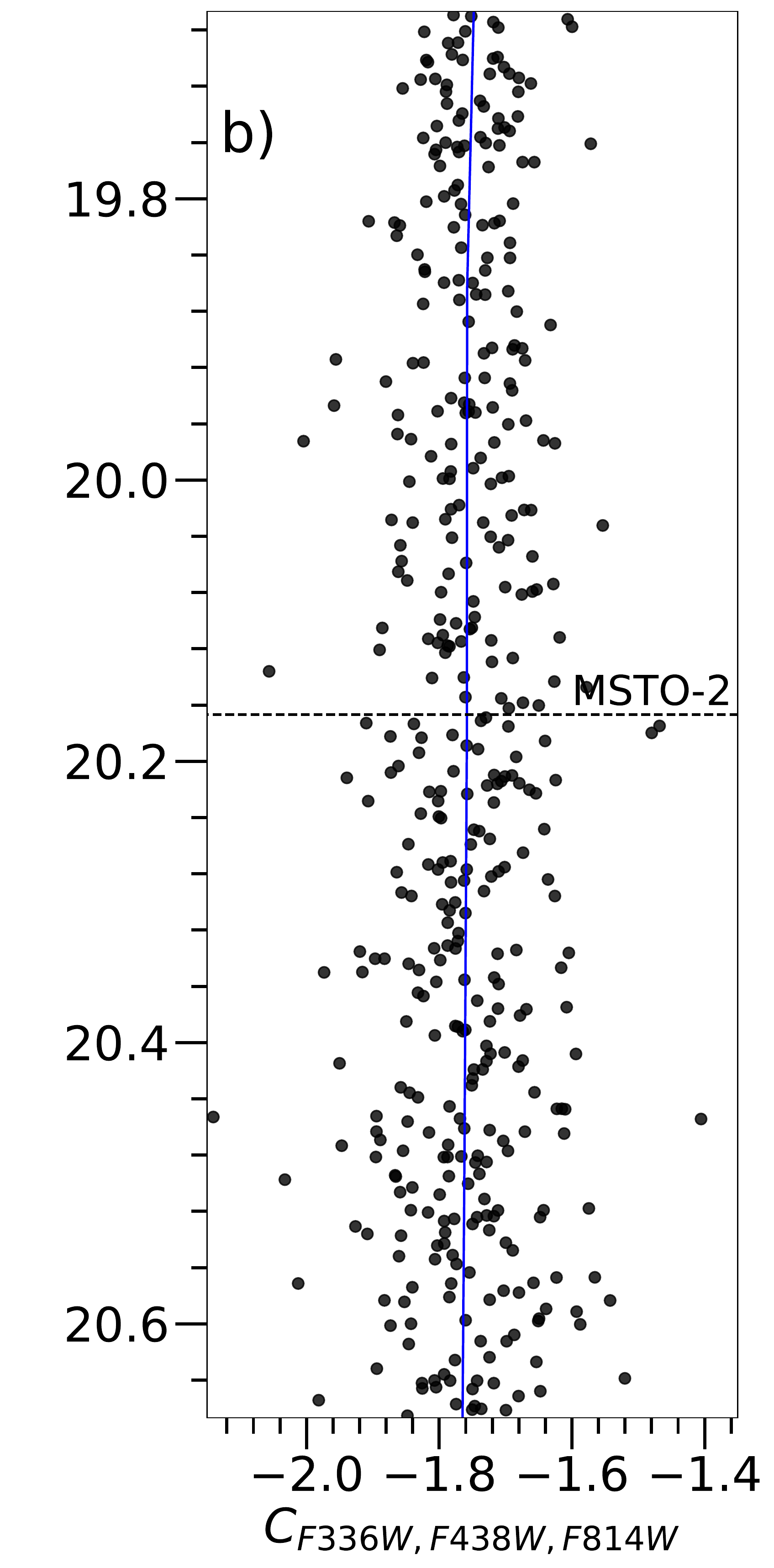}
 	\centering
 	\includegraphics[width=0.24602\linewidth, height=1.1\columnwidth]{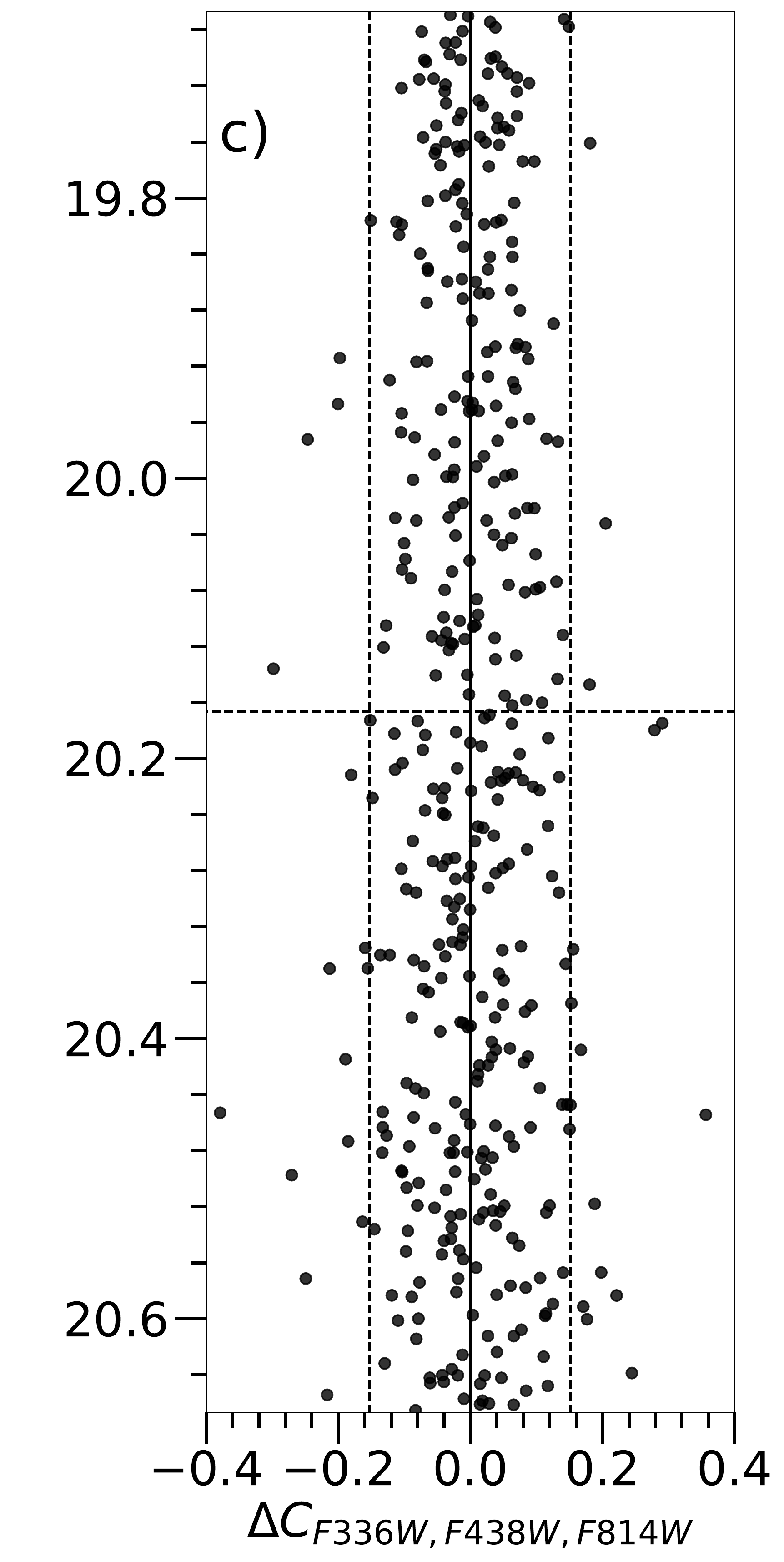}
 	\centering
 	\includegraphics[width=0.24602\linewidth, height=1.1\columnwidth]{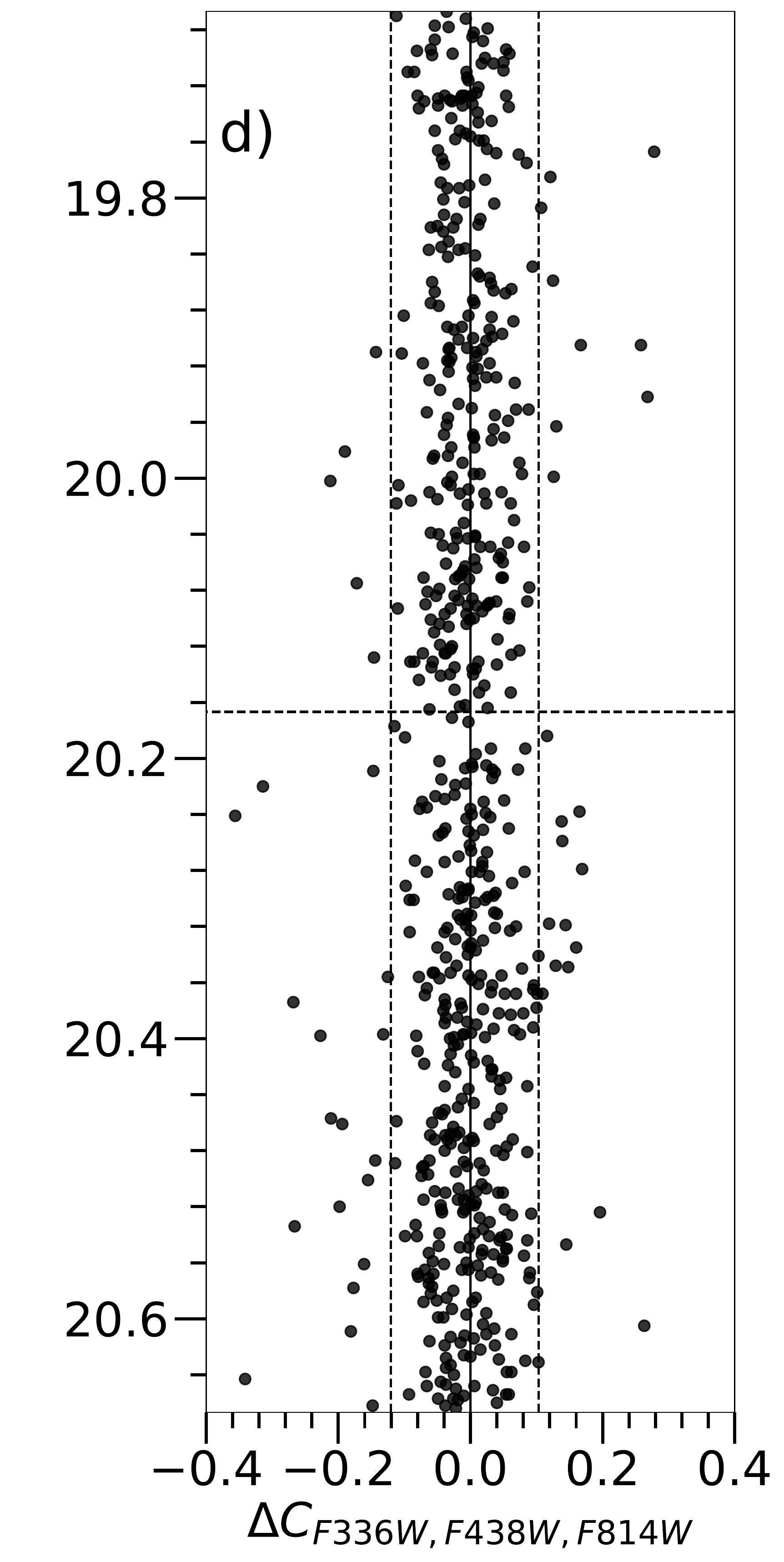}
 	\caption{Same as Figure \ref{fig:1898}, for NGC 1786}
 	\label{fig:1786}
\end{figure*}

Panel (a) of Figures~\ref{fig:1898} and~\ref{fig:1786} represents \textit{m\textsubscript{F814W}} vs \textit{C\textsubscript{F336W,F438W,F814W}} CMD with the stars represented by black dots and photometric error bars on the panel's right side. We started with measuring the main sequence turn-off (MSTO) point. This is done by using naive estimator as explained in \citet{silverman1986} and is summarized here. We divided the magnitude range (minimum mag to maximum mag in F814W filter) into 15 bins and found the median pseudo-colour and median magnitude in each bin. We repeated the procedure for different bin series by changing the initial point of the first bin by a quantity equal to a fraction of the predefined bin width. This was followed by box car averaging three adjacent points of the resulting median magnitudes and pseudo-colours, the purpose of which is to smoothen the curve connecting the points. The magnitude corresponding to the bluest pseudo-colour in the linear interpolation function of the boxcar averaged points was taken to be the MSTO (m\textsubscript{MSTO}). m\textsubscript{MSTO} is 22.100 mag for NGC 1898 and 22.166 mag for NGC 1786, in F814W filter. We then defined a luminosity interval of 1 mag centered at the reference line defined 2.0 F814W magnitude lesser than m\textsubscript{MSTO}. Then we calculated the fiducial line of RGB stars as described by the procedure above and got the pseudo-colour difference by subtracting the colour of each star from the fiducial line at the same F814W magnitude. We used moving average for NGC 121 and Lindsay 1 instead of box-car average to estimate the fiducial line since it gave us a better fit. Pseudo-colour difference in the RGB interval gives us the spread resulting from abundance variations and verticalizes the CMD. Panel (b) of Figures~\ref{fig:1898} and~\ref{fig:1786} shows the fiducial line of NGC 1898 and NGC 1786 respectively, represented by a blue curve. The difference betweeen the 4th percentile and 96th percentile of the distribution of pseudo-colour difference is taken as the observed RGB width (\textit{W\textsubscript{obs}}), as demonstrated in Panel (c) of Figures~\ref{fig:1898} and~\ref{fig:1786}. \textit{W\textsubscript{obs}} of NGC 1898 is 0.284 mag and that of NGC 1786 is 0.304 mag.

The error associated with the observed width was determined by bootstrapping test. This involves generating random copies of the observed stellar pseudo-colours in the selected RGB interval with replacement. First, we generated 1000 copies of the observed pseudo-colours in the selected interval, followed by a random extraction of subsample containing pseudo-colours equal to the number of stars in the selected interval and then calculated the RGB width for this extraction. We repeated this test 10,000 times. The difference between the observed width and the 68.27\textsuperscript{th} percentile of the 10,000 bootstrapping measurements of the RGB width gives us the standard error of the observed width. The associated error for NGC 1898 is 0.011 mag and for NGC 1786 is 0.018 mag.

The observed RGB width of a cluster is not its intrinsic RGB width since it is not accounted for the contribution by photometric errors to the RGB width. To filter this unwanted contribution and obtain the intrinsic RGB width of the cluster, we estimated the photometric error using ASs. The procedure to calculate is explained hereafter. We started with a random selection of a subsample from the AS catalogue equal to the number of observed stars in the analysed magnitude interval. The pseudo-colour difference was calculated by finding the difference between the input and output magnitude of each AS in the subsample. The RGB width of the AS subsample is the difference between the 4th percentile and 96th percentile of the subsample. It represents the contribution of photometric error to \textit{W\textsubscript{obs}}. This value was then subtracted in quadrature from the observed RGB width. We repeated this procedure 10,000 times and the average of those 10,000 measurements is the intrinsic RGB width (\textit{W\textsubscript{CF336W,F438W,F814W}}). Panel (d) of Figures~\ref{fig:1898} and~\ref{fig:1786} represents the error width obtained during one of the runs. The associated error of the RGB width from ASs was obtained by the bootstrapping method described above and it was then added in quadrature with the associated error of W\textsubscript{obs} and the resulting error was adopted as the total uncertainty of \textit{W\textsubscript{CF336W,F438W,F814W}}. The intrinsic RGB widths of NGC 1898 and NGC 1786 are 0.199 mag and 0.185 mag respectively with a total uncertainty of 0.014 mag and 0.019 mag respectively.

We applied the above procedure for the rest of the clusters analysed in this study. The results are tabulated in Table \ref{tab:table2}. For NGC 121 and Lindsay 1, the determined intrinsic RGB width is in line with the results of \citet{lagioia2019} within the estimated uncertainties. \citet{lagioia2019} determined the intrinsic RGB widths of NGC 121 and Lindsay 1 to be 0.157$\pm 0.008$ and 0.135$\pm 0.009$ respectively while our results are 0.169$\pm 0.015$ and 0.142$\pm 0.011$ respectively. Considering that ACS/WFC F814W data was not used for NGC 121, this agreement in the values of intrinsic RGB width indicates that F814W magnitude values obtained from both cameras (WFC3 and ACS) are inherently very similar to each other, in line with the synthetic photometry results of \citet{deustua2018}.

\begin{table*}
	\caption{Metallicity, current mass, initial mass, intrinsic and normalized RGB width values of the analysed globular clusters}
	\label{tab:table2}
	
	\resizebox{\textwidth}{!}{\begin{tabular}{c|c|c|c|c|c|c|c|c|c}
		\hline
		\textbf{ID}&\textbf{[Fe/H]}&\textbf{References}&\textbf{$M_{c}$ }&\textbf{References}&\textbf{$M_{ini}$}&\textbf{References}&\textbf{W\textsubscript{CF336W,F438W,F814W}}&\textbf{${\Delta}$W\textsubscript{CF336W,F438W,F814W}}\\
			& (dex)& &($M_{\sun}$)& &($M_{\sun}$)& &(mag)&(mag)&\\
		\hline
		NGC 1786&$-1.80$&1,2,3&3.72 $\times 10$\textsuperscript{5}&6&1.90 $\times 10$\textsuperscript{6}&9&0.185 $\pm 0.019$&0.068\\
		NGC 1898&$-1.32$&2,3,4&2.24 $\times 10$\textsuperscript{5}&6&4.50 $\times 10$\textsuperscript{5}&9&0.199 $\pm 0.014$&0.050\\
		NGC 121&$-1.30$&5&3.70 $\times 10$\textsuperscript{5}&6&7.08 $\times 10$\textsuperscript{5}&8&0.169 $\pm 0.015$&0.019\\
		Lindsay 1&$-1.14$&5&$\sim2 \times 10$\textsuperscript{5}&7&3.39 $\times 10$\textsuperscript{5}&8&0.142 $\pm 0.011$& $ -0.019$\\
		
		\hline
	\end{tabular}}
	\hspace{0.5in}1.\citet{mucciarelli2009} 2.\citet{beasley2002} 3.\citet{olszewski1991} 4. \citet{johnson2006} 5.\citet{lagioia20199}\hspace{0.5in} 6.\citet{mclaughlin2005} 7. \citet{glatt2011} 8.\citet{milone2020} 9.This paper
\end{table*}

\begin{table*}
	\caption{Results of Spearman Correlation Test}
	\label{tab:table3}
	
	\begin{tabular}{c|c|c|c|c|}
		
		\hline
		\textbf{Parameters} & \multicolumn{2}{c|}{\textbf{W\textsubscript{CF336W,F438W,F814W}}} & \multicolumn{2}{c|}{\textbf{${\Delta}$W\textsubscript{CF336W,F438W,F814W}}}\\
		\cline{2-5}
		& \textbf{R\textsubscript{s}} & \textbf{p-value} & \textbf{R\textsubscript{s}} & \textbf{p-value}\\
		
		\hline
		$M_{c}$&$0.291$&$>0.01$&$0.731$&$<0.01$\\ 
		$M_{ini}$&$0.400$&$<0.01$&$0.564$&$<0.01$\\
		Age&$-0.539$&$<0.01$&$0.116$&$>0.01$\\
		MLF&$-0.006$&$>0.01$&$-0.437$&$<0.01$\\
		MLA&$0.429$&$<0.01$&$0.424$&$<0.01$\\
		ML&$0.397$&$<0.01$&$0.444$&$<0.01$\\
	
		\hline
	\end{tabular}
\end{table*}

\begin{figure*}
	
	\centering
	\includegraphics[width=1\linewidth, height=1\columnwidth]{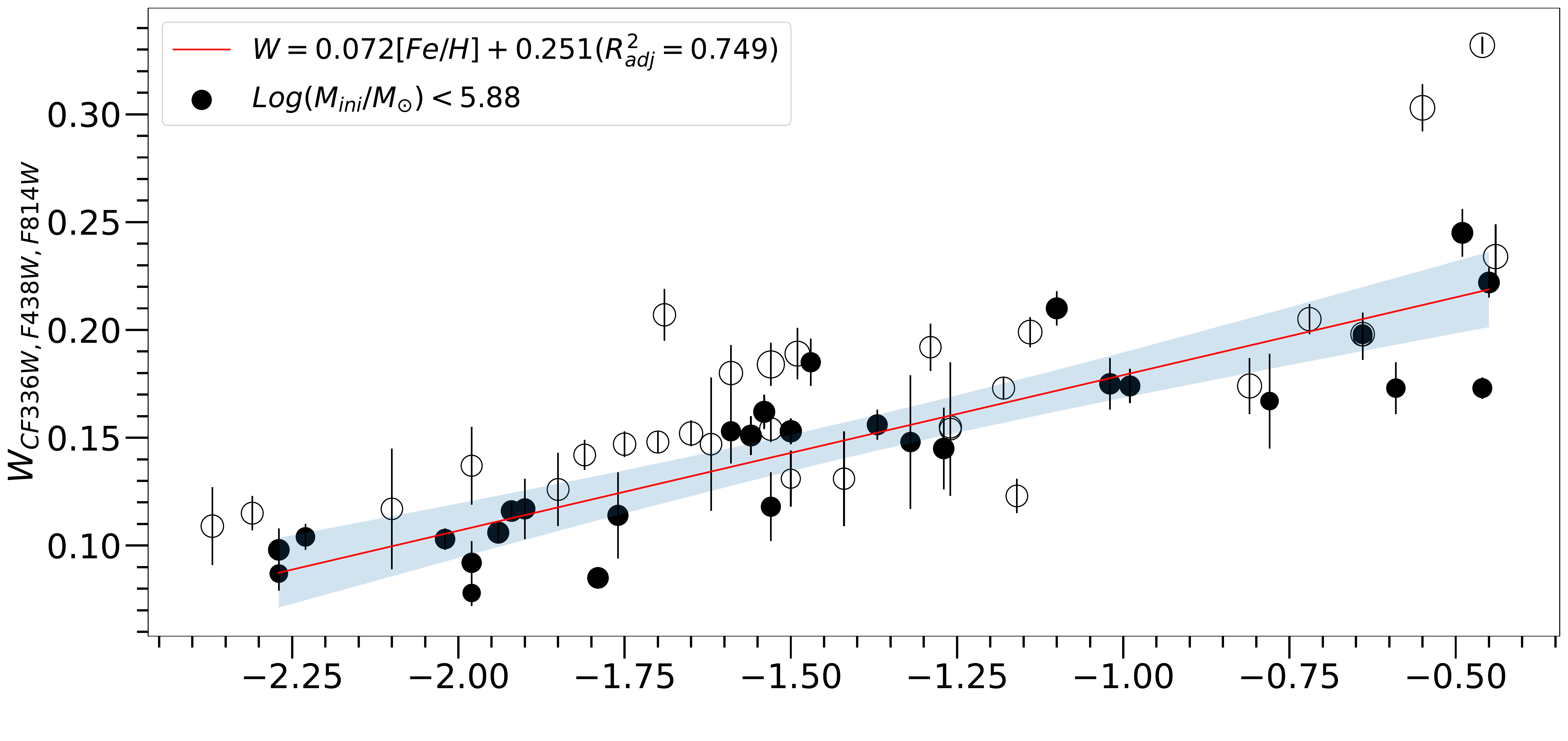}
	\centering
	\includegraphics[width=1\linewidth, height=1\columnwidth]{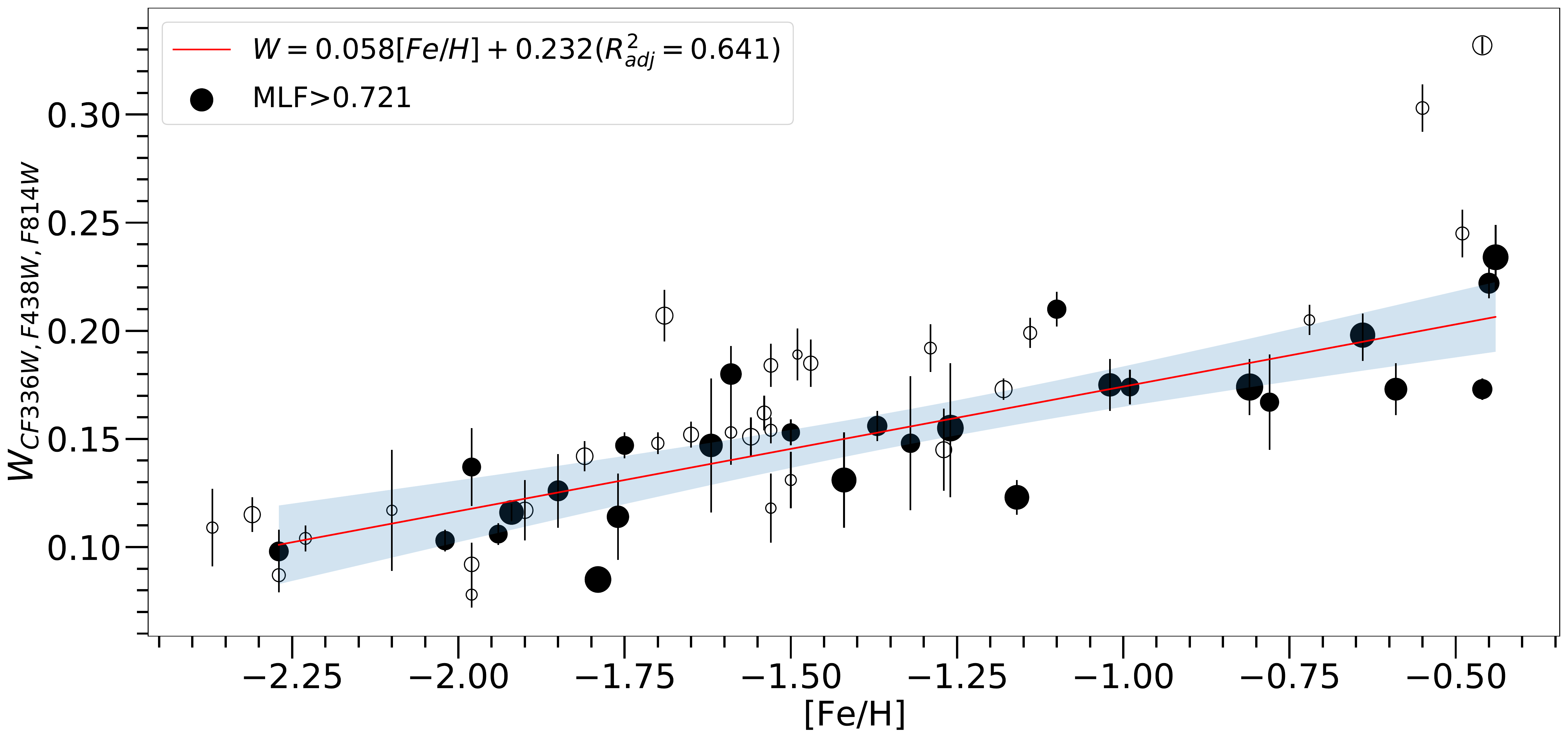}
	\caption{W\textsubscript{CF336W,F438W,F814W} vs [Fe/H] of the 58 Galactic GCs. The shaded GCs are the ones selected for deriving the regression line. The selection criteria for clusters in the top panel is based on $M_{ini}$ while for those in the bottom panel is based on MLF. The size of the clusters is proportional to $M_{ini}$ and MLF in the top and bottom panels respectively. The continuous red line is the regression line. The blue shaded region represents the 95\% confidence interval of the regression line. The equation of the line along with the adjusted determination coefficient is indicated in the legend.}
	\label{fig:metsub1}
\end{figure*}

\begin{figure*}
	
	\centering
	\includegraphics[width=1.05\linewidth,height=1.05\columnwidth]{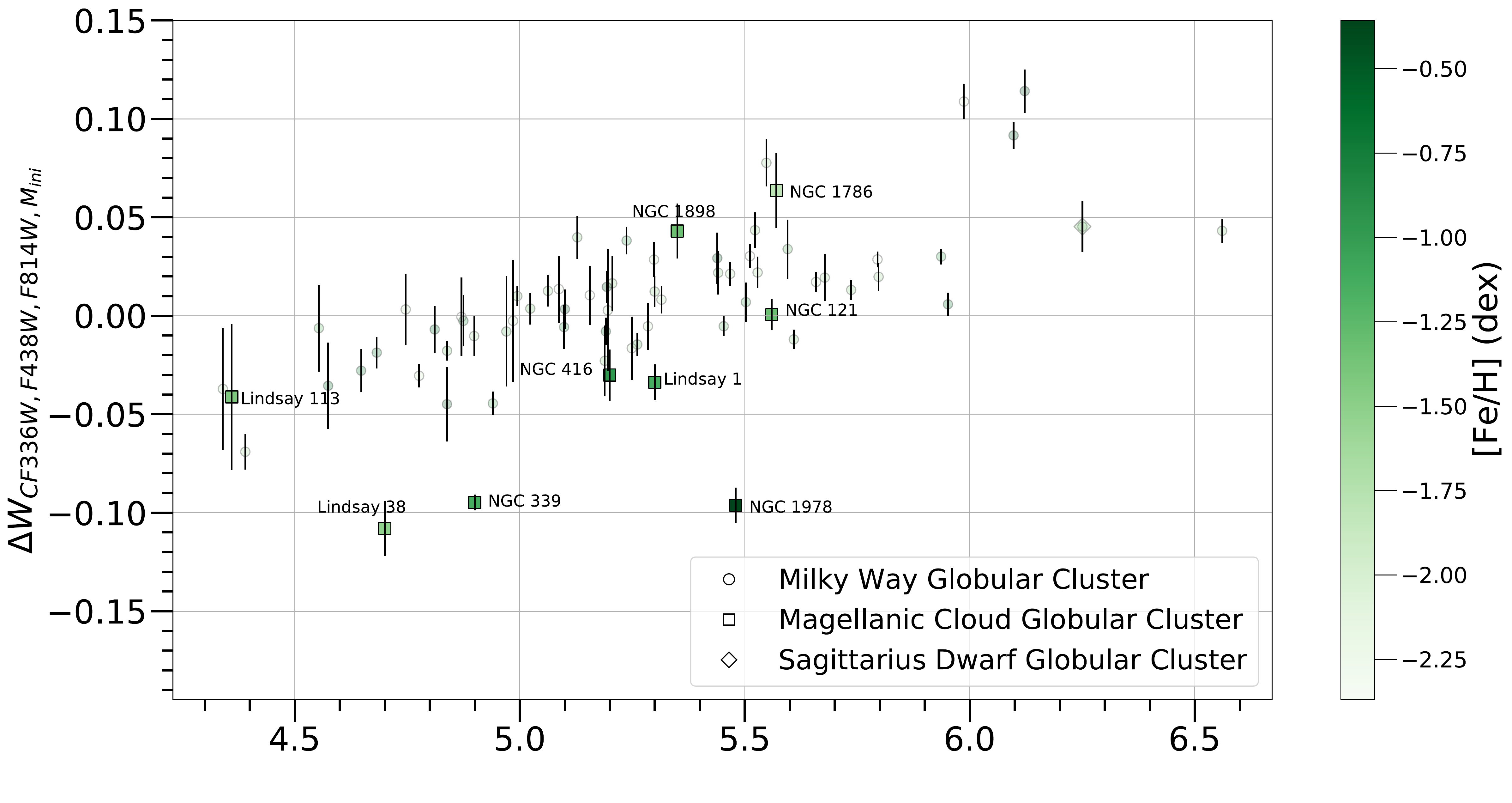}
	\centering
	\includegraphics[width=1.05\linewidth,height=1.05\columnwidth]{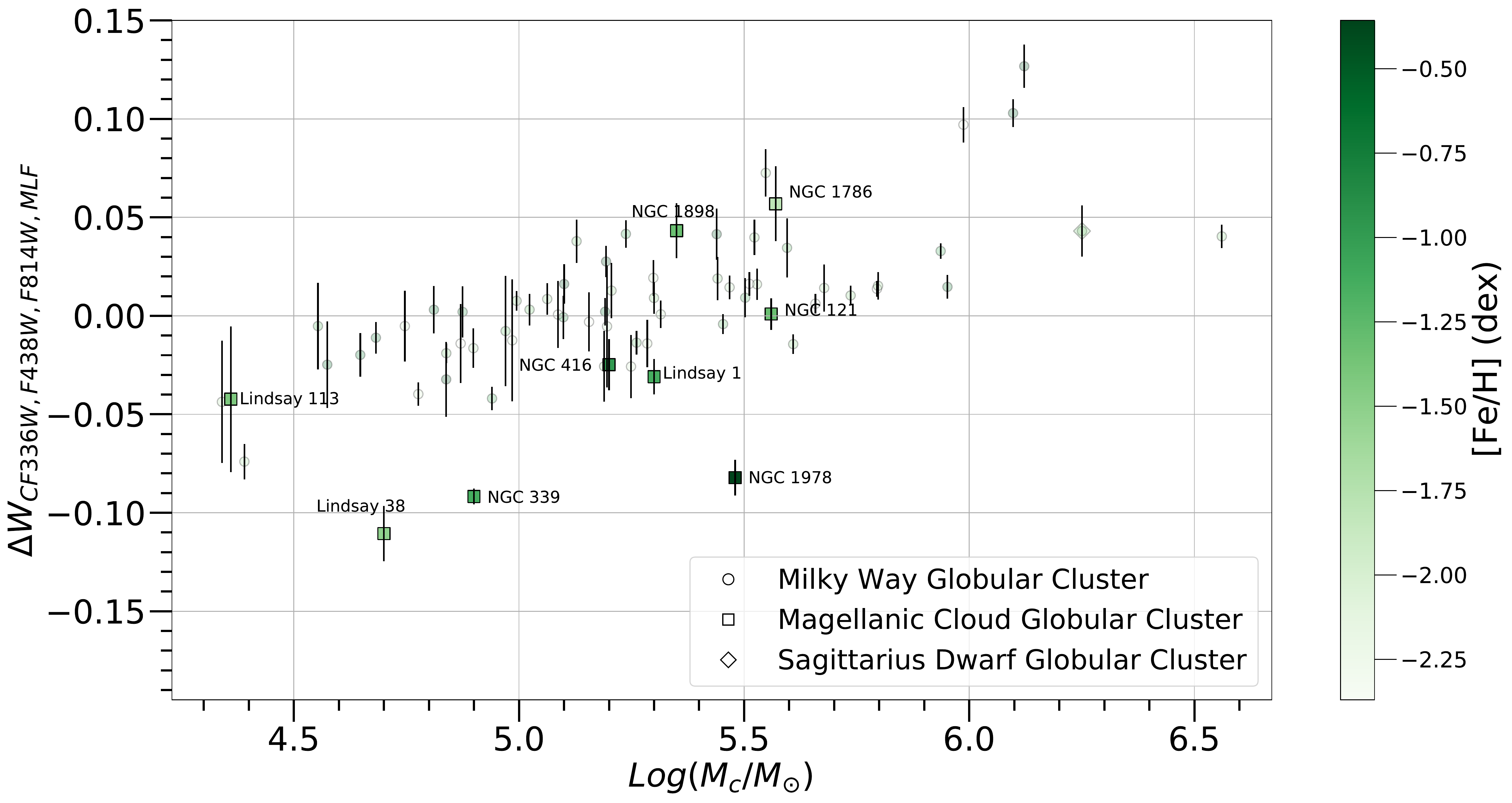}
	\caption{${\Delta}$\textit{W\textsubscript{CF336W,F438W,F814W,$M_{ini}$}}  and ${\Delta}$\textit{W\textsubscript{CF336W,F438W,F814W,MLF}}vs Log($M_{c}$/$M_{\sun}$) for the GCs. The colour of the cluster refers to the metallicity while the shape refers to the parent galaxy of the cluster. The vertical lines indicate the uncertainty associated with the RGB width of the clusters. The rest (apart from NGC 1786 and NGC 1898) of the data were obtained from \citet{lagioia2019}. Also marked are other MC clusters.}
	\label{fig:masss1}
\end{figure*}

\begin{figure*}
	
	\centering
	\includegraphics[width=1.05\linewidth,height=1.05\columnwidth]{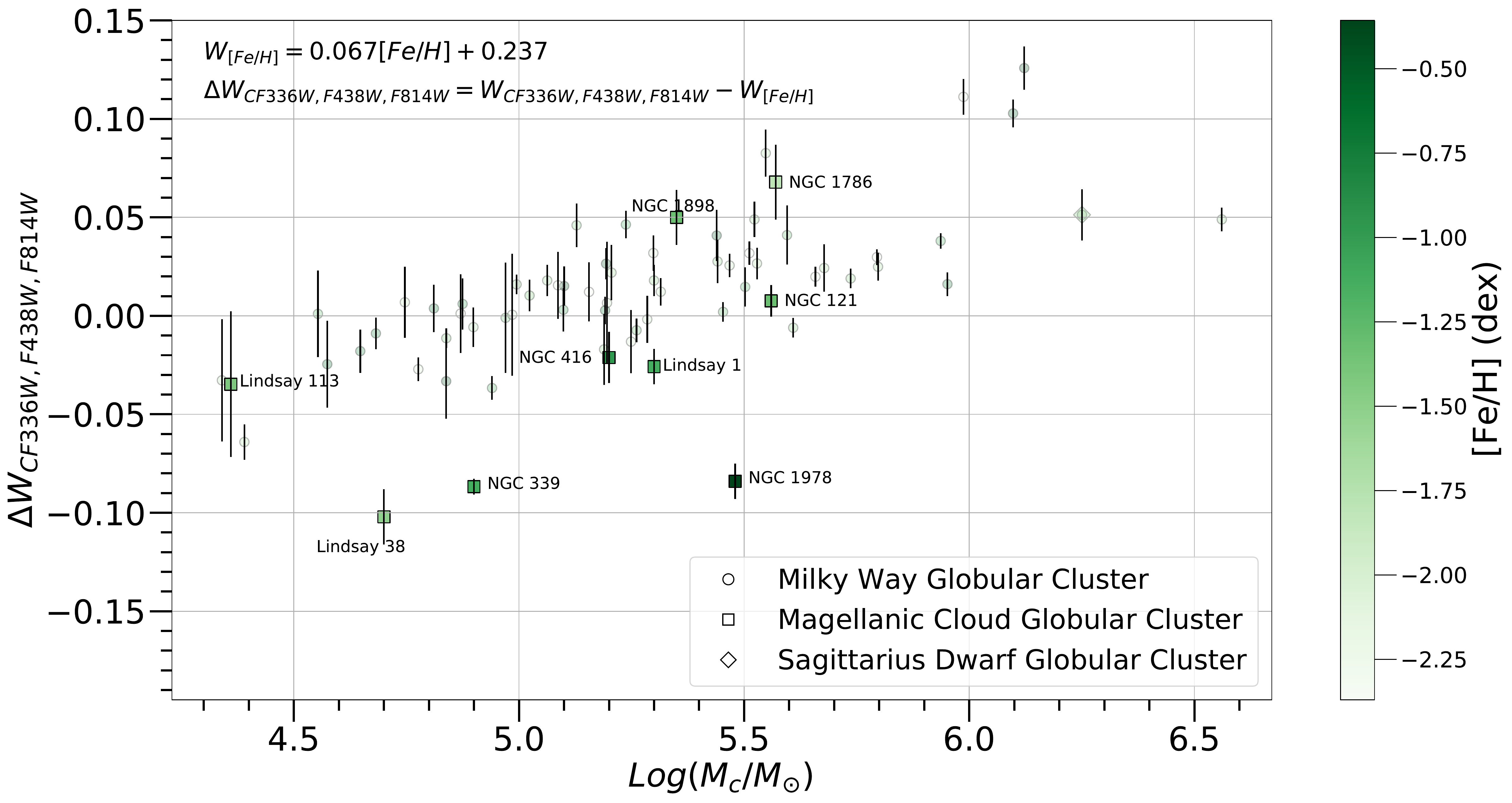}
	\centering
	\includegraphics[width=1.05\linewidth,height=1.05\columnwidth]{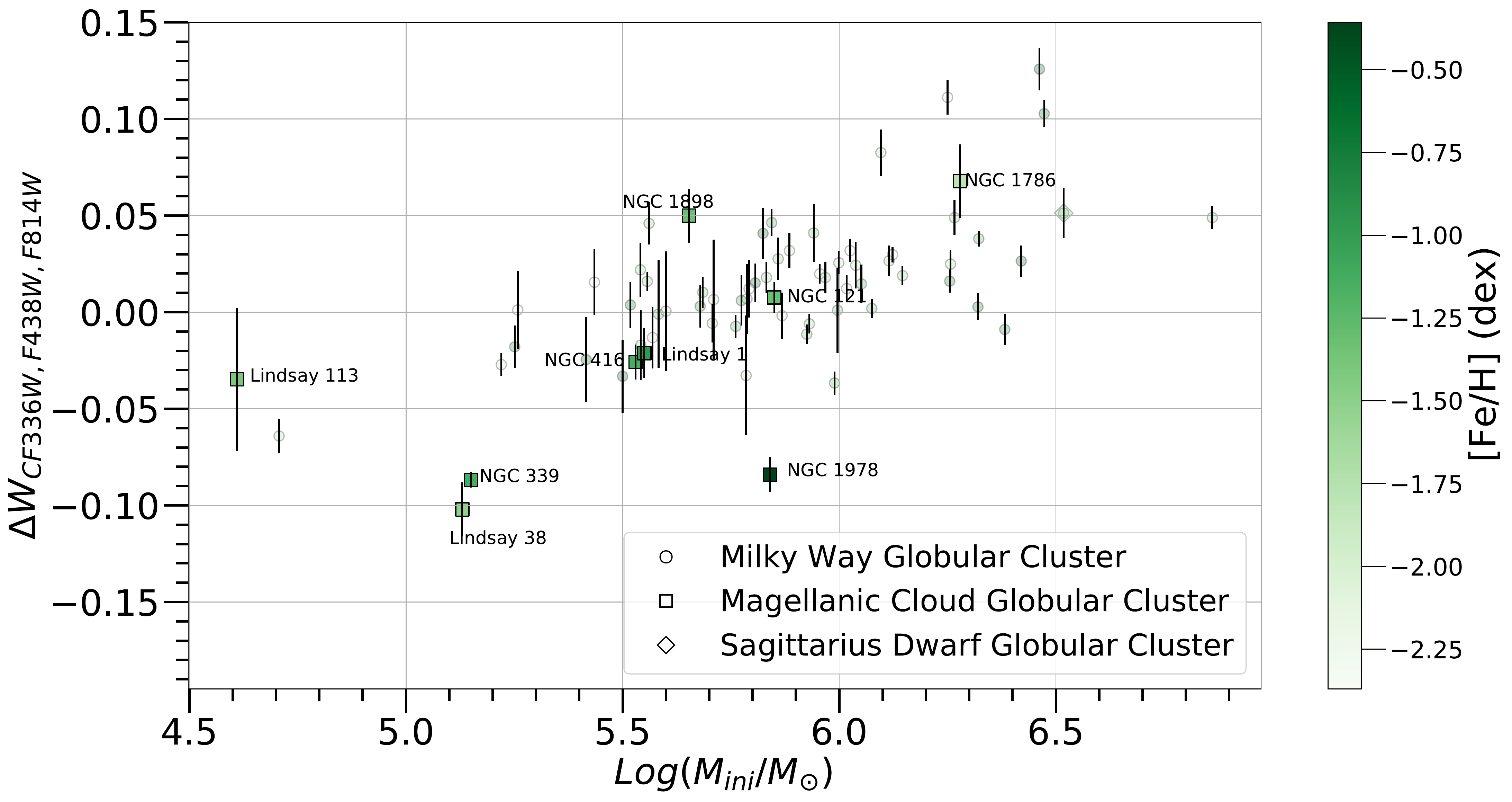}
	\caption{${\Delta}$\textit{W\textsubscript{CF336W,F438W,F814W}} vs Log($M_{c}$/$M_{\sun}$)  and Log($M_{ini}$/$M_{\sun}$)for the GCs. We can notice that NGC 1786 and NGC 1898 follow the same trend as the Galactic GCs.}
	\label{fig:masss}
\end{figure*}

\section{Effect of mass on RGB width} \label{section4}

\citet{lagioia2019} conducted a Spearman's rank correation test to find out the correlation between \textit{W\textsubscript{CF336W,F438W,F814W}} and the global parameters. The correlation is quantified by Spearman's rank correlation coefficient (R\textsubscript{s}) and the significance of the correlation is given by p-value. R\textsubscript{s} $=$ $-1$ represents perfect negative correlation while R\textsubscript{s} $=$ 1 represents perfect positive correlation. p-value represents the probability of finding a (R\textsubscript{s}) value equal or larger than the actual one. p $<0.05$ indicates strong evidence for correlation while p $<0.01$ indicates a highly significant correlation. \citet{lagioia2019} found a highly significant correlation between the \textit{W\textsubscript{CF336W,F438W,F814W}} and [Fe/H] with R\textsubscript{s} being 0.786, indicating a strong monotonicity with the intrinsic RGB width increasing with metallicity. There was also a highly significant and strong monotonic correlation (R\textsubscript{s}$=$0.924) between \textit{W\textsubscript{CF336W,F438W,F814W}} and \textit{W\textsubscript{CF275W,F336W,F438W}} \citep{milone2017} which shows that both filter combinations are almost equally sensitive to the variation of light elements in MPs. The R\textsubscript{s} values of the rest of the global parameters (slope of mass function, cluster core radius, age etc.) lie in the interval $\sim \pm 0.5$ which represents moderate to weak correlation. \citet{lagioia2019} took into account the effect of metallicity on the rest of the parameters to assess if the correlation with RGB width still holds.

\citet{lagioia2019} subtracted the contribution of metallicity to RGB width using then 58 Galactic GCs used by \citet{milone2018}. Since, massive clusters showed significant scatter in certain metallicity intervals, clusters with Log($M_{c}$/$M_{\sun}$) $<$ 5.22 were selected and a linear relation was computed for those clusters since those clusters show adequate linearity. The linear fit, which gives us the relation for RGB width contribution from metallicity, allows us to subtract the effect of metallicity on the RGB width of clusters. This width is called normalized RGB width (${\Delta}$\textit{W\textsubscript{CF336W,F438W,F814W}}). The linear relation is shown in Figure~\ref{fig:masss}. Spearman's rank correlation test was then carried out to find out the correlation of normalized RGB width with the global parameters. It is found that significant and strongly monotonic correlations emerge between intrinsic RGB width and mass and also other parameters such as central velocity dispersion and escape velocity, which are directly linked with mass. This indicates that metallicity and mass have a dominant effect on RGB width and that the observed variance of RGB width cannot be explained by only one of those two quantities. In the ${\Delta}$\textit{W\textsubscript{CF336W,F438W,F814W}} vs Log($M_{c}$/$M_{\sun}$) plot, it is found that the extragalactic GCs systematically lie beneath the Galactic GCs. There is a significant and monotonic correlation between ${\Delta}$\textit{W\textsubscript{CF336W,F438W,F814W}} and ${\Delta}$\textit{W\textsubscript{CF275W,F336W,F438W}}, the latter used by \citet{milone2017}, corroborating the earlier observation that both filter combinations are almost equally sensitive.

We used the updated catalogue of current masses and initial masses of Galactic GCs \citep{baumgardt2019} based on Gaia eDR3 release for our study. We performed Spearman's rank correlation test between the intrinsic RGB widths of Galactic GCs obtained by \citet{lagioia2019} and [Fe/H], $M_{c}$, $M_{ini}$, mass loss (ML), mass loss per Gyr (MLA), mass loss fraction (MLF) and age. The results of the correlation tests are tabulated in Table \ref{tab:table3}. \textit{W\textsubscript{CF336W,F438W,F814W}} scores a strong and significant correlation with [Fe/H], as observed by \citet{lagioia2019}. We see that the correlation between \textit{W\textsubscript{CF336W,F438W,F814W}} and $M_{c}$ is no longer significant (p>0.01) with the updated current mass data but we obtained a statistically significant correlation between \textit{W\textsubscript{CF336W,F438W,F814W}} and $M_{ini}$ (p<0.01, $ R\textsubscript{s}=0.400$). Significant correlations are also observed for MLA and ML. Since the equation of the weighted regression line is dependent on the clusters chosen for the fit and the cluster selection depends on the parameter of choice, we choose to derive two linear relations baseed on the selection of two parameters- $M_{ini}$ and MLF. In the [Fe/H] vs \textit{W\textsubscript{CF336W,F438W,F814W}} plot, we selected those Galactic GCs with initial mass less than the median of initial mass distribution of Galactic GCs since they show reasonable linearity, fit a linear relation to those GCs and obtained the metallicity contributed width (\textit{W}). The same procedure was followed to obtain linear relation by selecting clusters based on MLF. Figure \ref{fig:metsub1} shows [Fe/H] vs \textit{W\textsubscript{CF336W,F438W,F814W}} plot. The continuous red line represents the regression line while the selected clusters are marked with black filled circles. The regression line represents the width contributed by [Fe/H]. $R^{2}_{adj}$ is the adjusted determination coefficient and denotes the fraction of the selected clusters accounted for by the regression line. The clusters in the top panel are selected based on $M_{ini}$ while those in the bottom panel are selected based on MLF. The bottom panel shows adequate linearity among the selected clusters though there was no significant correlation between \textit{W\textsubscript{CF336W,F438W,F814W}} and MLF. We see that there is marked difference between the two slopes. This is important to consider because W increases with increase in slope and hence, the trends shown by RGB width after the subtraction of the effect of [Fe/H] could differ based on the linear relation used. We also derived linear relations using ML, MLA and $M_{c}$ but the regression line derived based on MLF and $M_{ini}$ had the lowest and highest slopes respectively. If there is a marked difference in the trend based on the slope of the regression line, it should be evident by comparing the results of these two scenarios. So, we subtracted \textit{W} of each cluster from \textit{W\textsubscript{CF336W,F438W,F814W}} to obtain the normalised RGB widths- ${\Delta}$\textit{W\textsubscript{CF336W,F438W,F814W,$M_{ini}$}} and ${\Delta}$\textit{W\textsubscript{CF336W,F438W,F814W,MLF}}. Figure \ref{fig:masss1} shows the plot of Log($M_{c}$/$M_{\sun}$) vs ${\Delta}$\textit{W\textsubscript{CF336W,F438W,F814W,$M_{ini}$}} and ${\Delta}$\textit{W\textsubscript{CF336W,F438W,F814W,MLF}}. We see that there is a difference in the trend of Galactic GCs whose Log($M_{c}$/$M_{\sun}$) is less than 5 between the two plots. Nevertheless there are no significant differences in the overall trend traced by the GCs and more importantly, no significant differences in the trend traced by the MC GCs with respect to the Galactic GCs which shows that the behaviour of GCs in this space is fairly stable. Since we have determined that there is no significant difference between the two scenarios, we think the choice of $M_{c}$ would be more appropriate since the estimates of $M_{c}$ are relatively much more reliable than $M_{ini}$ and to be consistent with \citet{lagioia2019}. We followed the same procedure to fit the regression line and estimate \textit{W} and subtracted it from  intrinsic RGB width to obtain the normalized RGB width (${\Delta}$\textit{W\textsubscript{CF336W,F438W,F814W}}). We then performed Spearman's correlation test between ${\Delta}$\textit{W\textsubscript{CF336W,F438W,F814W}} and the aforementioned parameters. The results are shown in Table \ref{tab:table3} and we see that new correlations emerge after the subtraction of the effect of metallicity. We see that there is a significant correlation between ${\Delta}$\textit{W\textsubscript{CF336W,F438W,F814W}} and all parameters associated with mass, among which $M_{c}$ scores the strongest correlation (R\textsubscript{s}$=$0.731), followed by $M_{ini}$. MLF, which showed no correlation with \textit{W\textsubscript{CF336W,F438W,F814W}}, shows a significant anticorrelation (R\textsubscript{s}$=-0.437$) with ${\Delta}$\textit{W\textsubscript{CF336W,F438W,F814W}}. The R\textsubscript{s} of MLA is almost the same before and after the subtraction of metallicity while that of ML increases marginally. The values are tabulated in Table \ref{tab:table2}.

 ${\Delta}$\textit{W\textsubscript{CF336W,F438W,F814W}} vs Log($M_{c}$/$M_{\sun}$)  and Log($M_{ini}$/$M_{\sun}$) graph is shown in Figure~\ref{fig:masss} but also Log(MLF), Log(MLA) and Log(ML) as shown in Appendix \ref{sec:plo}. Figure~\ref{fig:masss} shows the two clusters analysed in this study along with the cluster database used in \citet{lagioia2019}. We find that only two clusters-  NGC 339 and NGC 1978- lie below general trend of the globular clusters. There is no evidence for the presence of MPs in Lindsay 38 \citep{milone2020, martocchia2019} while Lindsay 113 may \citep{martocchia2019} or may not \citep{milone2020} be hosting MPs. While NGC 416 and Lindsay 1 occupy the lower end, NGC 121, NGC 1786 and NGC 1898 occupy the intermediate and higher ends of the general trend exhibited by the Galactic GCs. That is, five out of seven MC GCs hosting MPs follow the general trend exhibited by the Galactic GCs. The plots listed in the Appendix \ref{sec:plo} also corroborate this trend. This indicates that galaxy environment might only play a minor role in the formation of MPs in globular clusters. 

\section{Discussion} \label{section5}

\subsection{Effect of age on MPs}

In Figure~\ref{fig:masss}, all the three MC clusters occupying the intermediate and higher ends of the general trend exhibited by Galatic GCs are classicial GCs. Whether this indicates a possible role played by age or not remains to be seen. One important drawback is the underwhelming number of observations of MC GCs hosting MPs available in the filter combination used in this study. While the correlation tests didn't show any correlation between ${\Delta}$\textit{W\textsubscript{CF336W,F438W,F814W}} and age, it has to be taken into consideration that Galactic GCs are mostly classical GCs older than 10.5 Gyr. To explore the role of age, we need as many MC GCs across various age ranges as possible.

\subsubsection{Evidence suggesting the role of age}

There has been a lot of studies exploring the possible effect of age on the manifestation of MPs. \citet{martocchia2019} analysed the HST data of four MC GCs in F336W, F438W and F343N filters of WFC3/UVIS complemented with archival data in F555W and F814W filters of ACS/WFC to map the N spread of those clusters. Combined with the data and results of \citet{niederhofer2017a, niederhofer2017b} and \citet{martocchia2018b,martocchia2018a} totalling 16 MC GCs and 3 Galactic GCs, they found that the standard deviation of observed RGB widths of GCs in \textit{C\textsubscript{F336W,F438W,F343N}} colour and \textit{C\textsubscript{F343N,F438W,F814W}} pseudo-colour increase with their age, indicating that cluster age might play an important role alongside metallicity and mass. They also note that cluster age and cluster mass work simultaneously based on their analysis in which Lindsay 38 which has a similar age to NGC 416 and NGC 339, has a significantly smaller RGB width and lower mass than those two clusters. This is a possible explanation for the systematic low RGB widths of non-classical MC GCs in Figure~\ref{fig:masss}. \citet{saracino2020} investigated NGC 2121 ($~$2.5 Gyr) and NGC 1783 ($~$1.5 Gyr) using the pseudo-colour \textit{C\textsubscript{F275W,F343N,F435/8W}} and magnitude m\textsubscript{F275W,F814W} to construct `chromosome maps'(ChMs) \citep{milone2017}. They find a clear correlation between the RGB width in the pseudo-colour C\textsubscript{F275W,F343N,F435/8W} and age of the cluster. But they caution that the RGB width can't directly be linked with N abundance variations without accounting for the effects of first dredge-up \citep{salaris2020} which makes the determined values of RGB width lower limits to the actual values and that the effects are more pronounced for young GCs than old/intermediate ones.

\subsubsection{Evidence contradicting the role of age}

\citet{milone2020} calculated the fraction of 1G stars in various SMC and LMC clusters including the six SMC clusters and one LMC cluster investigated by \citet{lagioia2019}. They don't find any significant difference between the 1G fractions of Galactic GCs and ages, though they find that the 1G fractions of NGC 339 and NGC 1978 are significantly higher than those of Galactic GCs. This is corroborated by \citet{dondoglio2021} who find no correlation between the 1G fractions of GCs hosting MPs and their ages. But, as stated by \citet{milone2020}, no strong conclusion can be drawn from these findings since direct comparison of the values of Galactic GCs and MC GCs is rendered difficult owing to their different masses and ages.

\begin{figure*}
	
	\centering
	\includegraphics[width=1\linewidth,height=0.594\columnwidth]{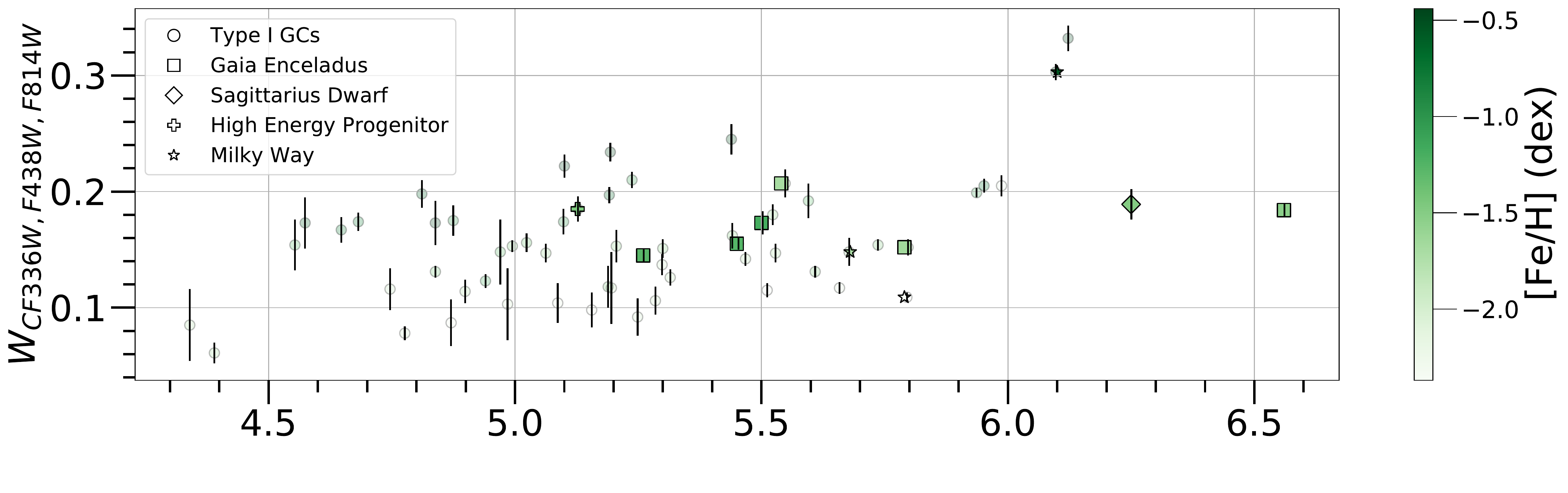}
	\centering
	\includegraphics[width=1\linewidth,height=0.594\columnwidth]{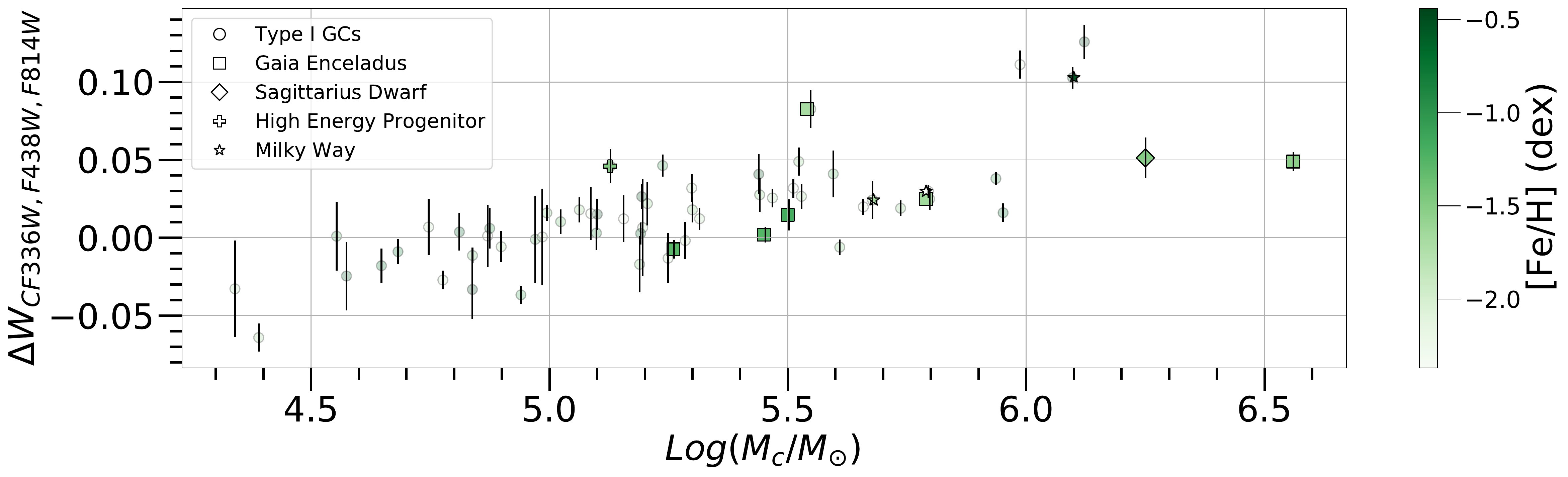}
	\caption{\textit{W\textsubscript{CF336W,F438W,F814W}} and ${\Delta}$\textit{W\textsubscript{CF336W,F438W,F814W}} vs Log($M_{c}$/$M_{\sun}$) for the Galactic GCs based on their type and progenitor. Different non-circular symbols represent different progenitors of Type II GCs. See section 3.2.5 in \citet{massari2019} for definitions of High Energy Progenitor and Low Energy Progenitor (see figure \ref{fig:milone2}).}
	\label{fig:milone1}
\end{figure*}

\begin{figure*}
	
	\centering
	\includegraphics[width=1\linewidth,height=0.594\columnwidth]{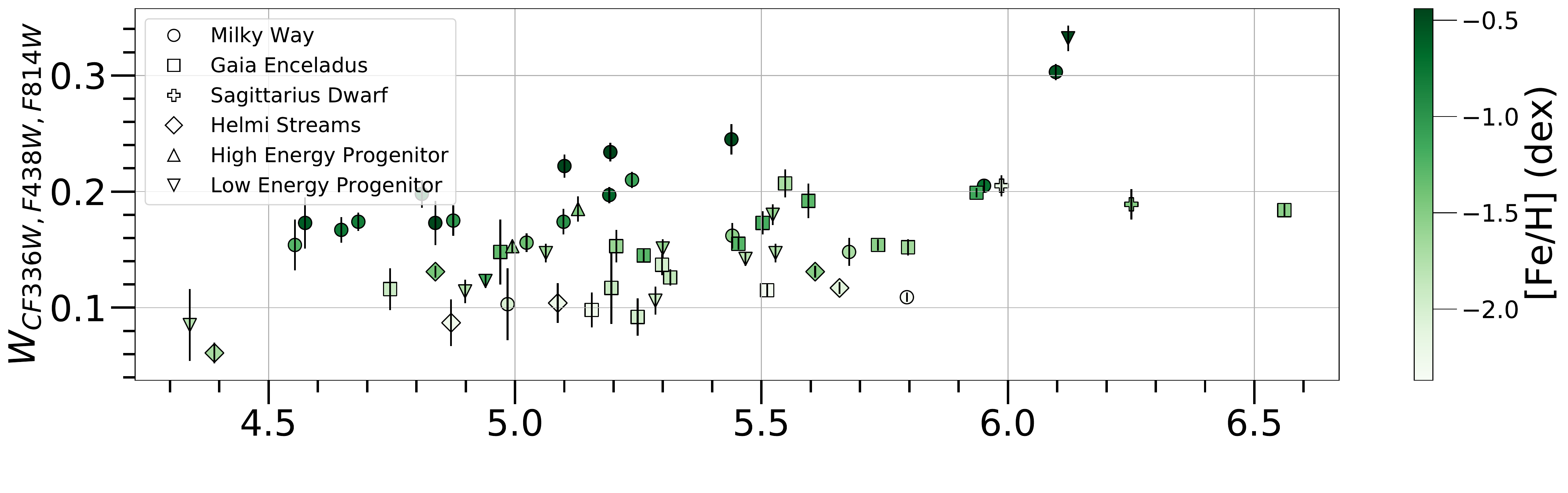}
	\centering
	\includegraphics[width=1\linewidth,height=0.594\columnwidth]{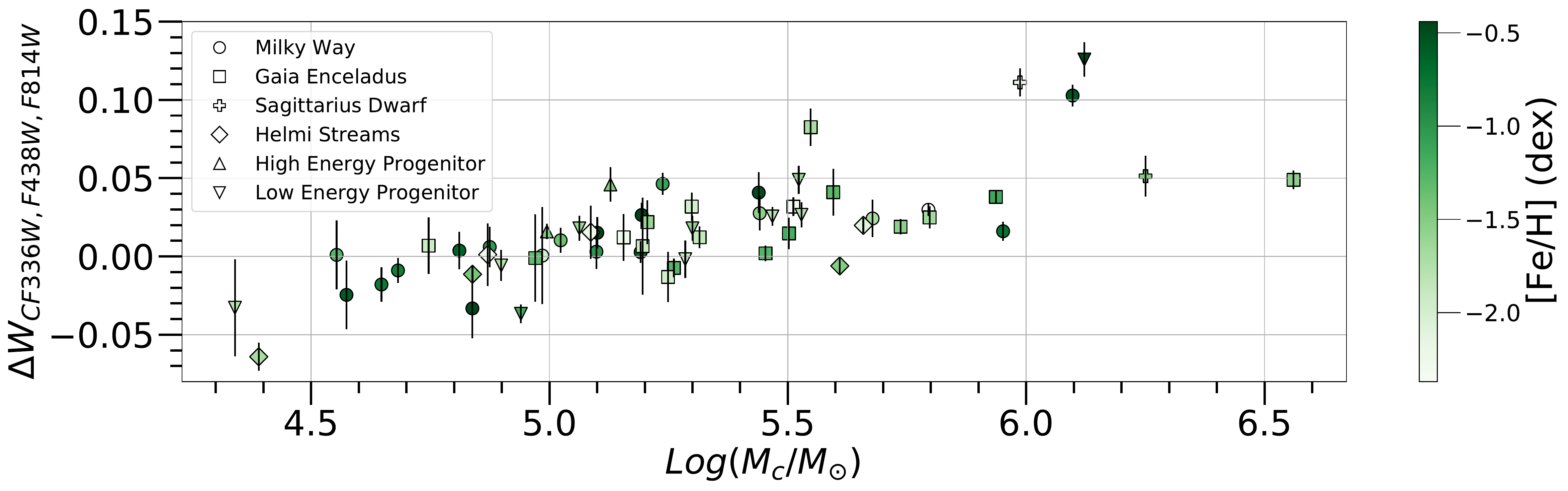}
	\caption{\textit{W\textsubscript{CF336W,F438W,F814W}} and ${\Delta}$\textit{W\textsubscript{CF336W,F438W,F814W}} vs Log($M_{c}$/$M_{\sun}$) for the Galactic GCs based on their progenitor. Different symbols represent different progenitors.}
	\label{fig:milone2}
\end{figure*}
\subsection{Effect of host galaxy on MPs}

\subsubsection{Evidence suggesting the role of host galaxy}

\citet{milone2017} and \citet{marino2019} classify GCs into two types- Type I and Type II- based on the ChMs of 58 GCs. Type I GCs are those which exhibit only a single sequence of 1G and 2G populations while Type II GCs exhibit multiple sequences of 1G and 2G populations with star-to-star variations in heavy elements like Fe and s-process elements with the two types present in 5:1 ratio. There are totally 13 known Type II GCs. \citet{massari2019} and \citet{milone2020} find that six (or possibly seven) of these clusters are found in a very specific region of the integrals of motion space and hence conclude that these clusters are possibly associated with the same parent galaxy, likely to be \textit{Gaia}-Enceladus. \citet{milone2020} concludes that since Type II GCs are not found in Magellanic Clouds and six of the known Galatic Type II GCs are associated with a single parent galaxy, their formation  is possibly dependent on their host galaxy.

At least four of the 13 known Type II GCs are linked with main bulge and main disc component of the Milky Way, hence we know Type II GCs are not exclusively of extragalactic origin. Classical MC GCs have similar metallicity spreads as Galactic GCs \citep{piatti2018}. The mass of \textit{Gaia}-Enceladus during collision with Milky Way was similar to SMC and based on the abundances of low metallicity stars of LMC, it could have been similar to LMC in its formative years \citep{helmi2018,hayes2018,marel2009}. The star formation rate of SMC is similar to that of \textit{Gaia}-Enceladus with more than one star formation event. \citep{sara2014,alvar2018}. These similarities between the galaxy environments make it reasonable to think that the apparent absence of Type II GCs in MCs could be due to the relatively lesser knowledge of MCs compared to Milky Way and that Type II GCs could be detected in MCs with further exploration. Though the environment of \textit{Gaia}-Enceladus could have been relatively more favourable for the formation of Type II GCs, given that 6 of the 13 known Type II GCs are linked to it, it may not have played a dominant role in the origin and manifestation of MPs given that the incidence of Type I GCs is far greater than that of Type II GCs.

We plotted \textit{W\textsubscript{CF336W,F438W,F814W}} and ${\Delta}$\textit{W\textsubscript{CF336W,F438W,F814W}} vs Log($M_{c}$/$M_{\sun}$) of Galactic GCs shape-coded for the type of GCs, as illustrated in figure \ref{fig:milone1} and the progenitor of GCs, as illustrated in figure \ref{fig:milone2} (see Table 1 in \citealt{massari2019}). Figure \ref{fig:milone1} neither shows any significant difference between the trends traced by Type I and Type II GCs nor by Type II GCs of different progenitors. Top panel of figure \ref{fig:milone2} shows that \textit{in-situ} Galactic GCs systematically have higher \textit{W\textsubscript{CF336W,F438W,F814W}} with higher [Fe/H] but the difference vanishes once the effect of [Fe/H] is subtracted and behaves similar to GCs of other progenitors, as shown in the lower panel of \ref{fig:milone2}. The trend is replicated when we replace $M_{c}$ with $M_{ini}$ as illustrated in Appendix \ref{sec:dp}. This corroborates our finding that galaxy environment may not play a dominant role in the origin and manifestation of MPs.

\subsubsection{Evidence contradicting the role of host galaxy}

\citet{saracino2019} investigated Lindsay 1 using ChM in \textit{${\Delta}$C\textsubscript{F275W,F336W,F438W}} pseudo-colour and \textit{m\textsubscript{F275W}-m\textsubscript{F814W}} colour and compared it with the Galactic GC NGC 288. Lindsay 1 and NGC 288 have similar metallicities but belong to different age ranges and galactic environments. They observe that the 1G and 2G populations have similar separations and their Helium enrichments are also similar \citep{chantereau2019,milone2018}. Hence, they conclude that the phenomenon of MPs is independent of age and galaxy environment. If this is true, the apparent increase in \textit{${\Delta}$W\textsubscript{CF336W,F438W,F814W}} of MC clusters with age in Figure \ref{fig:masss} may disappear with an increased sample of MC GCs in F336W, F438W and F814W filters.

\section{Conclusion} \label{section6}

Research over the last few decades has made it abundantly clear that globular clusters consist of multiple stellar populations rather than single stellar populations, thanks to advanced instrumentation. But the origin of MPs remains unclear. One aspect of the problem is to determine whether the galaxy environment plays a role in the formation of MPs. We compared GCs from Milky Way and LMC in the same age range to investigate if their RGB widths show any systematic variations from each other. To that extent, we analysed two classical LMC GCs namely, NGC 1786 and NGC 1898, and compared them with the available data on classical Galactic GCs. We used DOLPHOT to perform PSF photometry on the clusters, followed by selecting the stars from the photometry output by placing a few quality cuts, statistically removing the field stars and selecting only those stars that lie along the fiducial line of the cluster. The RGB width was then determined followed by estimating the uncertainty using bootstrapping and subtracting the photometric error using ASs test to get the intrinsic RGB width. As \citet{lagioia2019} found a strong monotonic correlation between the intrinsic RGB width and metallicity, the effect of metallicity was subtracted by computing a linear relation for those clusters with Log($M_{c}$/$M_{\sun}$) $<$ 5.22 since they showed adequate linearity. The results showed that significant and strong monotonic correlations arise between the metallicity subtracted RGB width and the total mass of the cluster. So we followed the same method and computed the normalized RGB width and plotted it against the mass parameters listed in Table \ref{tab:table2} of the two analysed clusters in addition to the cluster database used by \citet{lagioia2019}. We find that NGC 1786 and NGC 1898 follow the upper end of the general trend exhibited by Galactic GCs and Galactic GCs from different progenitors follow the same general trend as one another, thereby providing first evidence from an analysis of a large sample of Galactic and MC GCs that galaxy environment may not play a dominant role in the formation of MPs. But it should be taken into account that the sample size of classical MC GCs is small. There is a need for a much larger sample of classical LMC GCs in the filter combination used in this study and that of \citet{martocchia2019} and \citet{saracino2020} to verify our results and also explore the role of metallicity, mass and age in formation of MPs in greater detail. 

\section*{Acknowledgements}

FN acknowledges support from the European Research Council (ERC) under the European Horizon 2020 research and innovation programme (grant agreement no. 682115).

Based on observations made with the NASA/ESA Hubble Space Telescope, obtained from the data archive at the Space Telescope Science Institute. STScI is operated by the Association of Universities for Research in Astronomy, Inc. under NASA contract NAS 5-26555.

This research made use of Astropy, a community-developed core Python package for Astronomy \citep{astropy2013, astropy2018}, Matplotlib \citep{matplotlib}, NumPy \citep{numpyoliphant, numpywalt}, Scikit-learn \citep{scikit}, Statsmodels \citep{statsmodels}, Pandas \footnote{\url{https://doi.org/10.5281/zenodo.3509134}}\citep{pandaspaper} and SciPy \citep{scipy}. We thank In-Sung Jang for his tips and suggestions regarding the use of DOLPHOT. We thank Holger Baumgardt for graciously and promptly providing us with the initial cluster mass data of Galactic GCs. We thank Dalal El Youssoufi for her tips on manuscript editing. We thank the anonymous referee for their useful comments and suggestions that helped to improve the paper.

\section*{Data availability}

The HST imaging data are publicly available and can be retrieved from the MAST archive (\url{https://archive.stsci.edu/hst/}). The final data products (photometric and AS catalogues) underlying this article will be shared on reasonable request to the corresponding author.






\bibliographystyle{mnras}
\bibliography{final_manuscript.bib}


\appendix
\section{Normalized RGB width vs mass parameters}
\label{sec:plo}
\begin{figure*}
	
	\centering
	\includegraphics[width=1.05\linewidth]{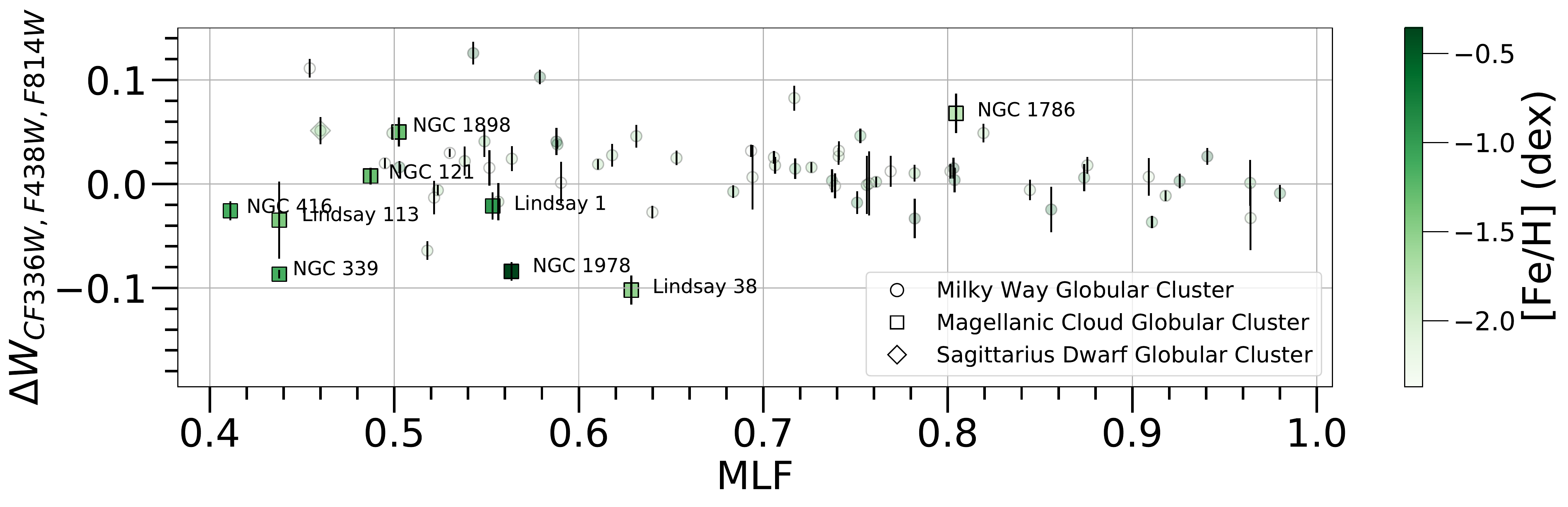}
	\caption{${\Delta}$\textit{W\textsubscript{CF336W,F438W,F814W}} vs MLF}
	\label{fig:masss2}
\end{figure*}

\begin{figure*}
	
	\centering
	\includegraphics[width=1.05\linewidth]{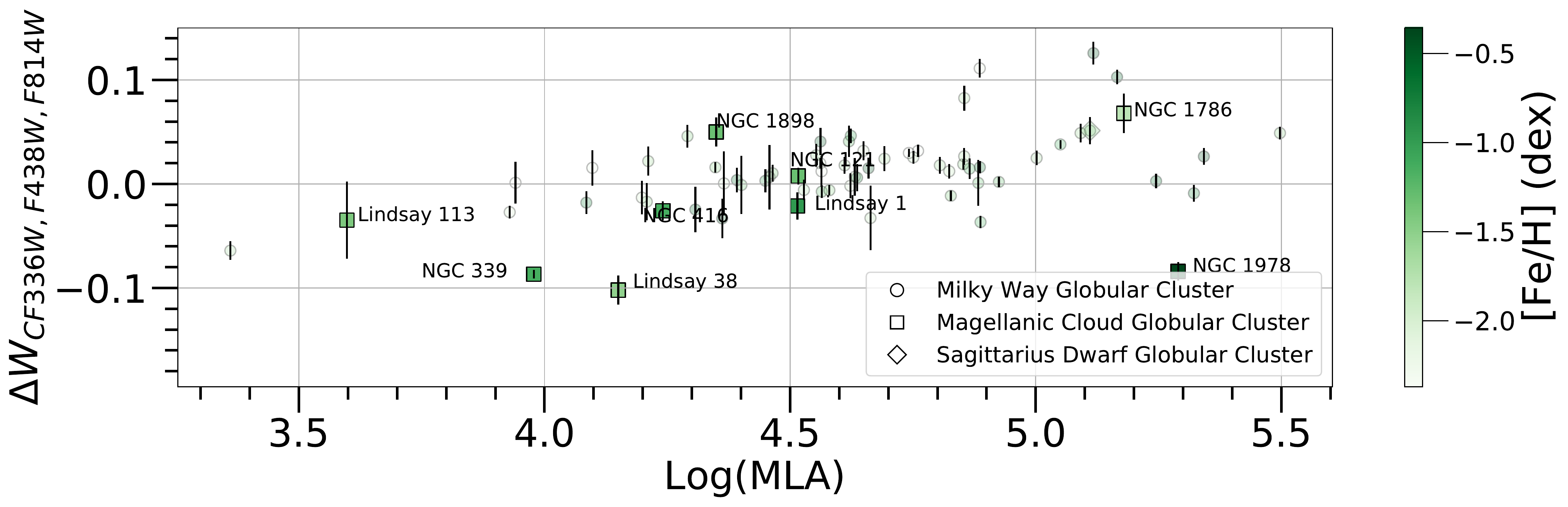}
	\caption{${\Delta}$\textit{W\textsubscript{CF336W,F438W,F814W}} vs Log(MLA)}
	\label{fig:masss3}
\end{figure*}

\begin{figure*}
	
	\centering
	\includegraphics[width=1.05\linewidth]{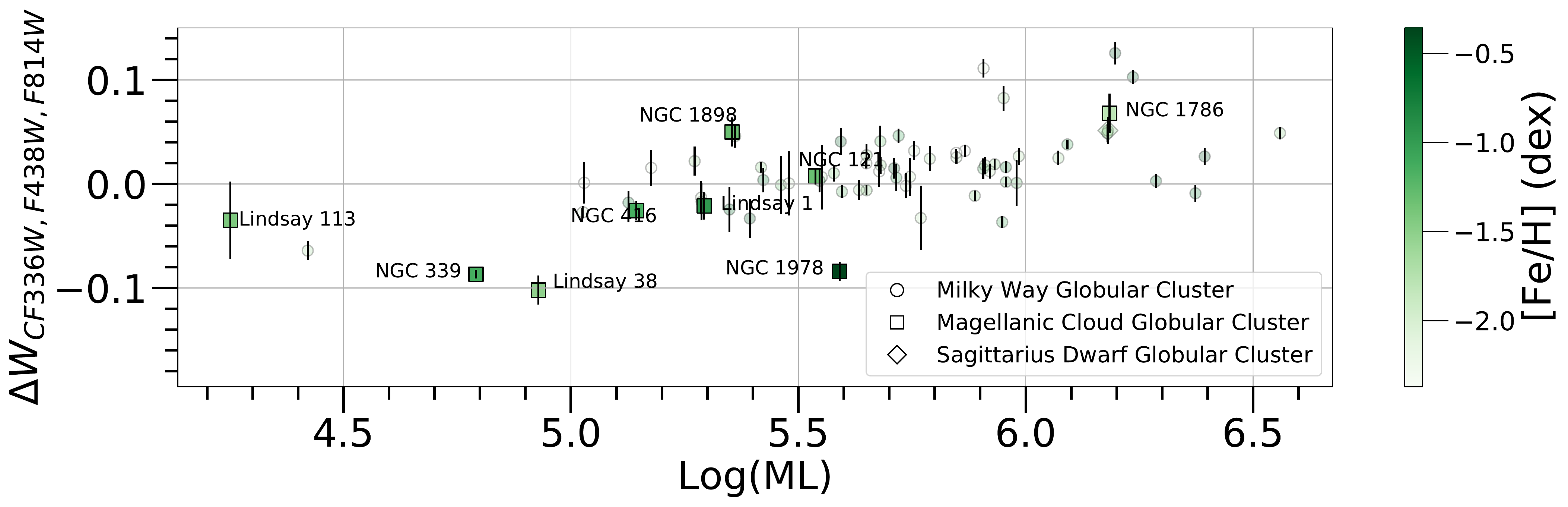}
	\caption{${\Delta}$\textit{W\textsubscript{CF336W,F438W,F814W}} vs Log(ML)}
	\label{fig:masss4}
\end{figure*}

\section{Normalized RGB width vs Initial mass for Galactic GCs of different types and progenitors}

\label{sec:dp}

\begin{figure*}
	
	\centering
	\includegraphics[width=1\linewidth,height=0.6\columnwidth]{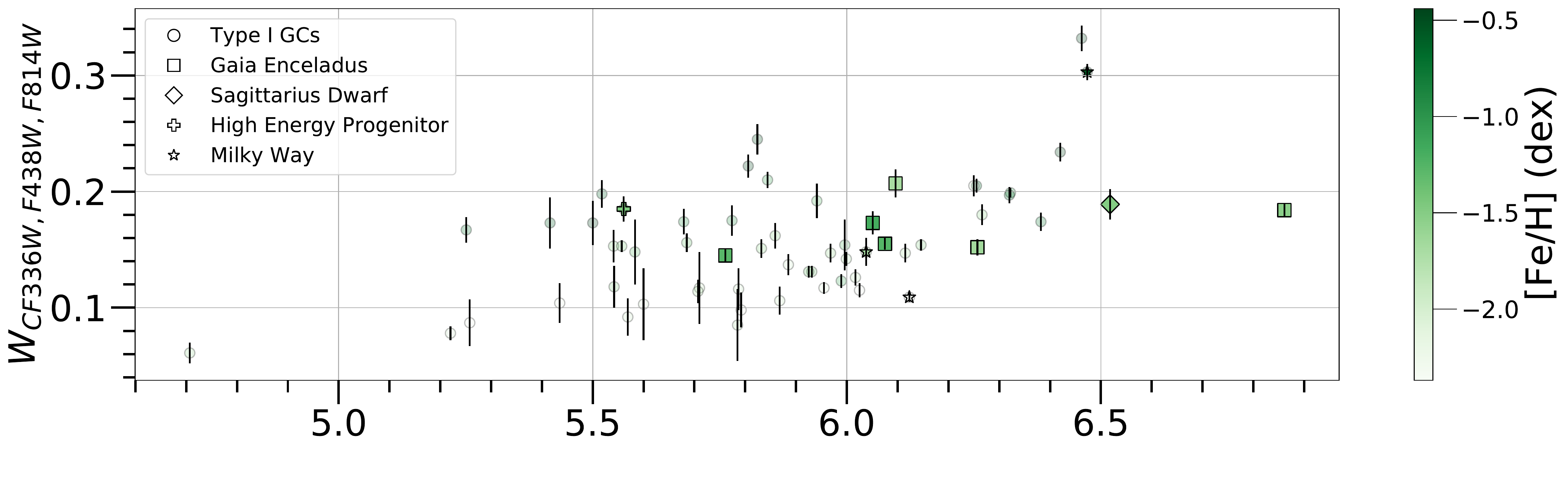}
	\centering
	\includegraphics[width=1\linewidth,height=0.6\columnwidth]{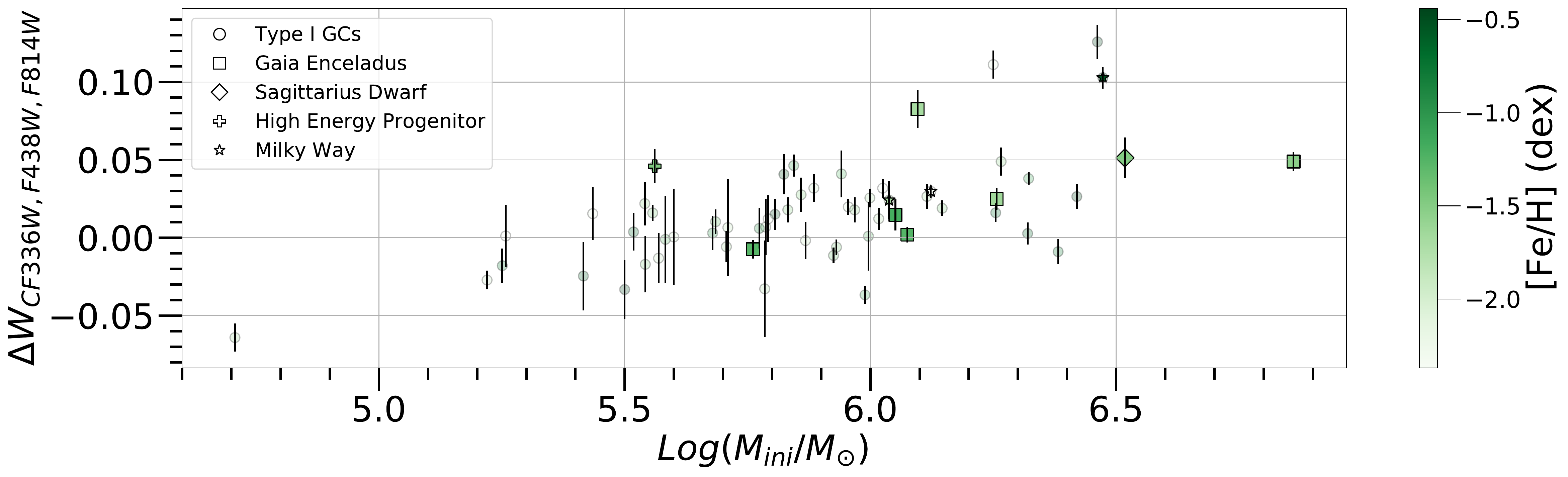}
	\caption{\textit{W\textsubscript{CF336W,F438W,F814W}} and ${\Delta}$\textit{W\textsubscript{CF336W,F438W,F814W}} and vs Log($M_{ini}$/$M_{\sun}$) for the Galactic GCs based on their type and progenitor. Different non-circular symbols represent different progenitors of Type II GCs.}
	\label{fig:milone3}
\end{figure*}

\begin{figure*}
	
	\centering
	\includegraphics[width=1\linewidth,height=0.6\columnwidth]{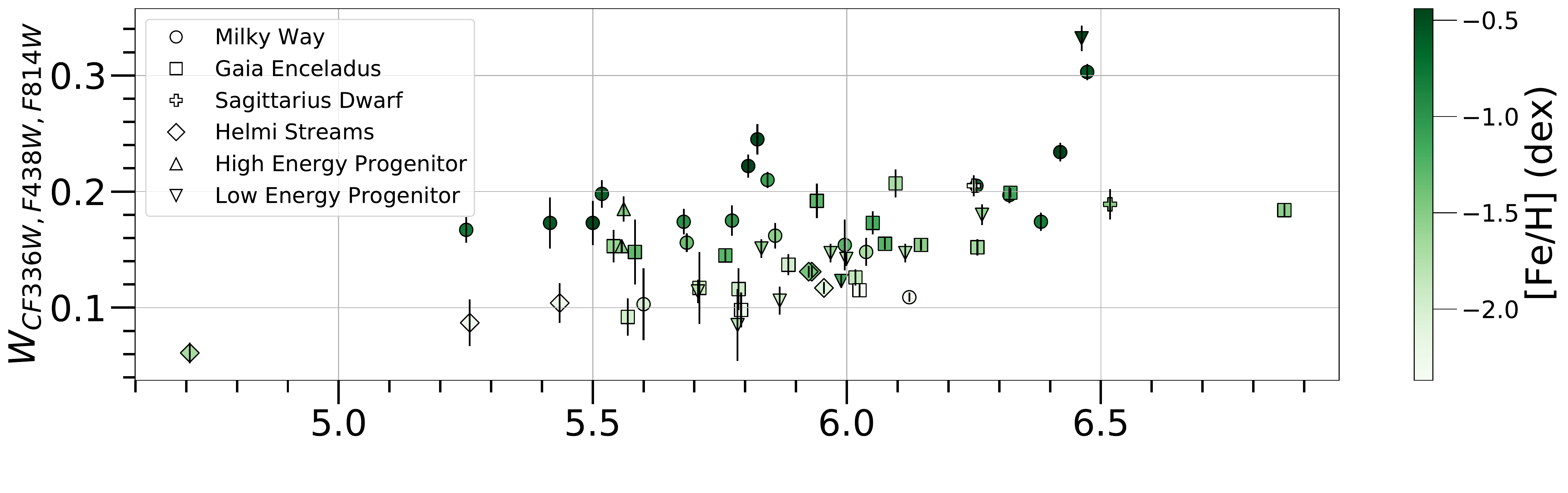}
	\centering
	\includegraphics[width=1\linewidth,height=0.6\columnwidth]{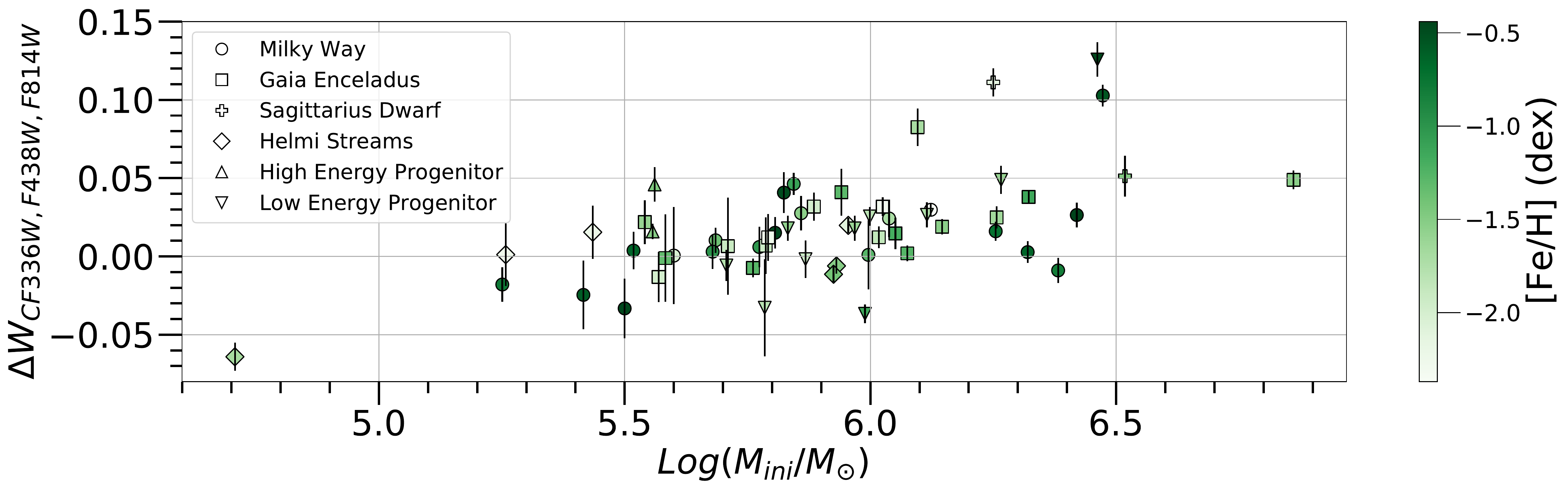}
	\caption{\textit{W\textsubscript{CF336W,F438W,F814W}} and ${\Delta}$\textit{W\textsubscript{CF336W,F438W,F814W}} vs Log($M_{ini}$/$M_{\sun}$) for the Galactic GCs based on their progenitor. Different symbols represent different progenitors.}
	\label{fig:milone4}
\end{figure*}



\bsp	
\label{lastpage}
\end{document}